\documentclass[useAMS]{aa}

\usepackage{times}
\usepackage{footmisc}
\usepackage{graphicx,multicol}
\usepackage{caption}
\usepackage{subcaption}
\usepackage{natbib}

\usepackage{amsmath,amsfonts,bm}
\usepackage{amsthm}
\bibpunct{(}{)}{;}{a}{}{,} 
\usepackage{txfonts}
\usepackage{relsize}
\usepackage[draft=False, colorlinks=true, citecolor=RoyalBlue, urlcolor=magenta, linkcolor=RoyalBlue, linktoc=all]{hyperref}
\usepackage{cleveref}
\usepackage[section]{placeins}
\usepackage[switch, mathlines]{lineno}
\usepackage[dvipsnames]{xcolor}

\newcommand\delad{\nabla_{\mathrm{ad}}}
\newcommand\delact{\nabla_{\mathrm{act}}}

\newcommand\Tstrut{\rule{0pt}{2.9ex}}         
\newcommand\Bstrut{\rule[-1.2ex]{0pt}{0pt}}   
\newcommand\TBstrut{\Tstrut\Bstrut}           

\newcommand{\cdbox}[1]{%
	{\color{blue}%
		\dbox{\color{black}#1}}%
}
\usepackage{dashbox,framed,color,ocg-p}
\newcommand{\ToggleLayer}[2]{%
	\leavevmode
	\pdfstartlink user {
		/Subtype /Link
		/Border [0 0 0]%
		/A <<
		/S/JavaScript
		/JS (
		var aOCGs = this.getOCGs(), Layer;
		var Layers = "#1".split(","), Active = -1, i, l;
		for (l=0; l<Layers.length; l++) {
			Layer = Layers[l];
			for (i=0; aOCGs && i<aOCGs.length; i++) {
				if (aOCGs[i].state && aOCGs[i].name == Layer) {
					Active = l;
					aOCGs[i].state = false;
				}
			}
			if (Active >= 0) break;
		}
		if (Active == -1) {
			for (l=0; l<Layers.length; l++) {
				if (Layers[l] == "") Active = l;
			}
		}
		Active = Active + 1;
		if (Active == Layers.length) Active = 0;
		Layer = Layers[Active];
		for (i=0; aOCGs && i<aOCGs.length; i++) {
			if (aOCGs[i].name == Layer) aOCGs[i].state = true;
		}
		)
		>>
	}#2%
	\pdfendlink
}

\begin{document}

\title{Birth of convective low-mass to high-mass second Larson cores}

\author{Asmita Bhandare$^{1,}${\thanks{Member of the International Max-Planck Research School for Astronomy and Cosmic Physics at the University of Heidelberg (IMPRS-HD), Germany} } 
\and Rolf Kuiper$^{2,1}$ 
\and Thomas Henning$^1$
\and Christian Fendt$^1$
\and Mario Flock$^1$ 
\and Gabriel-Dominique Marleau$^{2,3,1}$ 
}

\institute{$^1$ Max-Planck-Institut f\"ur Astronomie, K\"onigstuhl 17, D-69117 Heidelberg, Germany \\
\email{bhandare@mpia.de} \\
$^2$ Institut f\"ur Astronomie und Astrophysik, Universit\"at T\"ubingen, Auf der Morgenstelle 10, D-72076 T\"ubingen, Germany	\\
$^3$ Physikalisches Institut, Universit\"at Bern, Gesellschaftsstr.~6, CH-3012 Bern, Switzerland
}
 
\date{Received 31 October 2019 / Accepted 22 April 2020}

\abstract
{Stars form as an end product of the gravitational collapse of cold, dense gas in magnetized molecular clouds. This fundamentally multi-scale scenario occurs via the formation of two quasi-hydrostatic Larson cores and involves complex physical processes, which require a robust, self-consistent numerical treatment.}
{The primary aim of this study is to understand the formation and evolution of the second hydrostatic Larson core and the dependence of its properties on the initial cloud core mass.}
{We used the PLUTO code to perform high resolution, one- and two-dimensional radiation hydrodynamic (RHD) core collapse simulations. We include self-gravity and use a grey flux-limited diffusion approximation for the radiative transfer. Additionally, we use for the gas equation of state density- and temperature-dependent thermodynamic quantities (heat capacity, mean molecular weight, etc.) to account for the effects such as dissociation of molecular hydrogen, ionisation of atomic hydrogen and helium, and molecular vibrations and rotations. Properties of the second core are investigated using one-dimensional studies spanning a wide range of initial cloud core masses from 0.5~$M_{\odot}$ to 100~$M_{\odot}$. Furthermore, we expand to two-dimensional (2D) collapse simulations for a selected few cases of 1~$M_{\odot}$, 5~$M_{\odot}$, 10~$M_{\odot}$, and 20~$M_{\odot}$. We follow the evolution of the second core for $\geq$ 100 years after its formation, for each of these non-rotating cases.}	
{Our results indicate a dependence of several second core properties on the initial cloud core mass. Molecular cloud cores with a higher initial mass collapse faster to form bigger and more massive second cores. The high-mass second cores can accrete at a much faster rate of $\approx 10^{-2} M_{\odot}~yr^{-1}$ compared to the low-mass second cores, which have accretion rates as low as $10^{-5} M_{\odot}~yr^{-1}$. For the first time, owing to a resolution that has not been achieved before, our 2D non-rotating collapse studies indicate that convection is generated in the outer layers of the second core, which is formed due to the gravitational collapse of a 1~$M_{\odot}$ cloud core. Additionally, we find large-scale oscillations of the second accretion shock front triggered by the standing accretion shock instability (SASI), which has not been seen before in early evolutionary stages of stars. We predict that the physics within the second core would not be significantly influenced by the effects of magnetic fields or an initial cloud rotation.}
{In our 2D RHD simulations, we find convection being driven from the accretion shock towards the interior of the second Larson core. This supports an interesting possibility that dynamo-driven magnetic fields may be generated during the very early phases of low-mass star formation.}{}


\authorrunning{A. Bhandare et al.}

\maketitle

\graphicspath{{./graphics/}{./graphics2/}}


\section{Introduction}
\label{sec:intro}

Magnetized molecular clouds are known to be the birthplace of stars, which form by the gravitational collapse of dense, gaseous, and dusty cores within these clouds. Zooming in on the smallest scales in order to understand the complex physical processes such as hydrodynamics, radiative transfer, phase transition (in particular hydrogen dissociation), and magnetic fields has been challenging both theoretically and observationally \citep[e.g.][]{Nielbock2012, Launhardt2013, Dunham2014}. Among various fundamental questions which remain to be answered are the comprehensive characterisation of the core collapse \citep[see detailed reviews by][]{Larson2003,Mckee2007,Inutsuka2012, Teyssier2019, Pudritz2019}. It is thus crucial to unravel the role of various physical mechanisms involved during the transition of a molecular cloud core (i.e.~pre-stellar core) to the hydrostatic Larson cores, to understand how stars form. 

Several numerical studies using both grid-based \citep[][and references therein]{Bodenheimer1968, Winkler1980a, Winkler1980b, Stahler1980a, Stahler1980b, Stahler1981, Masunaga1998, Masunaga2000, Tomida2010b, Commercon2011, Vaytet2012, Tomida2013, Vaytet2013, Vaytet2017, Bhandare2018, Vaytet2018} and smoothed particle hydrodynamics (SPH) methods \citep{Bate1998, Whitehouse2006, Stamatellos2007, Bate2014, Wurster2018} have deduced the formation of a protostar to be a two-step process during which the non-homologous collapse leads to the presence of two quasi-hydrostatic cores, famously known as Larson's cores \citep{Larson1969}.  

The initially isothermal, optically thin cloud core collapses under its own gravity due to efficient cooling via thermal emission from dust grains. The collapse may be initiated either by the effects of non-ideal magnetohydrodynamics \mbox{\citep[e.g.][]{Shu1987, Mouschovias1991}} or the dissipation of turbulence, which reduces the effective sound speed in pre-stellar cores \citep[e.g.][]{Nakano1998}. It can also be triggered by an external shock wave crossing a previously stable cloud \citep{Masunaga2000} or a marginally stable Bonnor-Ebert sphere may also undergo a collapse. 

\begin{figure}[tp]
	\centering
	\includegraphics[width=\linewidth]{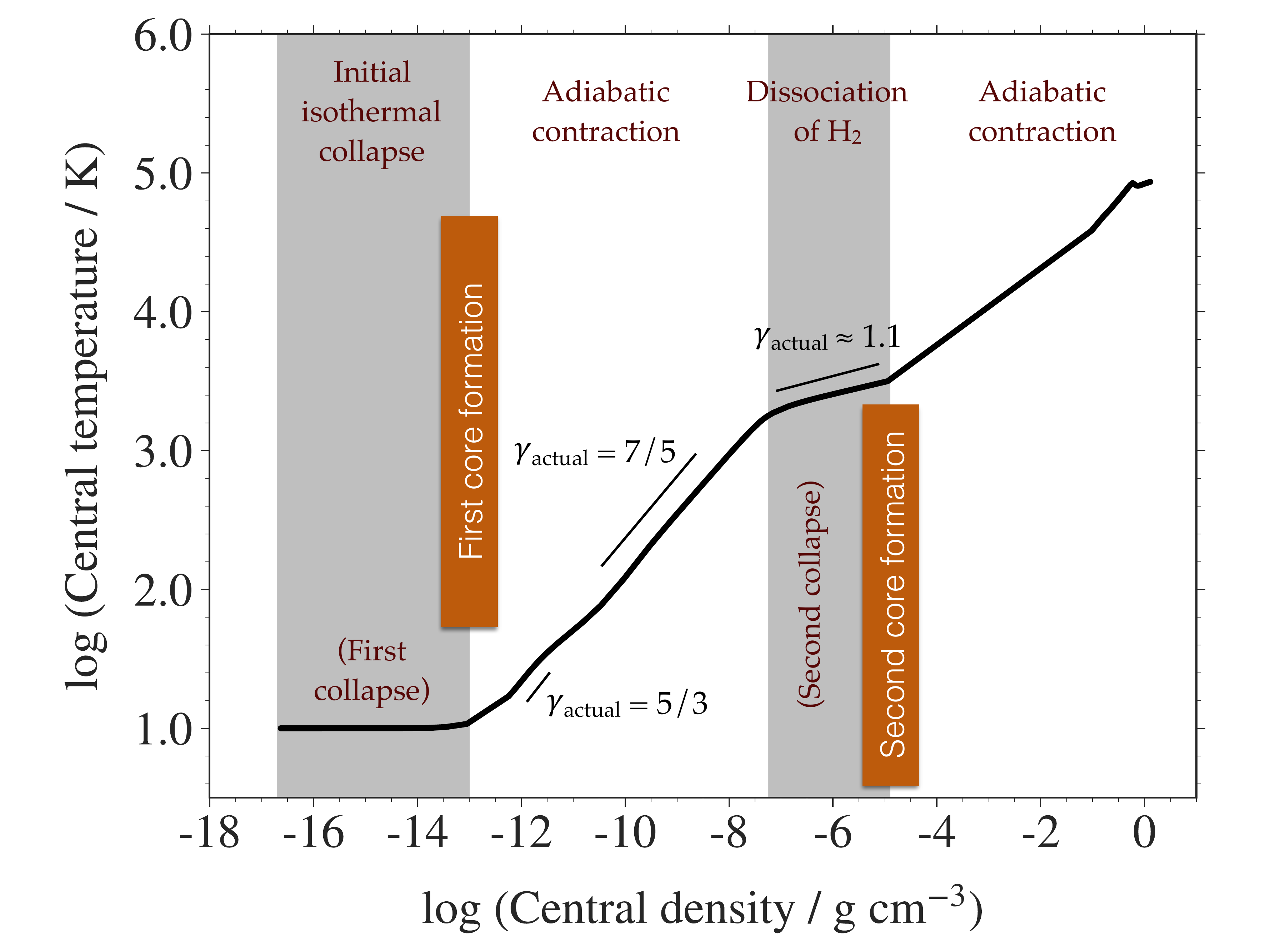}
	\caption{Thermal evolution showing the first and second collapse phase for a 1 $M_{\odot}$ cloud core. The change in adiabatic index $\mathrm{\gamma_{actual}}$ indicates the importance of using a realistic gas EOS.}
	\label{fig:thermalevolution}
\end{figure}

With time, the optical depth of the central collapsing region becomes greater than unity due to an increase in central density and the radiative cooling becomes inefficient. As the cloud core compresses, the temperature begins to rise in this dense region, which drives the first adiabatic collapse phase. This further leads to the formation of the first hydrostatic core, which eventually contracts adiabatically with an adiabatic index $\gamma_{\mathrm{actual}}$~$\approx$~5/3, where $\gamma_{\mathrm{actual}}$ is the change in the slope of the temperature evolution with density (see Fig. \ref{fig:thermalevolution}). As the temperature rises (T~$\gtrsim$~100~K), the rotational degree of freedom of the diatomic gas gets excited and the adiabatic index changes to $\approx$~7/5. The process of formation and evolution of the first core lasts for about $\mathrm{10^4}$ years, for the collapse of a 3000~au cloud core in marginal hydrostatic equilibrium, with an initial mass of 1~$M_\odot$ and 10 K temperature \citep[e.g.][]{Bhandare2018}.   

Hydrogen molecules ($\mathrm{H_2}$) begin to dissociate once the temperature reaches about 2000 K. This strongly endothermic process causes gravity to win over thermal pressure, which initiates the second collapse phase. During this phase the adiabatic index changes to $\approx$ 1.1, which is well below the stability limit of 4/3. The second hydrostatic core of atomic hydrogen is formed almost instantaneously once most of the $\mathrm{H_2}$ is dissociated and is then followed by a phase of adiabatic contraction. An increase in the thermal pressure eventually halts the collapse. The second core grows in mass as the surrounding envelope continues to accrete onto the central core. A star is born once the centre of the core reaches ignition temperatures \mbox{(T $\geq 10^6$ K)} for nuclear fusion, i.e.~deuterium and hydrogen burning. 

Our goal is to perform isolated molecular core collapse simulations, which involve detailed thermodynamical modelling along with radiation transport. We include in the equation of state (EOS) the dissociation of $\mathrm{H_2}$ and ionisation of atomic hydrogen and helium, and take the rotational and vibrational degrees of freedom into account. The numerical scheme and setup is described in Sect.~\ref{sec:Method}.  

Following the detailed analysis of the first core properties in \citet{Bhandare2018}, we first discuss the second core properties for a wide range of initial cloud core masses spanning from 0.5 $M_\odot$ to 100 $M_\odot$ in Sect. \ref{sec:2coreevolution}. Going a step beyond the one-dimensional (1D) studies, we further investigate the collapse for the cases of 1~$M_\odot$, 5~$M_\odot$, 10~$M_\odot$, and 20~$M_\odot$ initial non-rotating cloud cores using two-dimensional (2D) radiation hydrodynamic (RHD) simulations with a resolution that has not been achieved before. We follow the evolution of the second hydrostatic core for $\geq$~100 years after its formation, for each of these cases. Section~\ref{sec:results} onward we focus on the results from our 2D simulations. In Sect.~\ref{sec:2Dsims} we first describe the fiducial 1~$M_\odot$ case and discuss the properties of the second core. The highlight of our investigation is that we find indications for convection in the outer layers of the second hydrostatic core, triggered in these early stages of protostar formation. In contrast to fully convective stars, here, the energy is not generated at the stellar centre, but is provided by the accretion energy from outside the core. 

Furthermore, the dependence of the second core properties on the initial cloud core mass are presented in Sect.~\ref{sec:highmass}. For each of these 2D cases, in Sect.~\ref{sec:sasi}, we discuss the occurrence of the standing accretion shock instability (SASI), which could describe the observed large-scale oscillations of the second accretion shock. We state the limitations of our method in Sect.~\ref{sec:limitations} and provide comparisons with previous studies in Sect.~\ref{sec:comparison}. Section~\ref{sec:Summary} summarises the results from both our 1D and 2D core collapse studies presented in this paper.  

\section{Numerical method}
\label{sec:Method}

We perform 1D and 2D RHD calculations of a collapsing molecular cloud core using the (magneto) hydrodynamic code PLUTO \citep{Mignone2007, Mignone2012} combined with the self-gravity solver implemented and tested in \citet{Kuiper2010} and \citet{Kuiper2011}. We consider the approximation of local thermodynamic equilibrium (LTE) and a two-temperature (2T) approach for the gas and radiation for gas thermodynamics. The equations of hydrodynamics that account for the mass, momentum, and energy conservation and the time-dependent radiation transport equation are described in \citet{Bhandare2018}. We use the grey flux-limited diffusion (FLD) radiation transport module \mbox{MAKEMAKE}, the numerics of which is described in \citet{KuiperRT2010} and \mbox{Kuiper et al. (subm.)}. It has been argued by \cite{Vaytet2012, Vaytet2013} that the grey method proves to be sufficient in the early evolutionary stages of the collapse and the multi-group radiative transfer may become more important only in the later stages of protostellar evolution.

Additionally, we use a gas EOS from \citet{Dangelo2013} to account for effects such as dissociation of $\mathrm{H_2}$, ionisation of atomic hydrogen and helium, and molecular vibrations and rotations. This has been implemented in the PLUTO code by \citet{Vaidya2015} and updates to the radiation transport module to account for the realistic gas EOS are previously described in \citet{Bhandare2018}.

Tabulated dust opacities are used from \citet{Ossenkopf1994} whereas tabulated gas opacities are used from \citet{Malygin2014}. In order to account for the contribution from dust dominating at low temperatures, we updated the evaporation and sublimation module to consider a time-dependent evolution of the dust as previously discussed in \citet{Bhandare2018}. 

We make use of a conservative finite volume approach based on second-order Godunov-type schemes, that means a shock capturing Riemann solver implemented in PLUTO to solve the equations of hydrodynamics. Our basic configuration for the flux computation consists of the Harten-Lax-Van Leer approximate Riemann solver that restores the middle contact discontinuity (hllc), a monotonised central difference (MC) flux limiter using piecewise linear interpolation and we integrate with a Runge-Kutta second order (RK2) method. On the other hand, the FLD equation is solved in an implicit way using a standard generalised minimal residual solver with approximations to the error from previous restart cycles (LGMRES). A relative convergence tolerance value of $10^{-10}$ in terms of temperature is used. More details about the open-source solver library PETSc (Portable, Extensible Toolkit for Scientific Computation) can be found in \citet{petsc-efficient, petsc-user-ref, petsc-web-page}. 

\begin{table}[t]
	\centering
	\caption{Initial cloud core properties.}
	\resizebox{0.5\textwidth}{!}{
	\begin{tabular}[t]{cccccc}
		\hline
		$M_{0} ~(M_{\odot})$ & $R_\mathrm{cloud}$ (au) & $T_{\mathrm{0}}$ (K) & $T_{\mathrm{BE}}$ (K) & $M_{\mathrm{BE}}/M_{\mathrm{0}}$ & $\rho_\mathrm{c} ~(\mathrm{g ~cm^{-3}})$
		\TBstrut\\ \hline \hline
		0.5   & 3000   & 10.0  & 9.49   & 1.05e-00    & ~1.16e-17  \Tstrut \\
		1.0   & 3000   & 10.0  & 18.98   & 5.27e-01    & 2.33e-17   \\
		2.0   & 3000   & 10.0  & 37.96   & 2.64e-01    & 4.66e-17   \\
		5.0   & 3000   & 10.0  & 94.91   & 1.05e-01	  & 1.17e-16   \\
		8.0	  & 3000   & 10.0  & 151.85  & 6.58e-02    & 1.87e-16   \\
		10.0  & 3000   & 10.0  & 189.81  & 5.27e-02    & 2.33e-16   \\
		12.0  & 3000   & 10.0  & 227.78  & 4.39e-02	  & 2.80e-16   \\
		14.0  & 3000   & 10.0  & 265.74  & 3.76e-02	  & 3.26e-16   \\
		15.0  & 3000   & 10.0  & 284.72  & 3.51e-02    & 3.50e-16   \\
		16.0  & 3000   & 10.0  & 303.70  & 3.29e-02    & 3.73e-16   \\
		18.0  & 3000   & 10.0  & 341.66  & 2.93e-02    & 4.20e-16   \\
		20.0  & 3000   & 10.0  & 379.63  & 2.63e-02    & 4.66e-16   \\
		30.0  & 3000   & 10.0  & 569.44  & 1.76e-02    & 6.99e-16   \\
		40.0  & 3000   & 10.0  & 759.25  & 1.32e-02    & 9.33e-16   \\
		60.0  & 3000   & 10.0  & 1138.88 & 8.78e-03    & 1.40e-15   \\
		80.0  & 3000   & 10.0  & 1518.50 & 6.58e-03    & 1.87e-15   \\
		100.0 & 3000   & 10.0  & 1898.13 & 5.27e-03    & ~2.33e-15  \Bstrut \\ \hline
	\end{tabular}
	}
	\tablefoot{Listed above are the cloud core properties for runs with different initial cloud core mass $M_{0} ~(M_{\odot})$, outer radius $R_\mathrm{cloud}$~(au), temperature $T_{\mathrm{0}}$ (K), temperature $T_{\mathrm{BE}}$ (K) of a stable Bonnor-Ebert cloud core, stability parameter $M_{\mathrm{BE}}/M_{\mathrm{0}}$, and central density $\rho_\mathrm{c} ~(\mathrm{g ~cm^{-3}})$. }
	\label{tab:initialparams}
\end{table}

\subsection{Initial setup}
\label{sec:setup}

We use a stable Bonnor--Ebert \citep{Ebert1955, Bonnor1956} sphere like density profile as the initial density distribution, as also described in our previous work \citep{Bhandare2018}. Due to the fact that the hydrostatic equilibrium condition of a Bonnor--Ebert sphere does not have an analytical solution, the density profile $\rho(r)$ has to be solved for numerically. The initial outer $\rho_\mathrm{o}$ and central $\rho_\mathrm{c}$ densities are determined as
\begin{equation}
\begin{split}
&\rho_\mathrm{o} = \Bigg (\dfrac{1.18 ~c_\mathrm{s0}^3}{M_0 ~G^{3/2}} \Bigg)^2 \\
&\rho_\mathrm{c} = 14.1 ~\rho_\mathrm{o},
\end{split}
\end{equation}
where $G$ is the gravitational constant, $M_0$ is the initial cloud core mass, and the initial sound speed $c_\mathrm{s0}$ for a cloud core with radius $R_\mathrm{cloud}$ is computed as
\begin{align}
\mbox{$c_\mathrm{s0}^2 = \dfrac{G M_0}{\mathrm{ln}(14.1) ~R_\mathrm{cloud}} $}.
\end{align}

The density contrast between the centre and edge of the sphere corresponds to a dimensionless radius of $\xi$ = 6.45, where $\xi$ is defined as
\begin{align}
\mbox{$ \xi = \sqrt{\dfrac{4 \pi G \rho_\mathrm{o}}{c_\mathrm{s0}^2}} R_\mathrm{cloud} $}.
\end{align}

The integrated mass of the cloud core is the same as that of a critical Bonnor--Ebert sphere. The temperature $T_\mathrm{BE}$ for the stable sphere is computed as
\begin{align}
\mbox{$T_\mathrm{BE} = \dfrac{\mu ~c_\mathrm{s0}^2}{\gamma ~\Re}$},
\end{align}
where the mean molecular weight $\mu$ = 2.353, $\gamma$ = 5/3, and $\Re$ is the universal gas constant.  

The collapse is initiated by prescribing a thermal pressure using an initial temperature $T_\mathrm{0}$ of 10~K instead of $T_\mathrm{BE}$ from a stable Bonnor--Ebert sphere setup, which allows gravity to dominate. The radiation temperature is set to be initially in equilibrium with the gas temperature. The dust and gas temperatures are treated as perfectly coupled throughout the simulation. We do not include any initial cloud rotation in the simulations presented herein.

\subsection{One-dimensional setup}
\label{sec:1Dgrid}

The 1D studies span a wide range of initial cloud core masses from 0.5~$M_{\odot}$ to 100~$M_{\odot}$. The spatial domain extends from $10^{-4}$~au up to 3000~au for all the different cases, which implies that for a constant initial temperature of 10~K, the central density of the different Bonnor-Ebert spheres scales as a function of initial cloud core mass (see Table~\ref{tab:initialparams}). Convergence tests and effects of different initial cloud core properties have been discussed in our previous work \citep{Bhandare2018}.  

We use 320 uniformly spaced grid cells from $10^{-4}$~au to $10^{-2}$~au and 4096 logarithmically spaced grid cells from $10^{-2}$~au to 3000~au, with identical sizes for the last uniform cell and the first logarithmic cell. A logarithmic binning throughout the whole domain would lead to extremely small grid cells resulting in much smaller time steps. The domain was thus scaled differently in the inner dense regions to prevent the simulations from being computationally very expensive. Our computational grid thus comprises of 4416 cells in total, covering a dynamical range of seven orders of magnitude. The smallest radial cell size $\Delta x_\mathrm{min} = \Delta r = 3.09\times10^{-5} $~au. We use a minimum of 50 cells per Jeans length, which is estimated at the highest central density. Otherwise, we use $10^{3} - 10^{5}$ cells per Jeans length.

\subsection{Two-dimensional setup}
\label{sec:2Dgrid}
We expand our 1D studies for a few cases of 1~$M_\odot$, 5~$M_\odot$, 10~$M_\odot$, and 20~$M_\odot$ cloud cores, thus accounting for cases from the low-, intermediate-, and high-mass regimes. For this, we adopt a 2D spherical Eulerian grid with axial and midplane symmetry. The grid comprises of 1445 logarithmically spaced cells in the radial direction extending from \mbox{$10^{-2}$~au} to 3000~au. Thus spanning a dynamical range covering five orders of magnitude. The logarithmic spacing increases resolution in the central parts of the computational domain. In the polar direction, we use 180 uniformly spaced cells stretching from the pole ($\theta = 0^\circ$) to the midplane ($\theta = 90^\circ$). The number of cells is tuned in order to ensure equal spatial extent in the radial and polar direction with the smallest cell size $\Delta x_\mathrm{min} = \Delta r = r \Delta\theta = 8.77\times10^{-5}$~au. We use a minimum of 49 cells per Jeans length, which is estimated at the highest central density. Otherwise, we use 300~--~$10^{4}$ cells per Jeans length.

\subsection{Boundary conditions}
\label{sec:boundaryconditions}

We use a reflective boundary condition at the inner radial edge for the hydrodynamics and a zero gradient condition for the radiation energy (i.e.~no radiative flux can cross the inner boundary interface). At the outer radial edge, we use a Dirichlet boundary condition on the radiation temperature with a constant boundary value of $T_\mathrm{0}$ and an outflow--no-inflow condition for the hydrodynamics that includes a zero-gradient (i.e.~no force) boundary condition for 
the thermal pressure, the polar and the azimuthal velocity components. For the 2D runs, we use axisymmetric boundaries at the pole and mirror-symmetric boundaries at the equator. 

\section{Results: Second core in spherical symmetry}

In our previous work \citep{Bhandare2018}, we investigated the collapse of a molecular cloud core through the phase of the first hydrostatic core formation and tracked its evolution until the formation of the second core. In the studies presented herein, we follow the evolution of the second core for about 150 to 500 years after its formation, depending on the initial cloud core mass. 

Figure~\ref{fig:thermalevolution} shows the different evolutionary stages during the collapse of a fiducial 1 $M_{\odot}$ cloud core in one of our simulation runs. The collapse of an initial isothermal molecular cloud core proceeds via a first phase of adiabatic compression and contraction, which forms the first hydrostatic core. This is then followed by a second collapse triggered due to dissociation of $\mathrm{H_2}$ forming the second core, which undergoes another phase of adiabatic contraction as summarised in Sect.~\ref{sec:intro}. The phase transitions at the different stages indicates the importance of using a realistic gas EOS as detailed in Sect.~\ref{sec:Method}. 

\subsection{Dependence on initial cloud core mass}
\label{sec:2coreevolution}

\begin{figure*}[]
	\centering
	\hspace*{-1cm}	
	\begin{subfigure}{0.243\textwidth}
		\includegraphics[width= 1.2\textwidth]{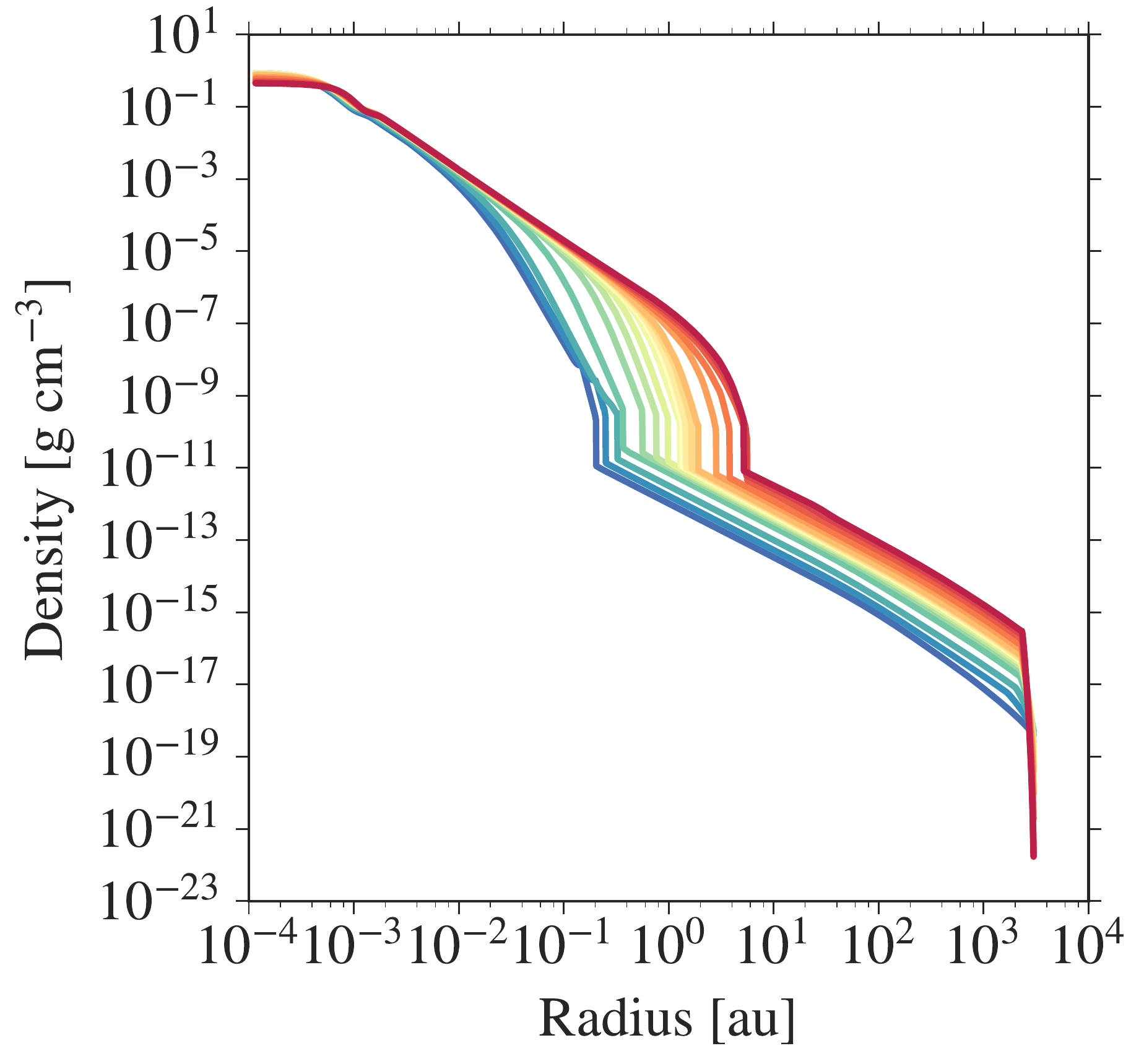}
	\end{subfigure}
	\hspace{0.5in}
	\begin{subfigure}{0.243\textwidth}
		\includegraphics[width= 1.2\textwidth]{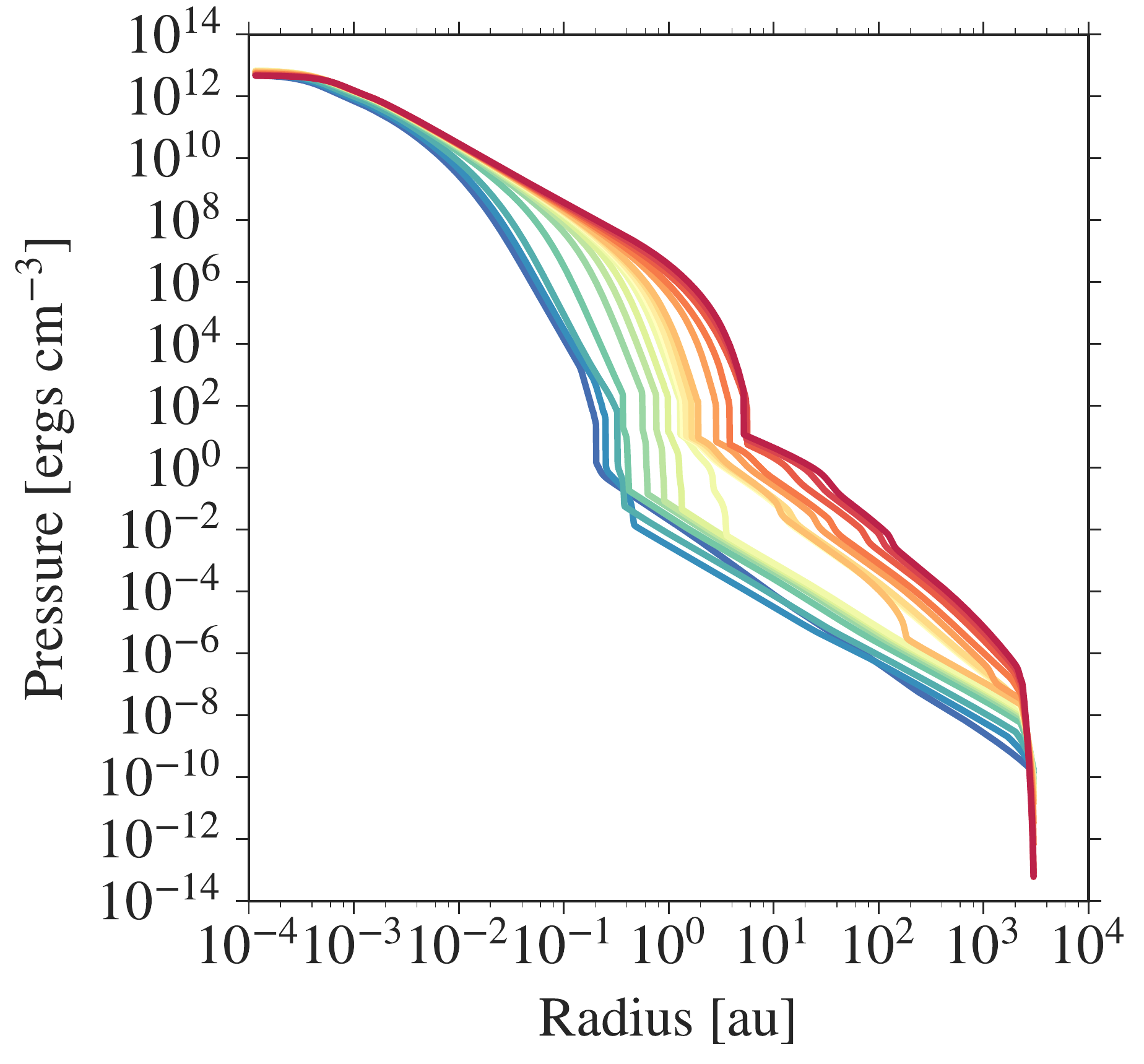}
	\end{subfigure}
	\hspace{0.5in}
	\begin{subfigure}{0.243\textwidth}
		\includegraphics[width= 1.2\textwidth]{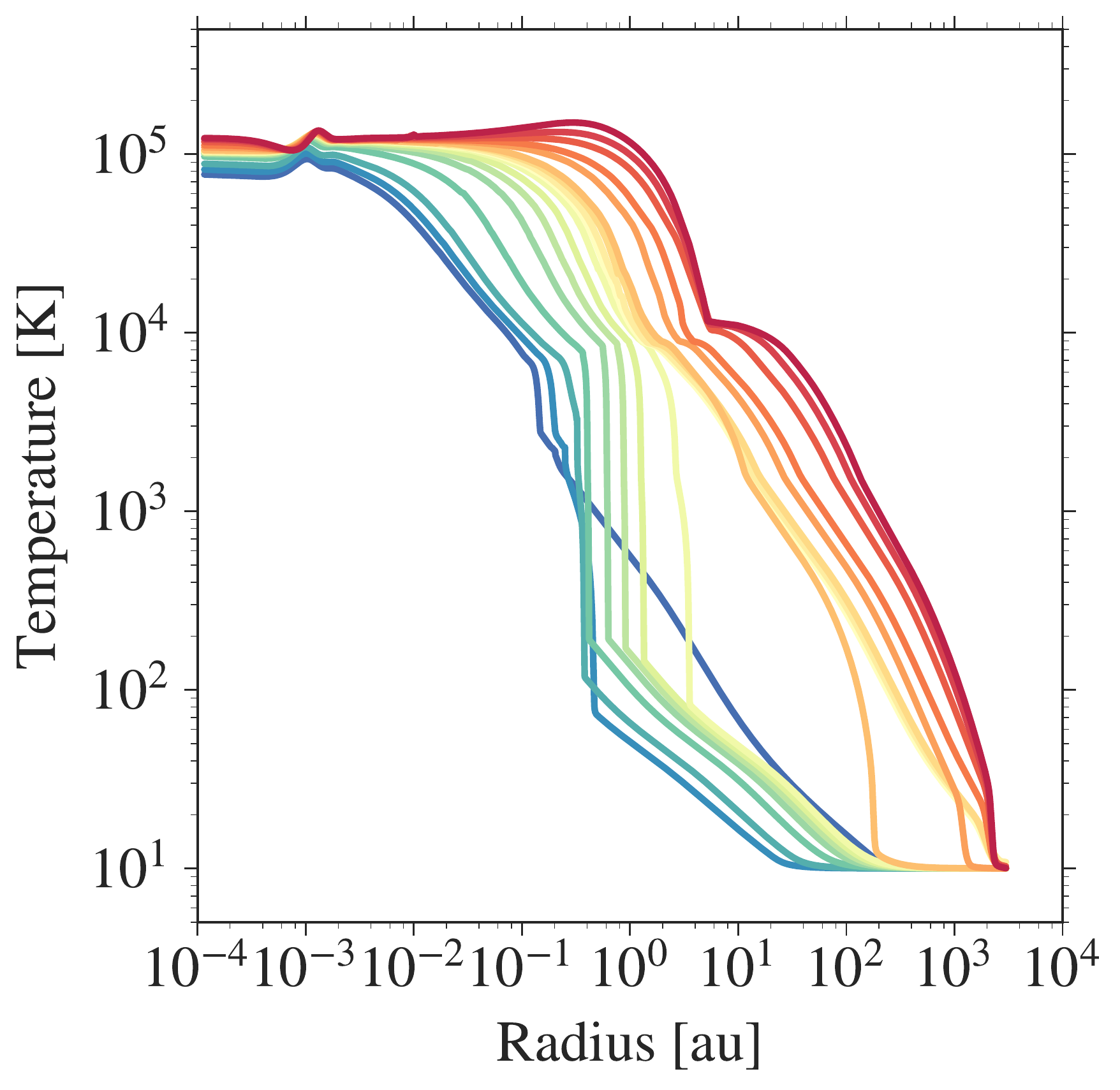}
	\end{subfigure}
	\hspace*{-1cm}	
	\begin{subfigure}{0.243\textwidth}
		\includegraphics[width= 1.2\textwidth]{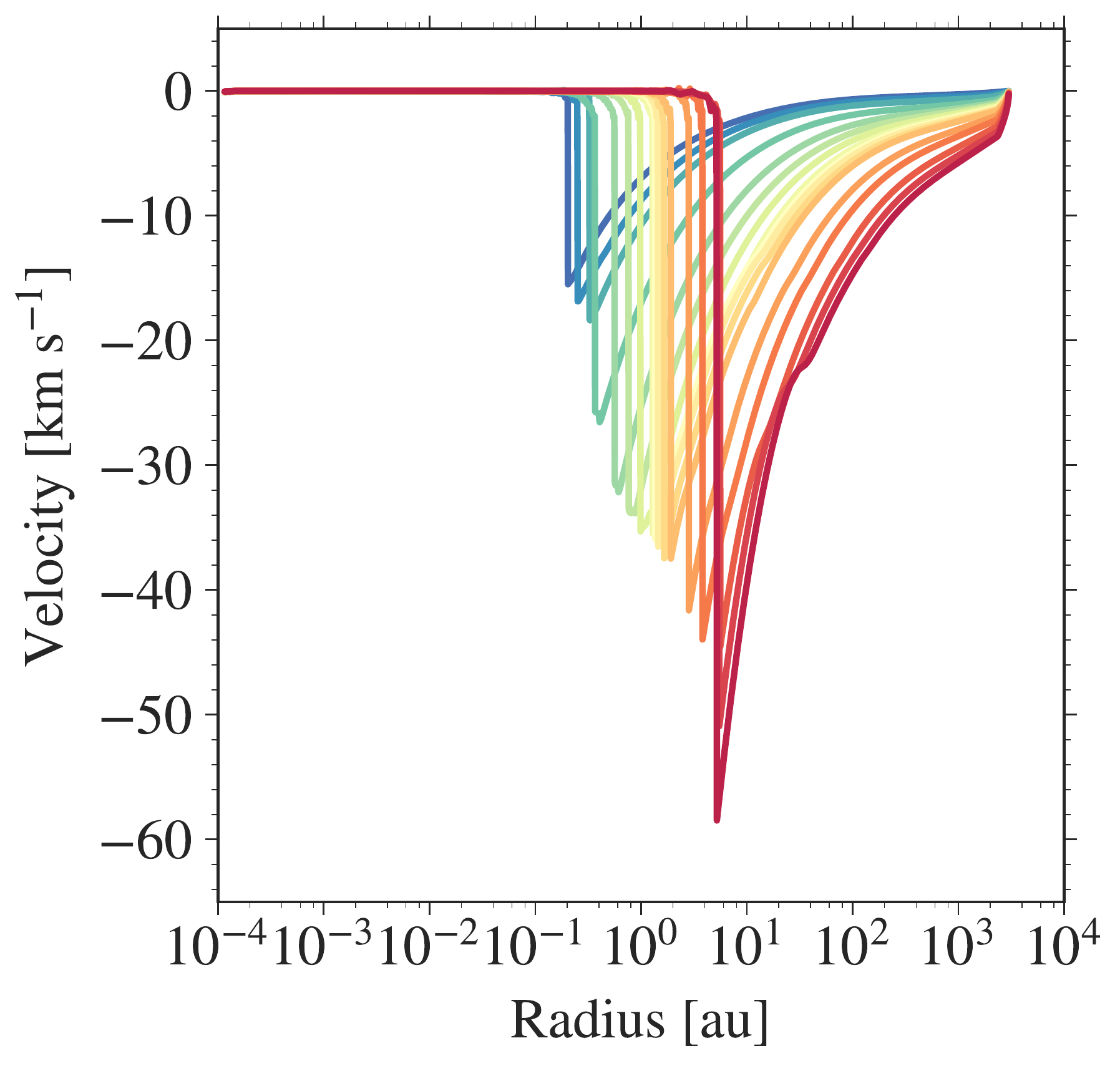}
	\end{subfigure}
	\hspace{0.5in}
	\begin{subfigure}{0.243\textwidth}
		\includegraphics[width= 1.2\textwidth]{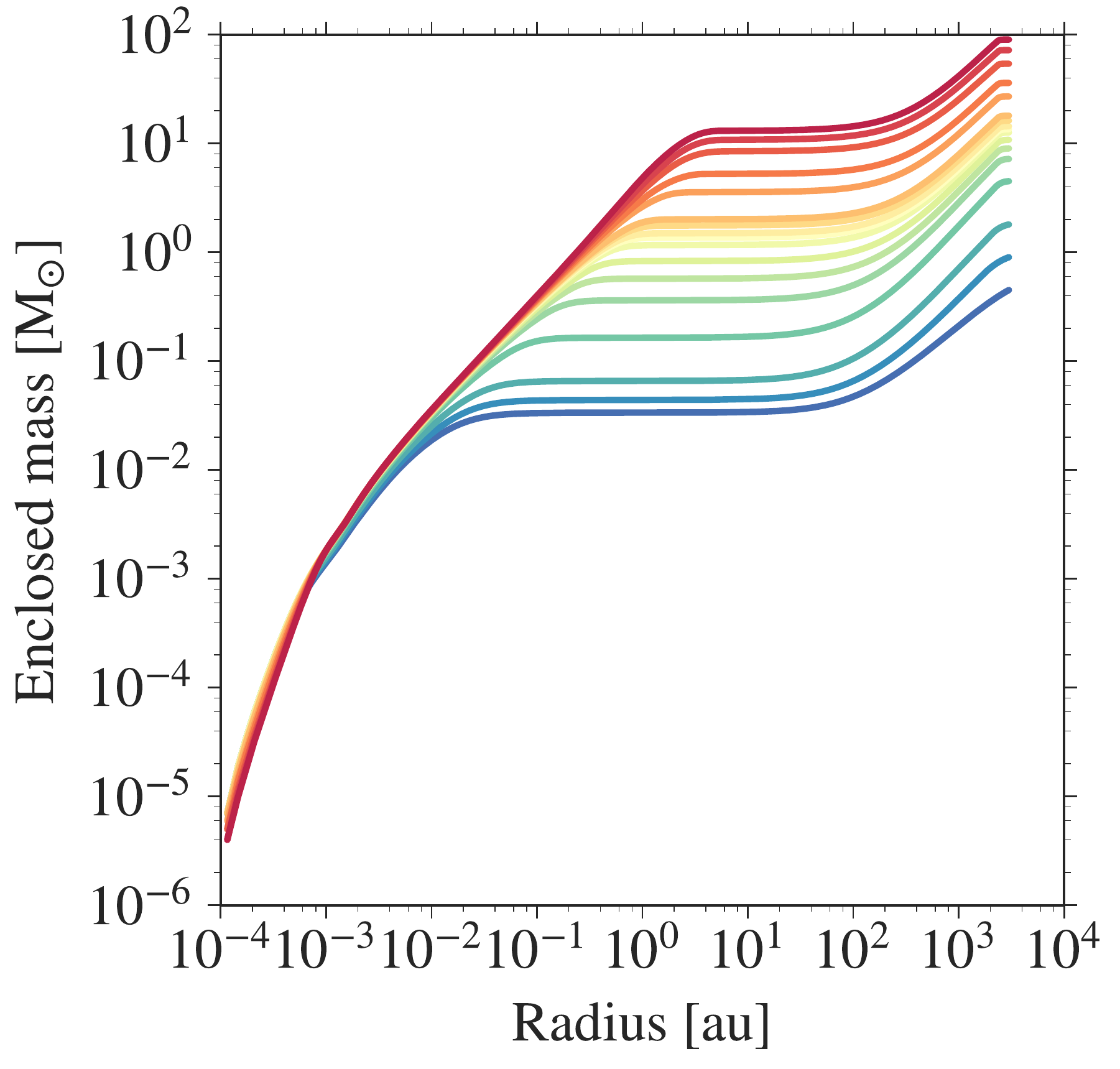}
	\end{subfigure}
	\hspace{0.5in}
	\begin{subfigure}{0.243\textwidth}
		\includegraphics[width= 1.15\textwidth]{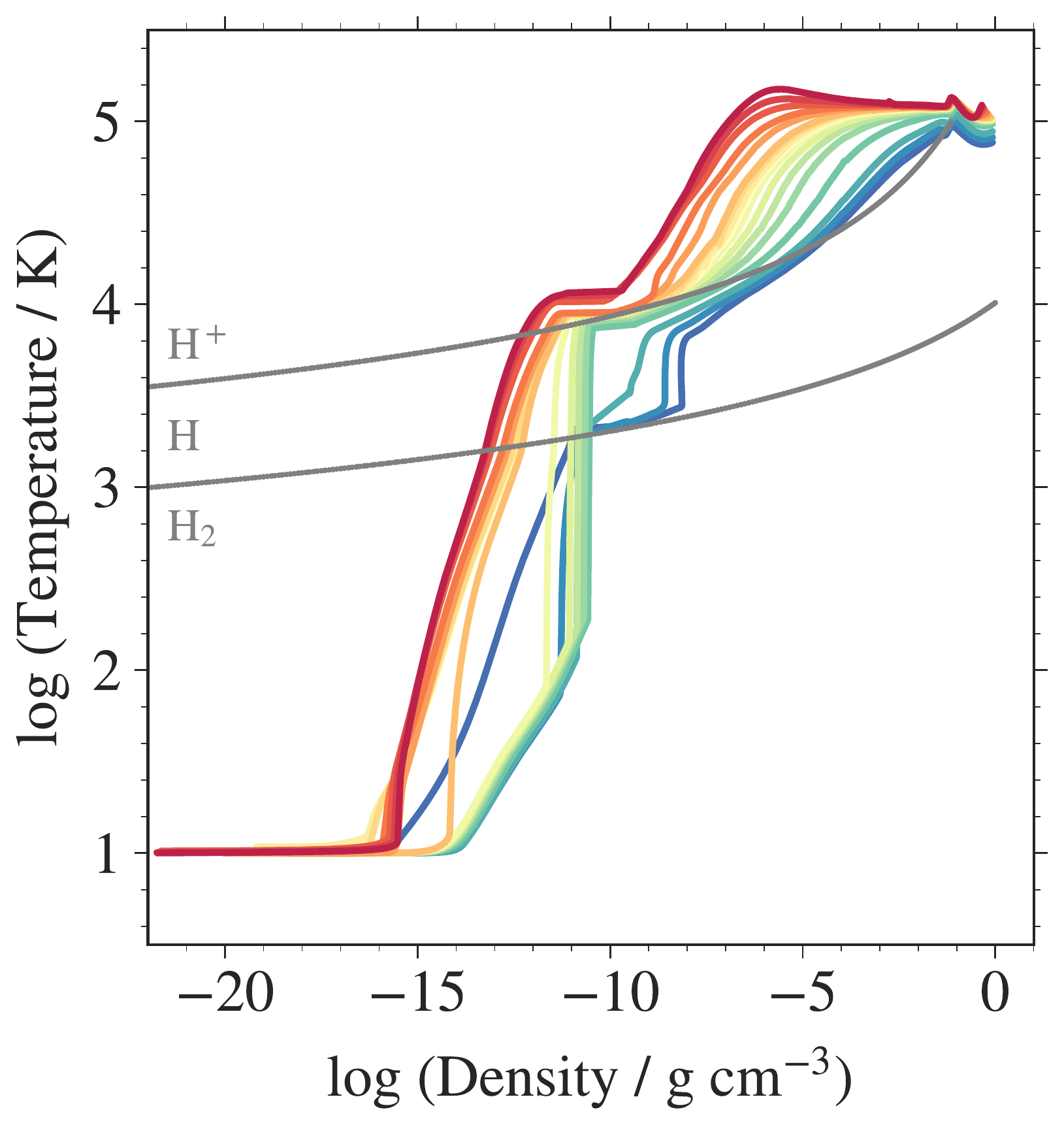}
	\end{subfigure}
	\hspace*{-1cm}	
	\begin{subfigure}{0.243\textwidth}
		\includegraphics[width= 1.2\textwidth]{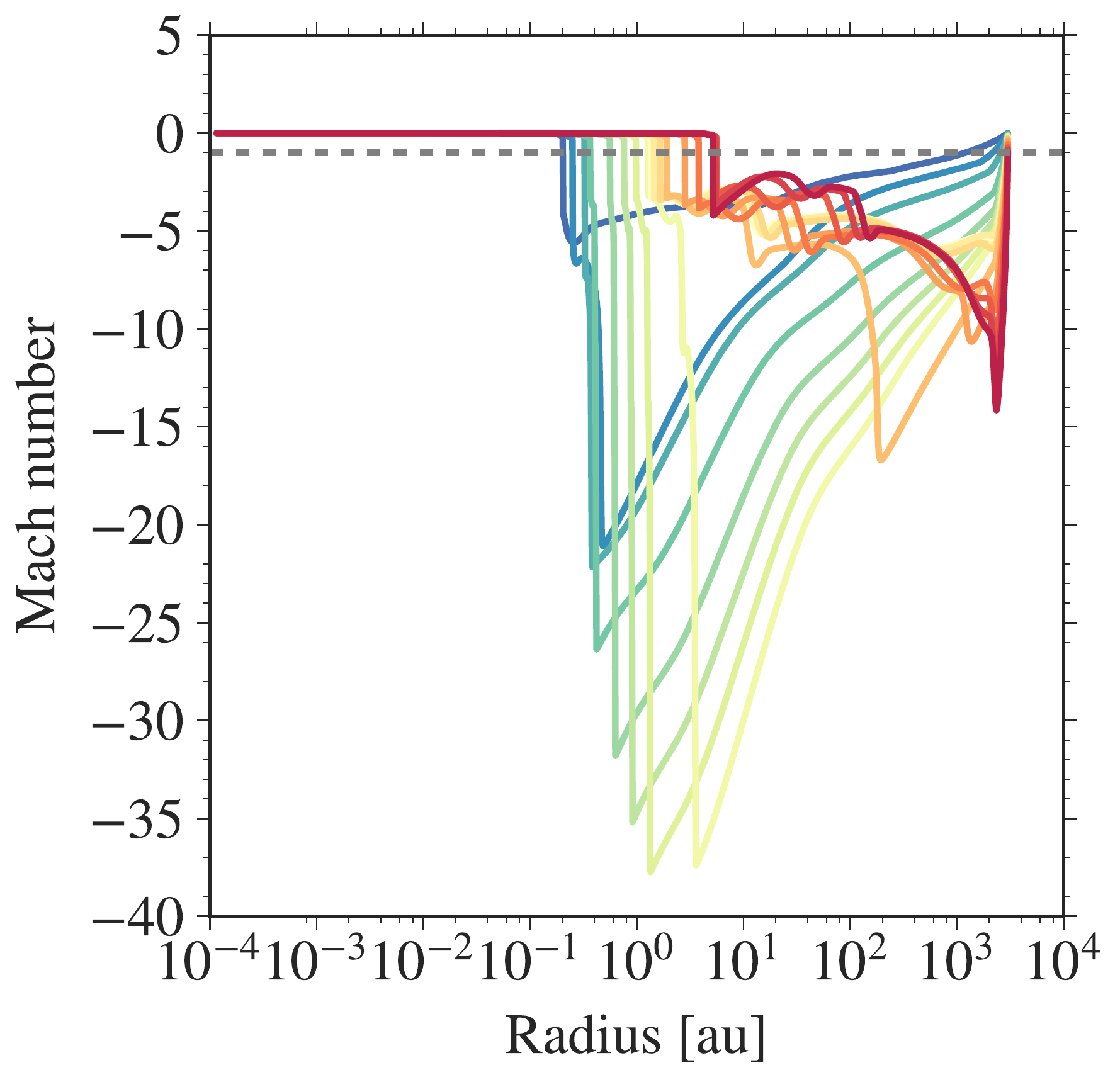}
	\end{subfigure}
	\hspace{0.5in}
	\begin{subfigure}{0.243\textwidth}
		\includegraphics[width= 1.2\textwidth]{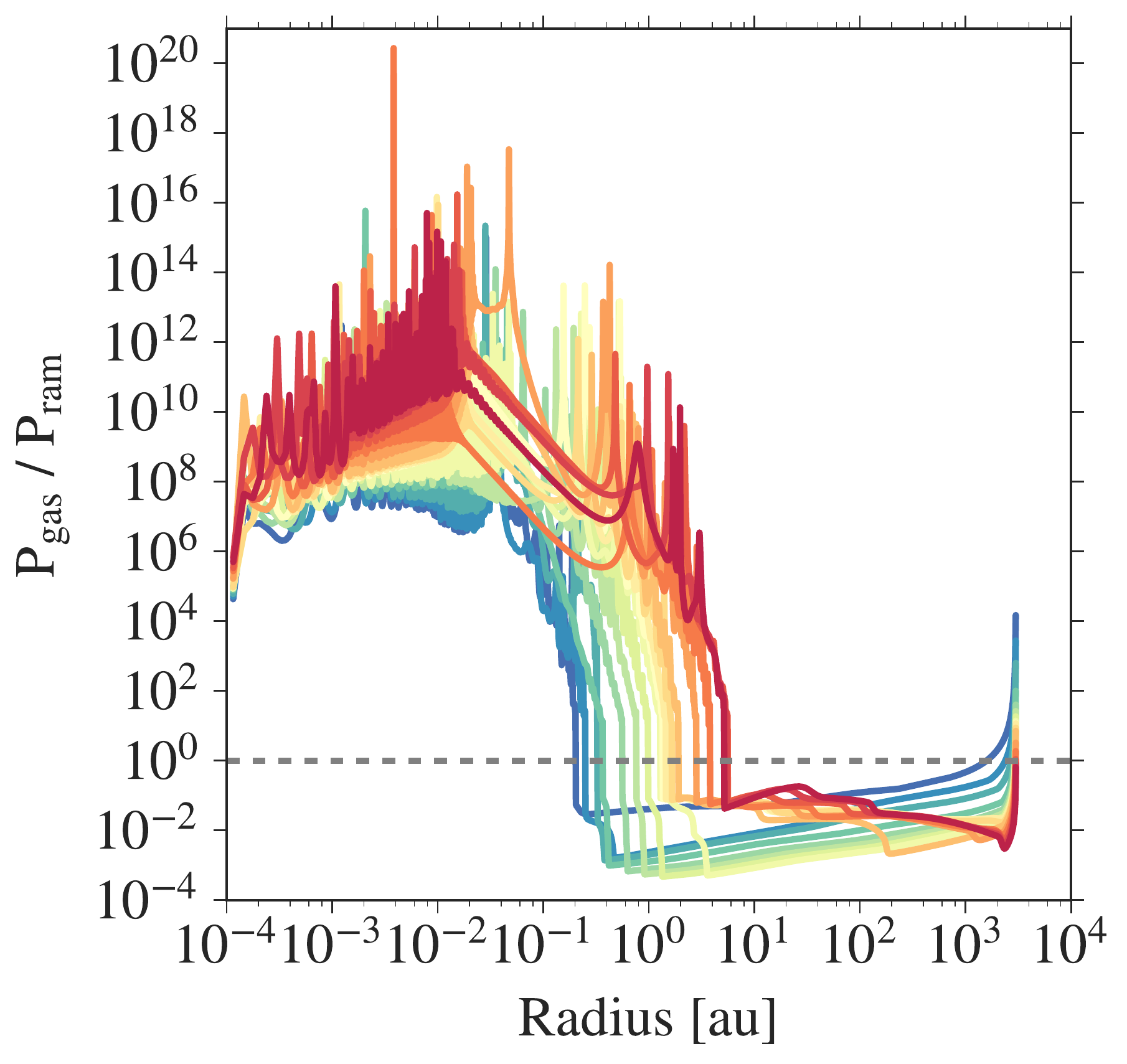}
	\end{subfigure}
	\hspace{0.5in}
	\begin{subfigure}{0.243\textwidth}
		\includegraphics[width= 1.2\textwidth]{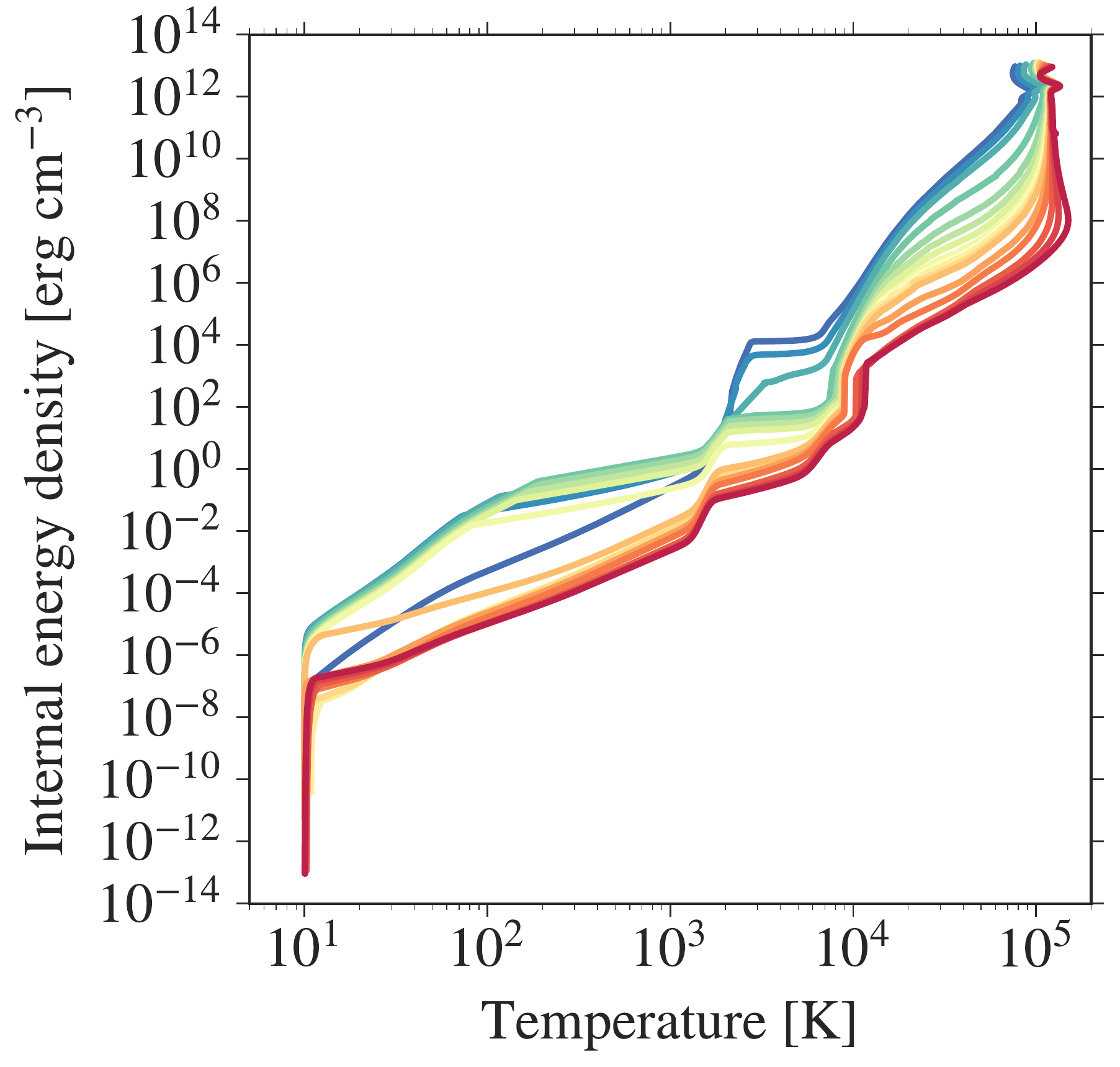}
	\end{subfigure}
	\hspace*{-1cm}	
	\begin{subfigure}{0.243\textwidth}
		\includegraphics[width=1.2\textwidth]{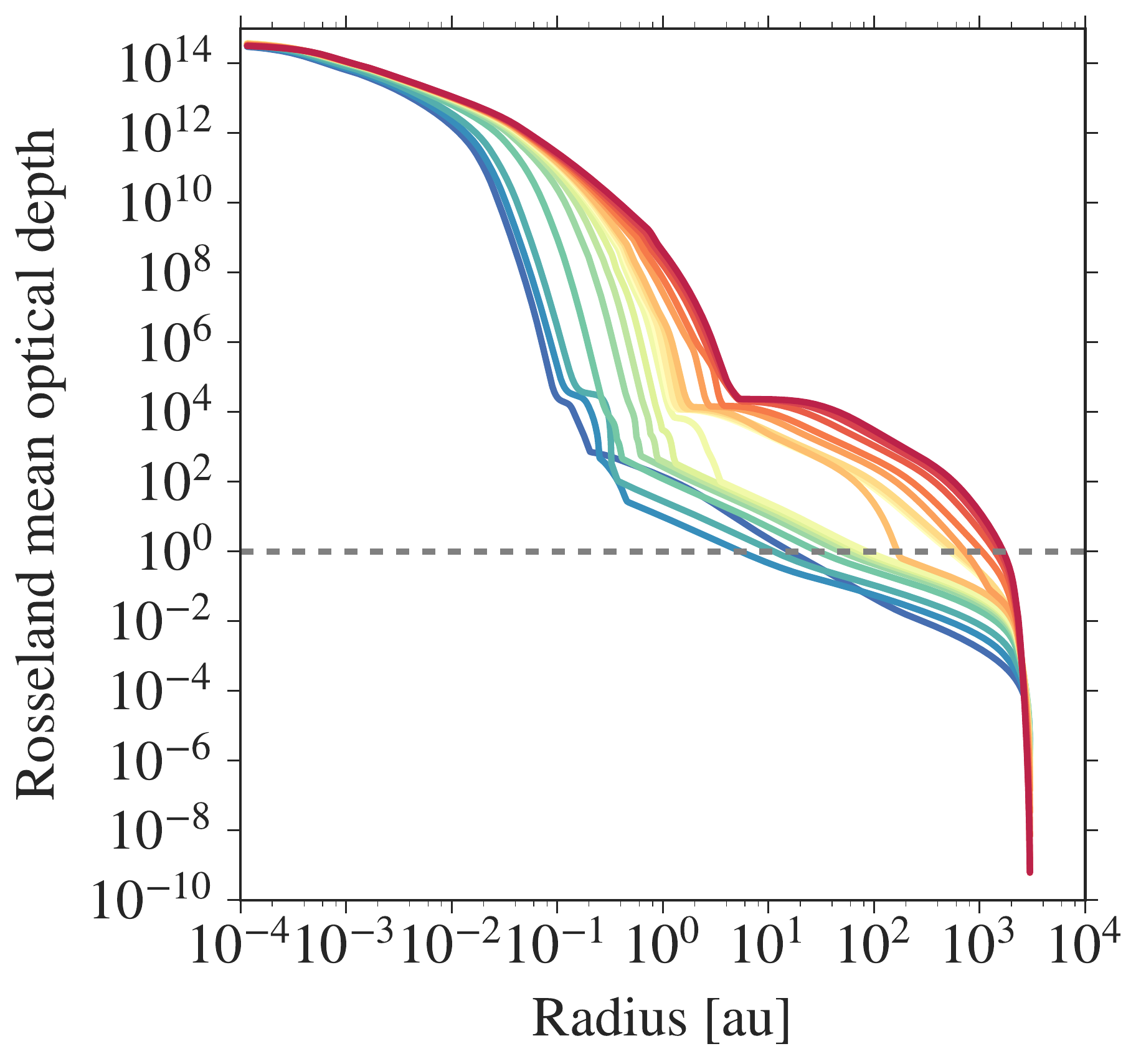}
	\end{subfigure}
	\hspace{0.5in}
	\begin{subfigure}{0.243\textwidth}
		\includegraphics[width=1.2\textwidth]{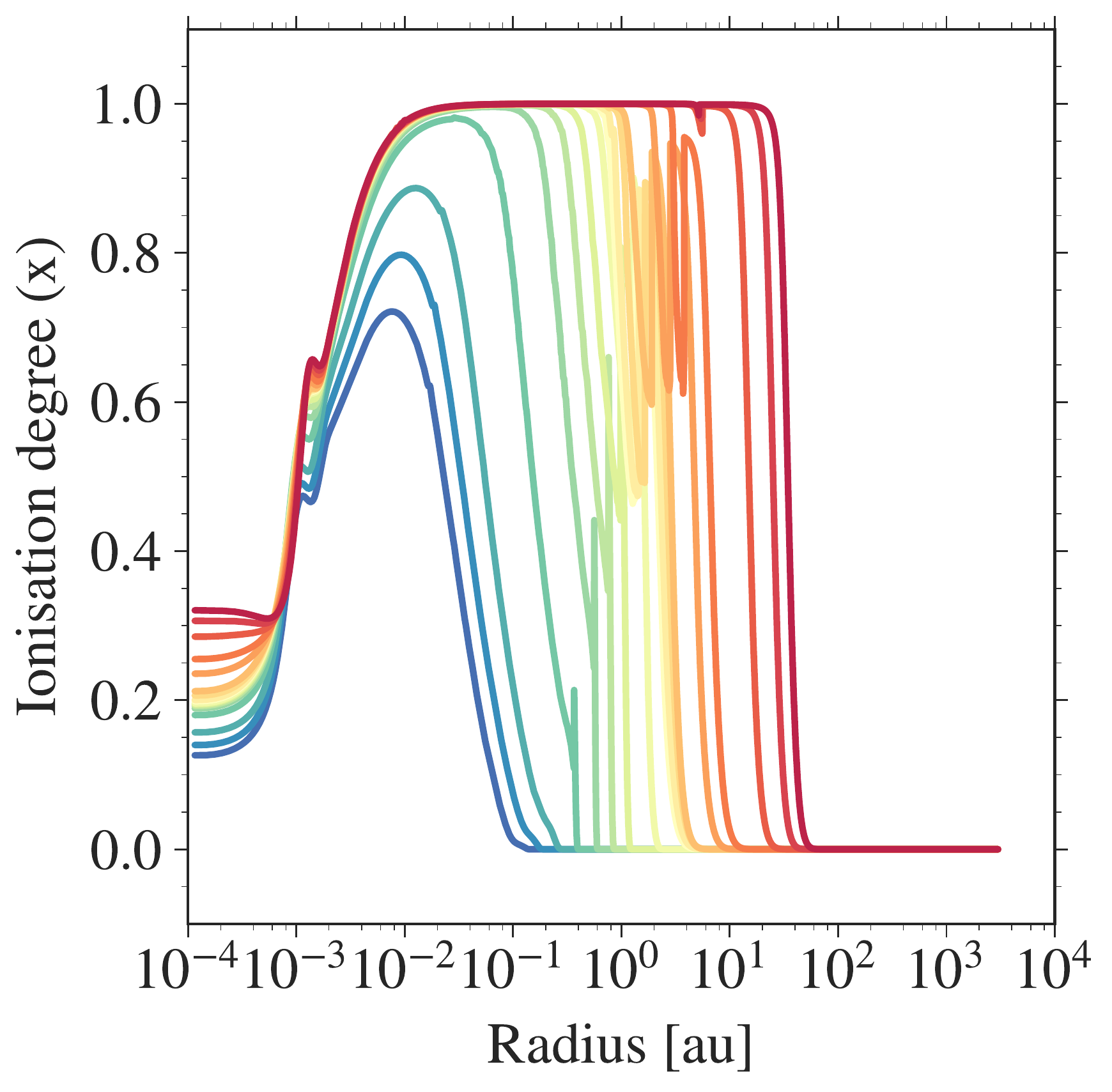}
	\end{subfigure}
	\hspace{0.5in}
	\begin{subfigure}{0.243\textwidth}
		\includegraphics[width= 1.2\textwidth]{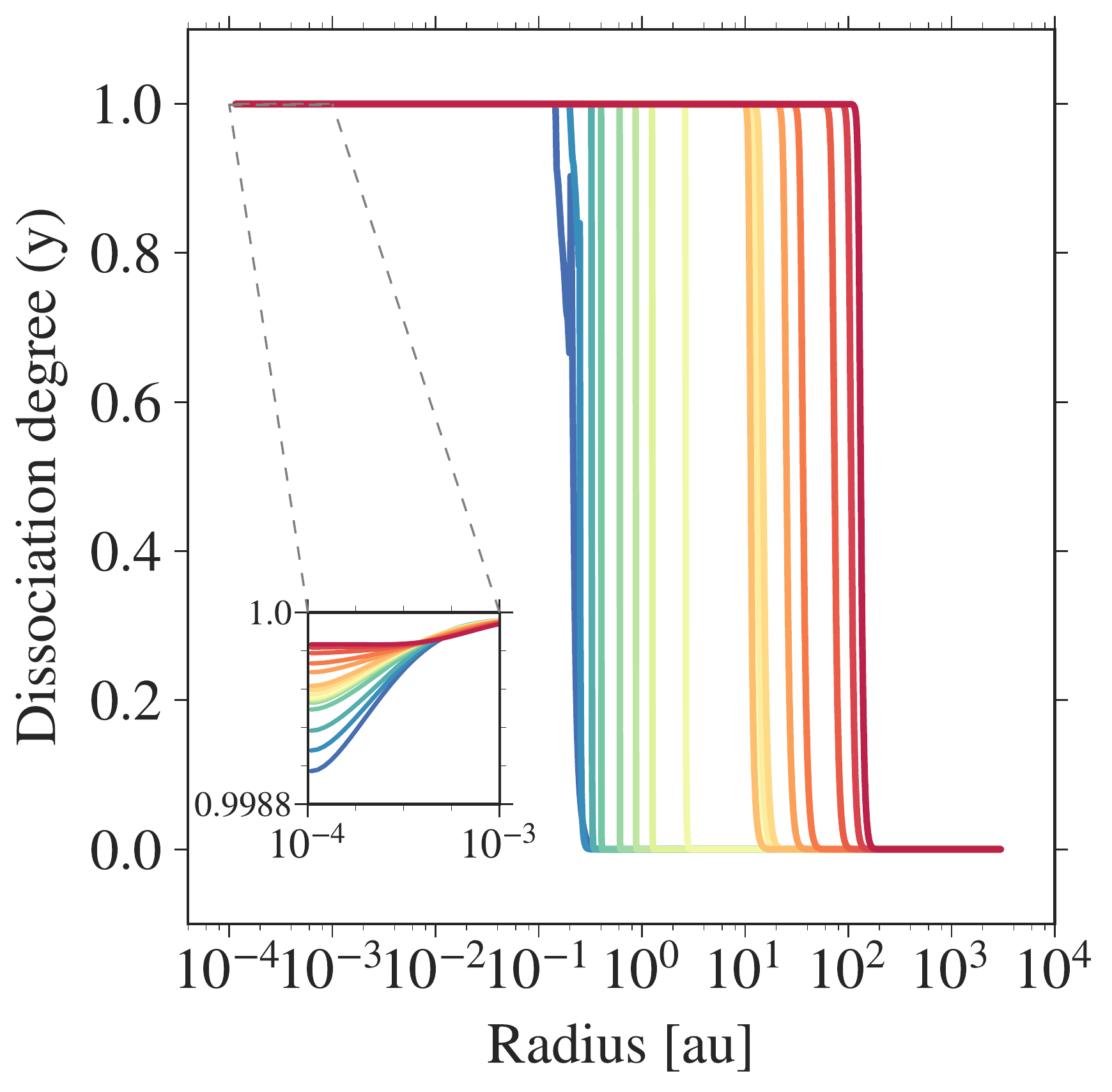}
	\end{subfigure}
	\hspace{0.5in}
	\hspace*{-1cm}	
	\begin{subfigure}{0.7\textwidth}
		\includegraphics[width=1.1\textwidth]{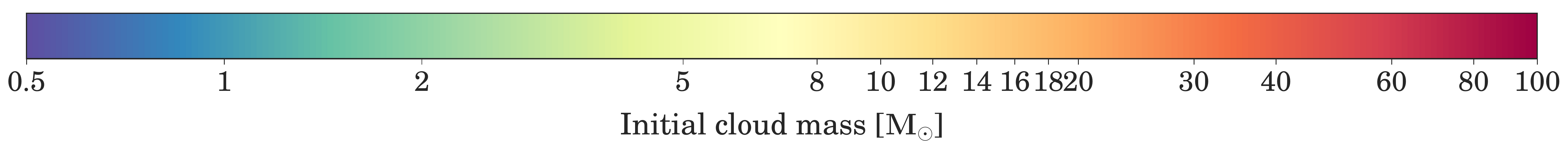}
	\end{subfigure}
	\caption{Radial profiles (across and down) of \mbox{\bf a)}~density, \mbox{\bf b)}~pressure, \mbox{\bf c)}~gas temperature (= radiation temperature), \mbox{\bf d)}~velocity, and \mbox{\bf e)}~enclosed mass as well as the \mbox{\bf f)}~thermal structure, \mbox{\bf g)}~Mach number, \mbox{\bf h)}~ratio of gas to ram pressure $P_\mathrm{ram} = \rho u^2$, \mbox{\bf i)}~internal energy density as a function of temperature, \mbox{\bf j)}~optical depth, \mbox{\bf k)}~degree of ionisation \citep[][Eq.~16]{Bhandare2018}, and \mbox{\bf l)}~degree of dissociation \citep[][Eq.~17]{Bhandare2018} for all the cases at the final simulation snapshot (final times for different cases indicated in Table~\ref{tab:SCproperties}) in our simulations when the central density is roughly 0.5 -- 0.8 $\mathrm{g ~cm^{-3}}$. Different colours indicate cloud cores with different initial masses as seen in the colour bar. Grey lines in the thermal structure plot show the 50\,\%\ dissociation and ionisation curves.}
	\label{fig:masscomparison}
\end{figure*}

In order to investigate the dependence of the second core properties on the initial cloud core mass, we further span a wide range of masses from 0.5~$M_{\odot}$ to 100~$M_{\odot}$. Figure~\ref{fig:masscomparison} shows the radial profiles of a variety of physical properties for all the different masses at a snapshot in time when the second core has evolved further and the first core no longer exists. Most of the material from the first core has been accreted on to the second core until this evolutionary stage. The accretion shock front in the radial velocity profile, which also coincides with the discontinuity in the density profile, defines the second core radius. This second shock is seen to be subcritical, that means the post-shock temperature is higher than the pre-shock temperature, suggesting that the accretion energy is transferred onto the second core and not radiated away (see discussions in \citealt{Vaytet2013} and \citealt{Bhandare2018}, and compare to the planetary case in \citealt{Marleau2017,Marleau2019}). 

We compare the different cases at a point in time when the central densities are around 0.5 -- 0.8~$\mathrm{g ~cm^{-3}}$ and the central temperatures are roughly $10^5$~K. We note that the evolutionary timescales for the cloud cores with different initial masses are not the same, which indicates that high-mass cloud cores collapse faster than the low-mass ones. Most significant differences due to initial cloud core masses are visible outwards from the second core. The thermal structure indicates that the initially similar isothermal cloud cores eventually heat up at different densities. This has a significant effect on the formation timescale and the lifetime of the first and second cores. 

The radial temperature profile shows an off-centred peak. The location of this peak corresponds to the radial position of the plateau seen in the density profile. This off-centred peak is also seen in studies by \citet[][see their Fig.~4]{Masunaga2000} and \citet[][see their Fig.~2]{Tomida2013} (and also \citealt{Winkler1980a,Winkler1980b} and \citealt{Stahler1980a}). Both these studies make use of the gas EOS by \citet{Saumon1995}, which accounts for the effects of Fermi energy of the (partially) degenerate electrons. \citet{Masunaga2000} suggest that the off-centred peak seen in the temperature profile is due to partially degenerate electrons in the central region. In the off-centred region, the densities are not the highest and the thermal energy dominates the Fermi energy. There, the gas pressure is therefore more sensitive to the temperature. In the innermost regions, $P_\mathrm{gas} \ll P_\mathrm{deg}$ and hence there is no temperature rise. In other words, the off-centred temperature peak, for example in \citet{Masunaga2000} can be interpreted as a temperature depression in the centre. However, since the effects of Fermi energy are not included in the (ideal) gas EOS used in our models \citep{Dangelo2013}, the reason for the off-centred peak in the temperature profile at the highest mass densities must be a different one in this case. The steady increase in density towards the centre of the second core due to the self-gravity of the core results in lower fraction of thermal ionisation (displayed in Fig.~\ref{fig:masscomparison}k). The associated release of energy yields the peak in the radial temperature profile. Furthermore, in this density--temperature regime, the gas also departs from being fully thermally dissociated in hydrogen (see inset in Fig.~\ref{fig:masscomparison}l), which possibly plays a role in producing this peak.

Towards the high-mass regime, the temperature profiles of the second core becomes flat (i.e.~isothermal) where the shock around 0.1 -- 1~au is hotter than in the core centre. In case this behaviour would hold as the core evolves further (in the high-mass case, accretion happens up to the ignition of hydrogen burning and longer), this will severely affect its internal evolution. The effect of the outer boundary conditions is visible in some of the profiles. For example, the sharp discontinuity seen in the temperature profiles for the high-mass cases is because the temperature at the outer boundary is set to a fixed value of 10~K. However, this does not have any significant effects on the evolution and properties of the hydrostatic cores on the smaller scales. The bumps seen in the radial Mach number profiles for the high-mass cases are due to the behaviour of the adiabatic index $\Gamma_1$ during the dissociation and ionisation phase. In the high-mass regime, the temperature at the bump positions in the radial temperature profiles corresponds to the required temperatures for thermal hydrogen dissociation and ionisation. The radial velocity in the inner core regions, which are in hydrostatic equilibrium, fluctuates around the zero value. This effect is visible as the noise or spikes in the radial profiles of the ratio of gas to ram pressure in Fig.~\ref{fig:masscomparison}h. The radial profiles of the optical depth indicates that the second cores are optically thick, which makes it extremely difficult to detect them observationally and trace this evolutionary stage. 

The region of fully atomic hydrogen in the dissociation profile extends far beyond the second core radius. The underlying reason is that the infalling gas in front of the second core radius (indicated by the shock) is already heated to temperatures beyond the dissociation temperature. This effect becomes especially clear for cases of higher accretion rates (i.e.~higher accretion energy).

\begin{table*}[ht!]
	\centering
	\caption{Properties of the second core estimated at the final simulation snapshot $\mathrm{t_{final}}$ when central density $\mathrm{\rho_{c, final}}$ reaches $\approx$ 0.5 -- 0.8~$\mathrm{g ~cm^{-3}}$, for different initial cloud core masses $M_{0}$ with a fixed outer radius $R_\mathrm{cloud} = 3000$ au and an initial temperature $T_{\mathrm{0}}$~=~10~K. }
		\begin{tabular}{cccccccc}
			\hline
			$M_{0} ~(M_{\odot})$ & $\mathrm{\rho_{c, final}} ~(\mathrm{g ~cm^{-3}})$ & $\mathrm{t_{final}}$ (yr) & $R_{\mathrm{sc}}$ (au)  & $M_{\mathrm{sc}} ~(M_{\odot})$ & $T_{\mathrm{sc}}$ (K)  & $\dot{M}_\mathrm{sc} ~(M_{\odot}~yr^{-1})$ & ${L}_\mathrm{acc} ~(L_{\odot})$  \TBstrut\\ \hline \hline
			0.5 & 0.854 & 38039.76 & 0.203 & 3.36e-02 & 2.07e+03 & 3.23e-05 &  ~7.78e-01  \Tstrut \\
			1.0 & 0.851 & 18635.00 & 0.253 & 4.40e-02 & 1.79e+03 & 6.52e-05 & 1.65e+00 \\
			2.0 & 0.853 & 11543.76 & 0.329 & 6.58e-02 & 1.67e+03 & 1.52e-04 & 4.43e+00 \\
			5.0 & 0.849 & 6860.80 & 0.405 & 1.65e-01 & 1.54e+03 & 5.53e-04 & 3.27e+01 \\
			8.0 & 0.848 & 5433.00 & 0.613 & 3.62e-01 & 1.50e+03 & 1.18e-03 & 1.01e+02 \\
			10.0 & 0.853 & 4909.00 & 0.876 & 5.73e-01 & 1.56e+03 & 1.71e-03 & 1.64e+02 \\
			12.0 & 0.849 & 4530.70 & 0.987 & 8.31e-03 & 8.62e+03 & 2.32e-03 & 2.85e+02 \\
			14.0 & 0.848 & 4251.80 & 1.290 & 1.17e+00 & 8.89e+03 & 3.01e-03 & 3.97e+02 \\
			15.0 & 0.835 & 4138.62 & 1.359 & 1.36e+00 & 8.77e+03 & 3.26e-03 & 4.77e+02 \\
			16.0 & 0.825 & 4015.29 & 1.455 & 1.50e+00 & 8.77e+03 & 3.61e-03 & 5.41e+02 \\
			18.0 & 0.792 & 3799.47 & 1.656 & 1.77e+00 & 8.77e+03 & 4.34e-03 & 6.75e+02 \\
			20.0 & 0.761 & 3610.66 & 1.914 & 2.01e+00 & 8.99e+03 & 5.17e-03 & 7.93e+02 \\
			30.0 & 0.662 & 2996.06 & 2.829 & 3.57e+00 & 8.97e+03 & 9.64e-03 & 1.77e+03 \\
			40.0 & 0.596 & 2623.36 & 3.803 & 5.25e+00 & 8.96e+03 & 1.50e-02 & 3.02e+03 \\
			60.0 & 0.519 & 2160.45 & 5.605 & 8.48e+00 & 1.04e+04 & 2.77e-02 & 6.11e+03 \\
			80.0 & 0.474 & 1859.02 & 5.486 & 1.08e+01 & 1.11e+04 & 4.20e-02 & 1.21e+04 \\
			100.0 & 0.448 & 1655.55 & 5.206 & 1.31e+01 & 1.16e+04 & 5.78e-02 & ~2.12e+04 \Bstrut \\ \hline
		\end{tabular}
	\tablefoot{The properties listed are the second core radius $R_{\mathrm{sc}}$, mass $M_{\mathrm{sc}}$, temperature $T_{\mathrm{sc}}$, accretion rate $\dot{M}_\mathrm{sc}$, and accretion luminosity~${L}_\mathrm{acc}$.}
	\label{tab:SCproperties}
\end{table*}

\subsection{Second core properties}
\label{sec:1Dsecondcore}

\begin{figure}[!htp]
	\centering
	\begin{subfigure}{0.47\textwidth}
		\includegraphics[width=\textwidth]{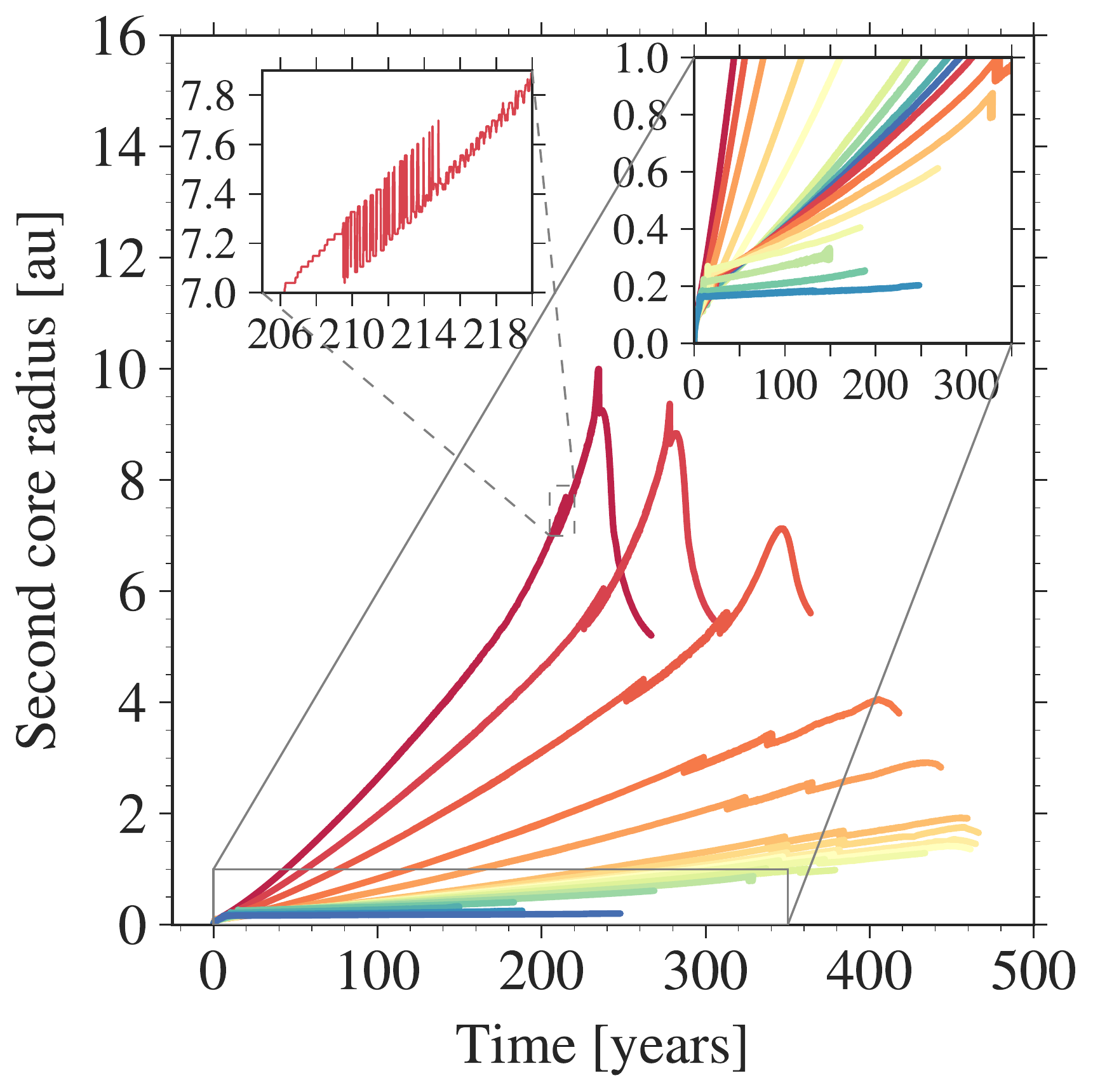}
	\end{subfigure}
	\begin{subfigure}{0.4\textwidth}
		\hspace{0.4cm}
		\includegraphics[width=\textwidth]{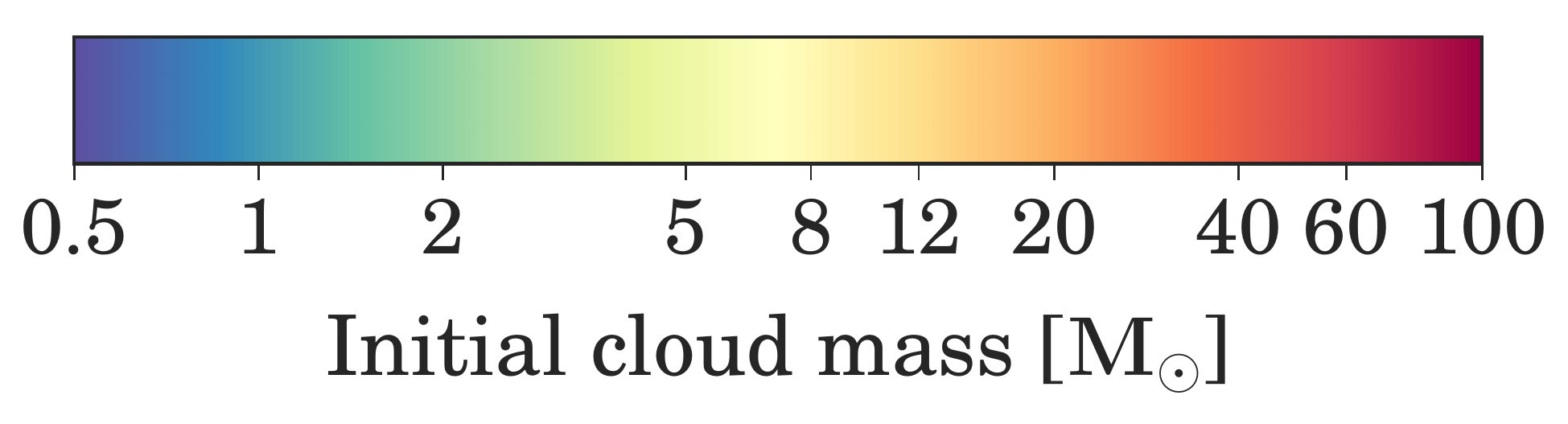}
		\vspace{0.3cm}
	\end{subfigure}	
	\begin{subfigure}{0.47\textwidth}
		\includegraphics[width=\textwidth]{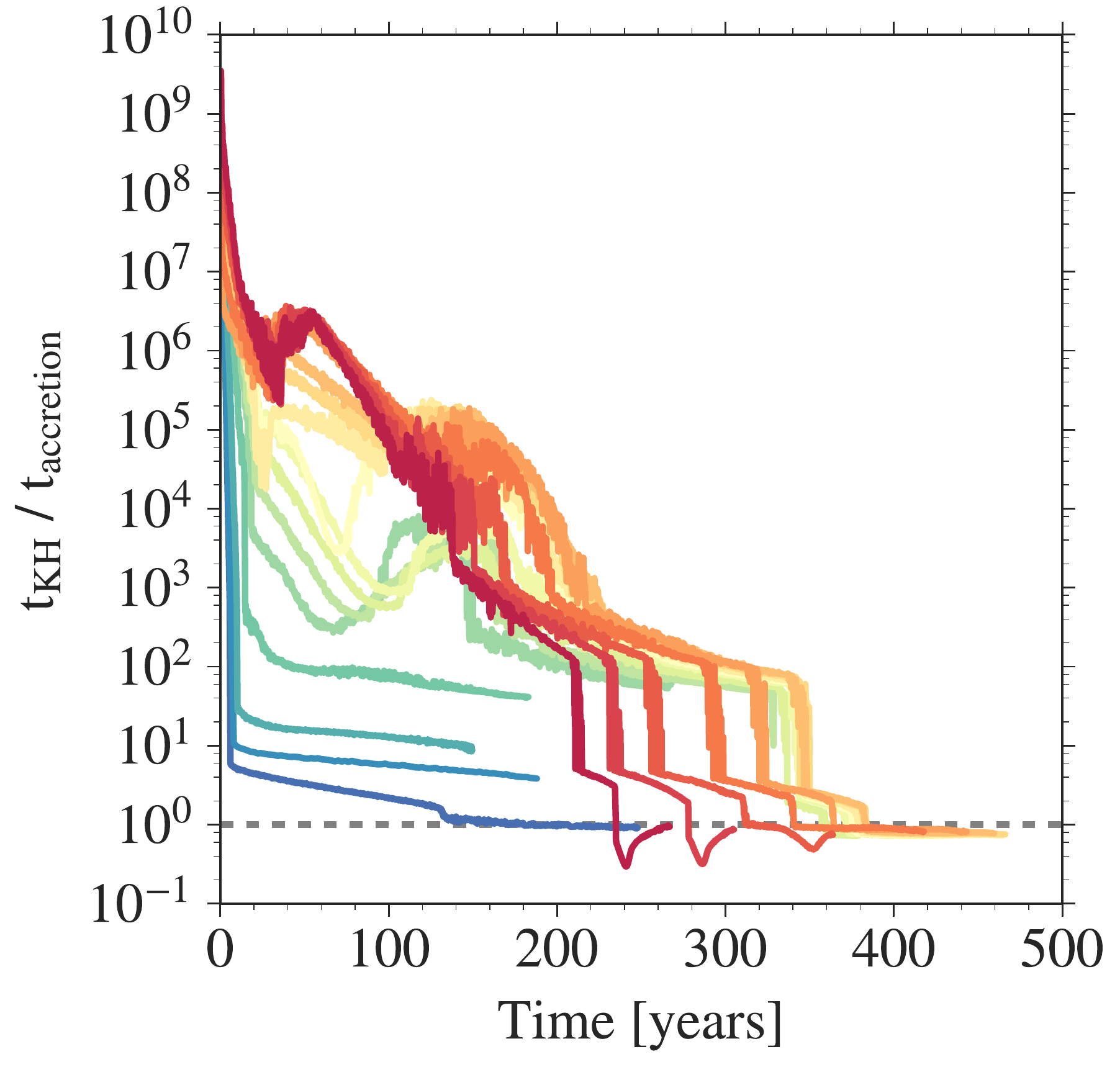}
	\end{subfigure}	
	\caption{\textit{Top:} Spatial evolution of the second core radius as a function of time for all the collapse scenarios with different initial cloud core masses ranging from 0.5~$M_\odot$ to 100~$M_\odot$ as indicated in the colour bar. The evolution is traced from the onset of the second core formation until the central density reaches $\approx$ 0.5 -- 0.8 $\mathrm{g ~cm^{-3}}$. The inset in the upper left zooms in on the back and forth behaviour for one of the curves. This results from the jump between the two local minima in the velocity profile of the accretion shock, which is used to define the second core radius.
	\textit{Bottom:} Comparison of the Kelvin--Helmholtz and accretion timescales. A large (small) ratio is associated with an expansion (contraction) phase of the second hydrostatic core (see top panel).}
	\label{fig:SCevolution}
\end{figure}

In this section, we discuss the dependence of the second core properties on the initial cloud core mass. Figure~\ref{fig:SCevolution} shows the spatial evolution of the second core radius over a period of time from the onset of the second core formation until the central density reaches $\approx$ 0.5 -- 0.8 $\mathrm{g ~cm^{-3}}$. The second core radius is defined using the position of the accretion shock in the velocity profile, which is similar to the position of the discontinuity or sharp rise in the density profile. In this work, the onset of formation of the second core is defined as the time when a prominent second accretion shock is visible in the velocity profile. The central density is greater than $\mathrm{10^{-2} ~g ~cm^{-3}}$ at this time snapshot in our simulations. 

In the high-mass regime, we find that, initially, the second core gradually expands with time, thus growing in size. This initial expansion occurs since the Kelvin--Helmholtz timescale\footnote{The Kelvin--Helmholtz and accretion timescales are computed using
\begin{align*}
\mbox{$t_\mathrm{KH} = \dfrac{G M_\mathrm{sc}^2}{L_\mathrm{sc} ~R_\mathrm{sc}}$} ~\mathrm{and}~ 
\mbox{$t_\mathrm{accretion} = \dfrac{M_\mathrm{sc}}{\dot{M}_\mathrm{sc}}$},
\end{align*}
respectively. Here, the luminosity $L_\mathrm{sc} = 4 \pi R_\mathrm{sc}^2 F_\mathrm{rad}$, where $F_\mathrm{rad}$ is the radiative flux just \textit{outside} the second core radius, that means it includes the cooling flux from the second core as well as the accretion luminosity from the accretion shock.} 
is much greater than the accretion timescale during this phase, as seen in the lower panel in Fig.~\ref{fig:SCevolution}. After reaching a maximum radius, the second core undergoes a phase of contraction during which the accretion timescale dominates. We note a similar behaviour in the low-mass regime where the collapse proceeds relatively slowly.

There is a back and forth behaviour seen in the evolution of the second core radius for the intermediate- and high-mass cases (see top panel in Fig.~\ref{fig:SCevolution}). This effect results from the jump between the two close local minima in the velocity profile of the accretion shock, which is used to define the second core radius. Another contributing factor for this behaviour are the small-scale oscillations of the second accretion shock. Both these effects are resolved due to a high time resolution and do not affect the overall behaviour of the second core radius.

\begin{figure}[!htp]
	\centering
	\begin{subfigure}{0.4\textwidth}
		\includegraphics[width=\textwidth]{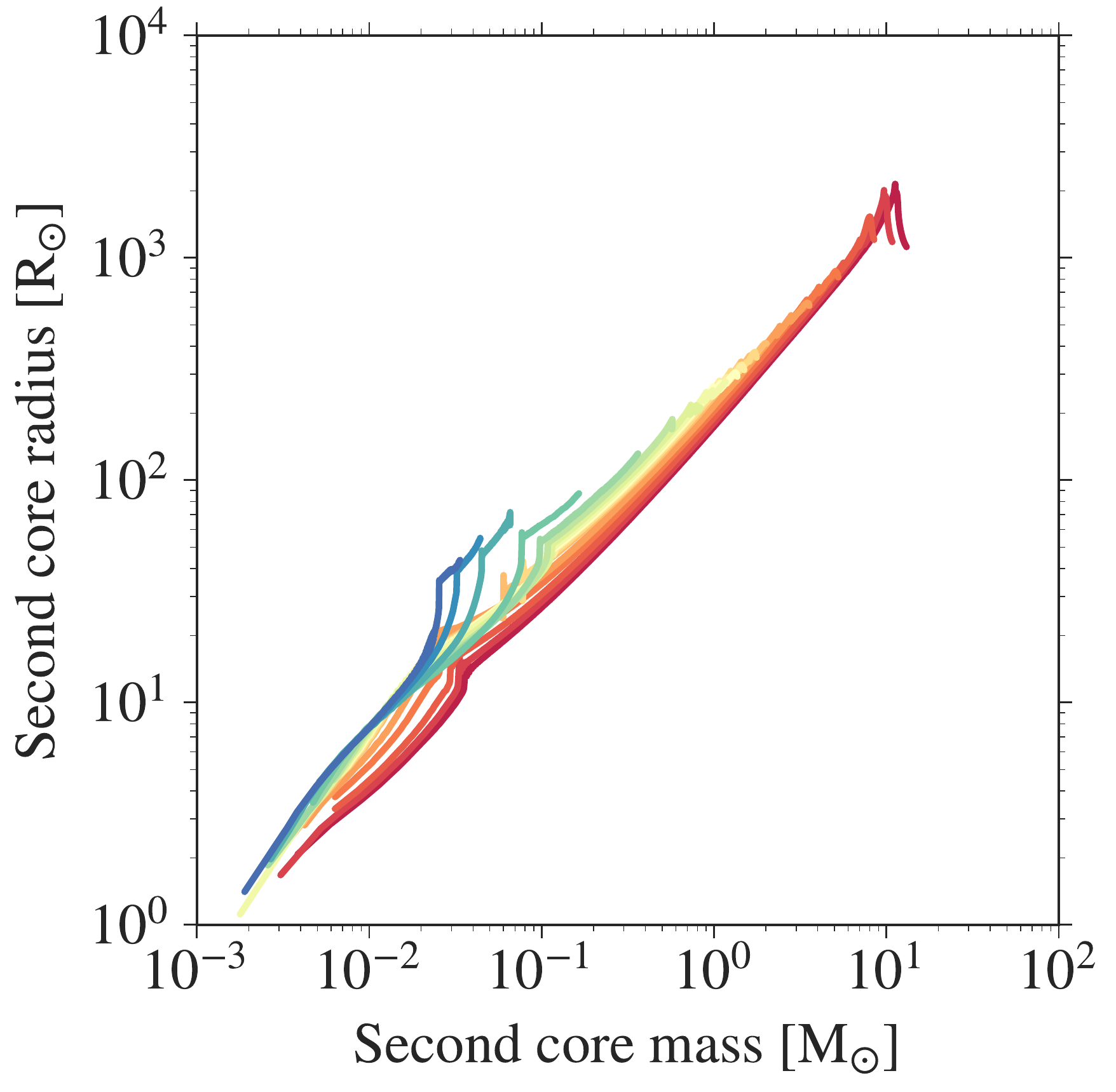}
		\vspace{0.2cm}
	\end{subfigure}
	\begin{subfigure}{0.4\textwidth}
		\includegraphics[width=\textwidth]{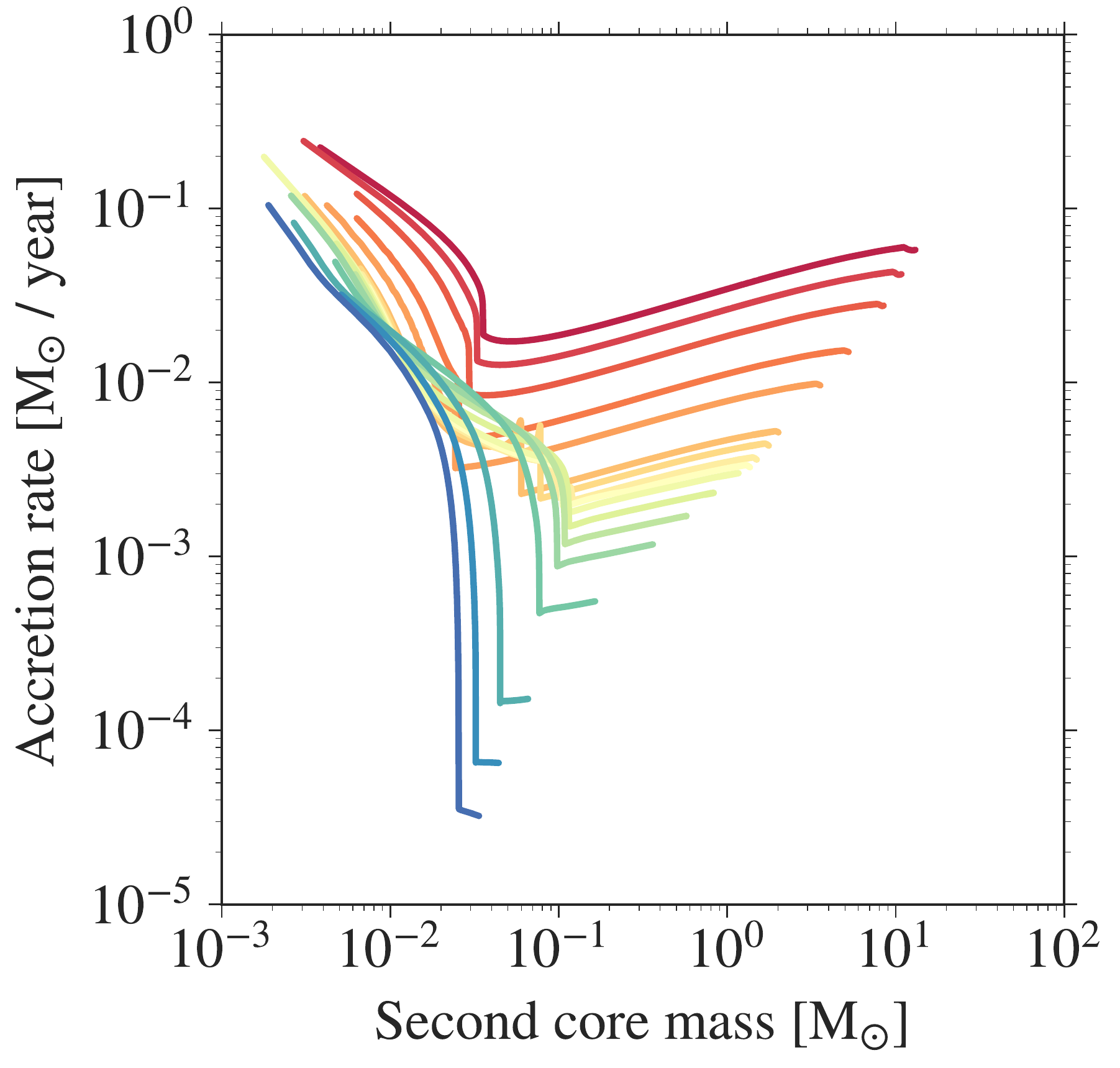}
		\vspace{0.2cm}		
	\end{subfigure}
	\begin{subfigure}{0.4\textwidth}
		\includegraphics[width=\textwidth]{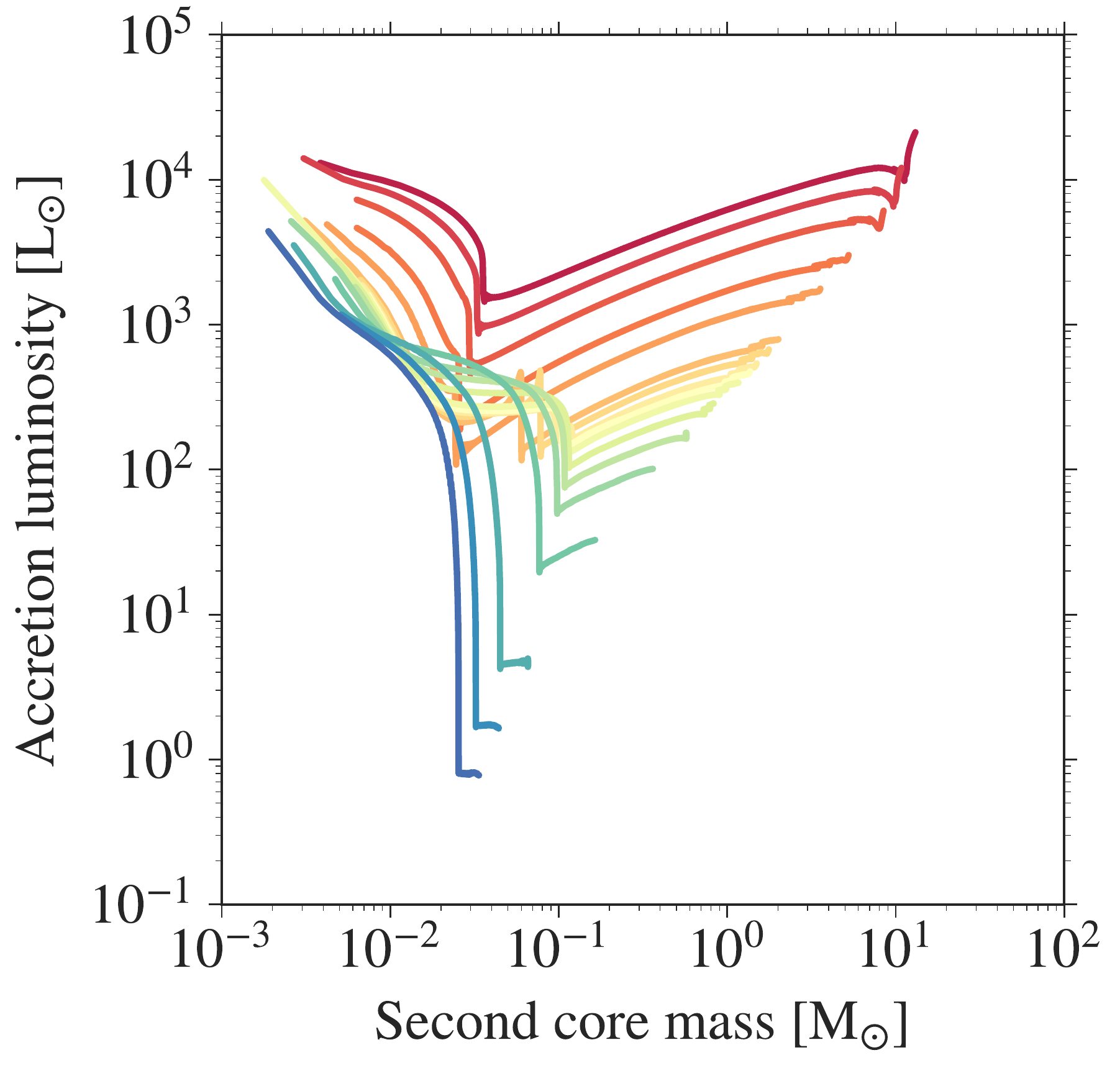}		
	\end{subfigure}		
	\caption{Second core radius (top), accretion rate (middle), and accretion luminosity (bottom) as a function of the enclosed mass for all the collapse scenarios with different initial cloud core masses ranging from 0.5~$M_\odot$ (blue) to 100~$M_\odot$ (red). The evolution is traced from the onset of the second core formation until the central density reaches $\approx$ 0.5 -- 0.8~$\mathrm{g ~cm^{-3}}$. }
	\label{fig:SCproperties}
\end{figure}

For a more quantitative comparison, Fig.~\ref{fig:SCproperties} shows the second core radius, accretion rate, and accretion luminosity as a function of the enclosed mass for different initial cloud core masses. These second core properties are displayed over a period of time from the onset of the second core formation until the central density reaches $\approx$ 0.5 -- 0.8 $\mathrm{g ~cm^{-3}}$ as listed in Table~\ref{tab:SCproperties}. The accretion rate $\dot{M}_\mathrm{sc}$ is estimated as
\begin{align}
\mbox{$ \dot{M}_\mathrm{sc} = 4 \pi ~{R_\mathrm{sc}^2} ~\rho_\mathrm{sc} ~u_\mathrm{sc} $},
\end{align}
where $\rho_\mathrm{sc}$ and $u_\mathrm{sc}$ are the density and velocity at the second core radius $R_\mathrm{sc}$, respectively. The accretion luminosity ${L}_\mathrm{acc}$ computed using the accretion rate and mass $M_\mathrm{sc}$ enclosed within the second core radius is given as
\begin{align}
\mbox{${L}_\mathrm{acc} = \dfrac{G M_\mathrm{sc} \dot{M}_\mathrm{sc}}{R_\mathrm{sc}} $}.
\end{align}

\begin{figure*}[h]
	\centering
	\includegraphics[width=0.85\textwidth]{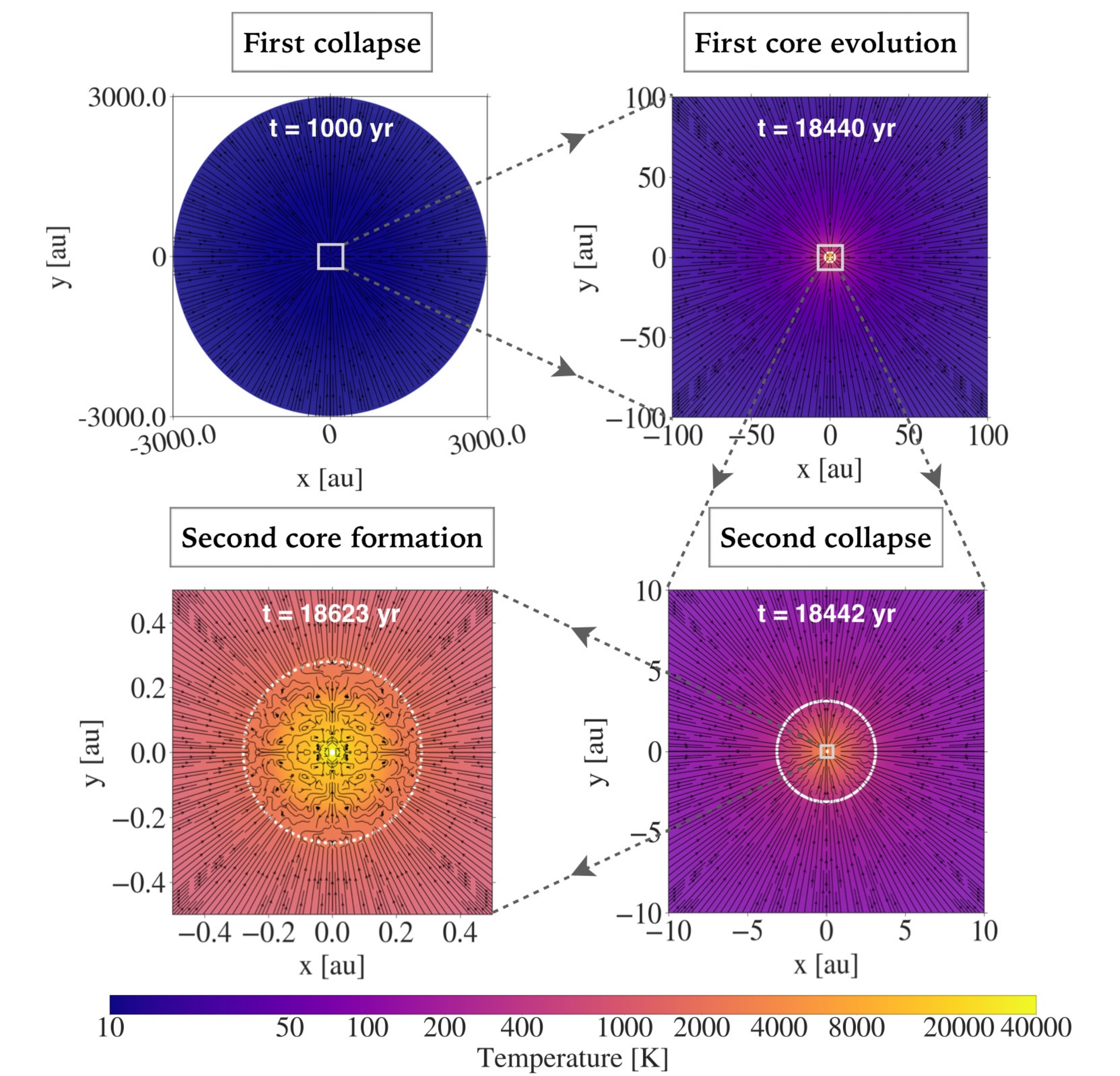}
	\caption{2D temperature snapshots zooming down to sub-au scales showing the evolution of a 1~$M_{\odot}$ cloud core with an initial temperature of 10~K and outer radius of 3000~au. The velocity streamlines in black indicate the infalling material and also the mixing within the second core at the last snapshot. The solid and dashed white contours indicate the first and second accretion shocks, respectively. Note the different spatial scales from top left to bottom left.}
	\label{fig:snapshots}
\end{figure*}

\noindent The second core mass gradually increases as the core evolves through the expansion and contraction phases. As expected, higher initial cloud core masses lead to more massive second cores. Initially, material from the first core accretes onto the second core at a much faster rate of $\approx 10^{-2}~M_{\odot}~yr^{-1}$. The accretion rate slows down over time and can decrease to roughly a few times $10^{-5}~M_{\odot}~yr^{-1}$ in the low-mass end. The accretion luminosity in the high-mass regime is much higher than in the low-mass regime. Various properties of the second core are listed in Table~\ref{tab:SCproperties}. 

\section{Results: a 2D view of the second core}
\label{sec:results}

Results from our 1D studies of formation and evolution of the second hydrostatic core, for a wide range of initial cloud core masses, were discussed in Sect.~\ref{sec:2coreevolution}. In this section, we further expand our investigation to 2D collapse for selected initial cloud core masses of 1~$M_{\odot}$, 5~$M_{\odot}$, 10~$M_{\odot}$, and 20~$M_{\odot}$. The main aim of this study is to resolve the second core using a resolution that has not been achieved before. The details of the computational grid and the resolution are discussed in Sect.~\ref{sec:2Dgrid}.

The 2D simulations presented herein are basically a scaled-up version of the 1D runs; cloud rotation is not included as the cloud core is initialised as being at rest. We use the same initial conditions as discussed in Sect.~\ref{sec:setup} with a fixed outer radius of 3000~au and a constant initial temperature of 10~K. We account for the effects of self-gravity, radiation, and phase transitions on the evolution of these pre-stellar cores. 

\begin{figure*}[!htp]
	\centering
	\hspace*{-1cm}	
	\begin{subfigure}{0.243\textwidth}
		\includegraphics[width= 1.2\textwidth]{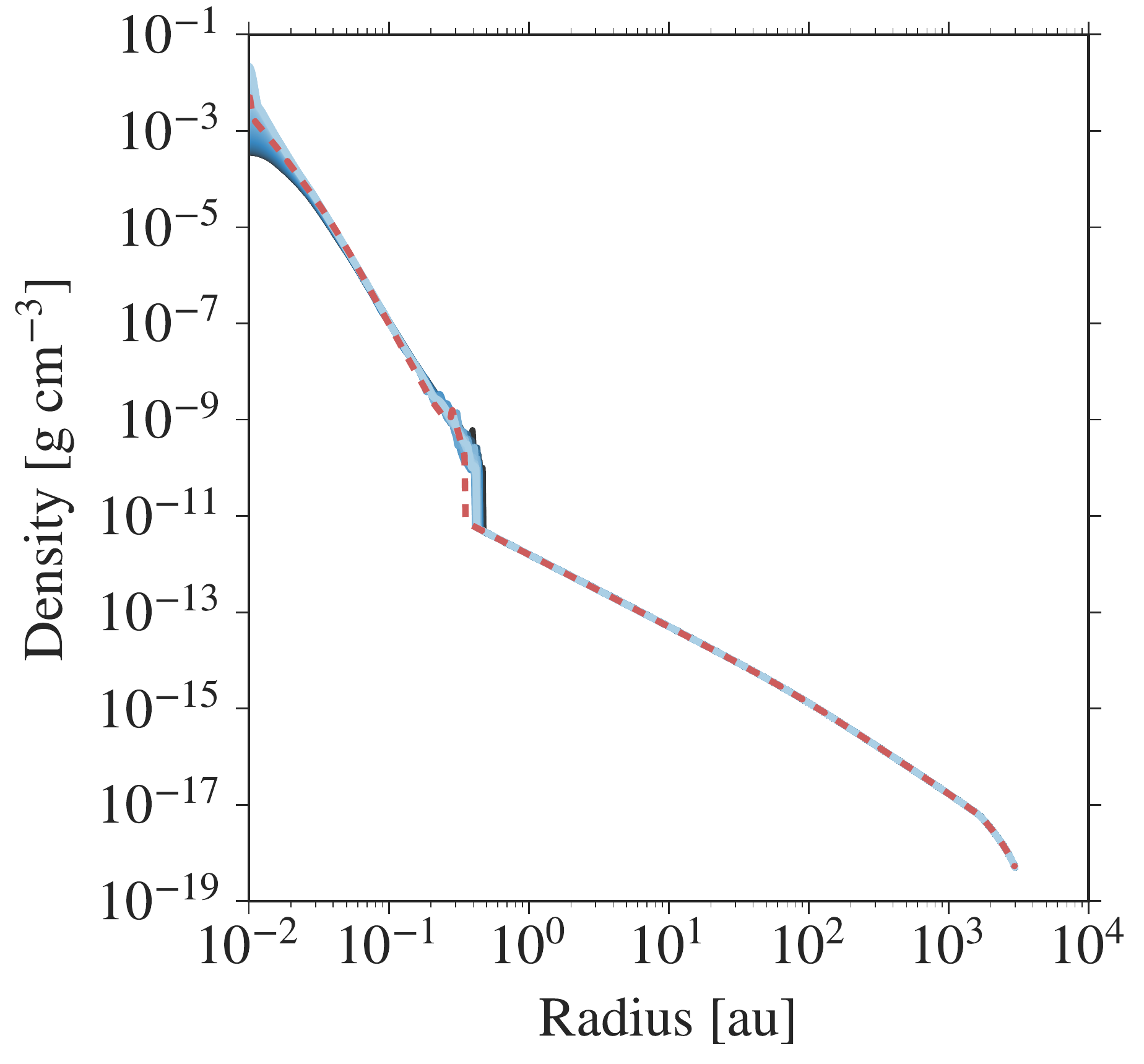}
	\end{subfigure}
	\hspace{0.5in}
	\begin{subfigure}{0.243\textwidth}
		\includegraphics[width= 1.2\textwidth]{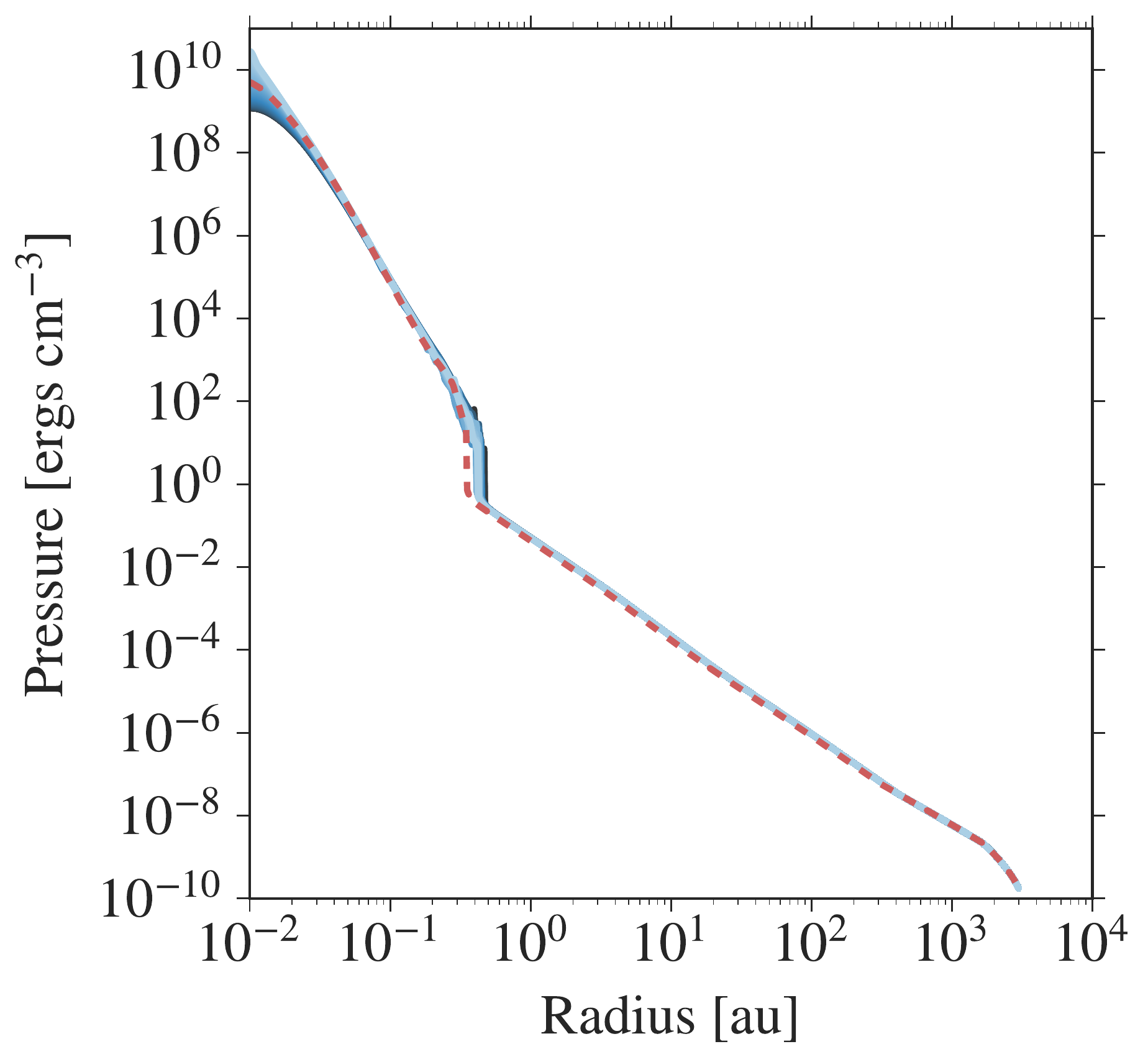}
	\end{subfigure}
	\hspace{0.5in}
	\begin{subfigure}{0.243\textwidth}
		\includegraphics[width= 1.2\textwidth]{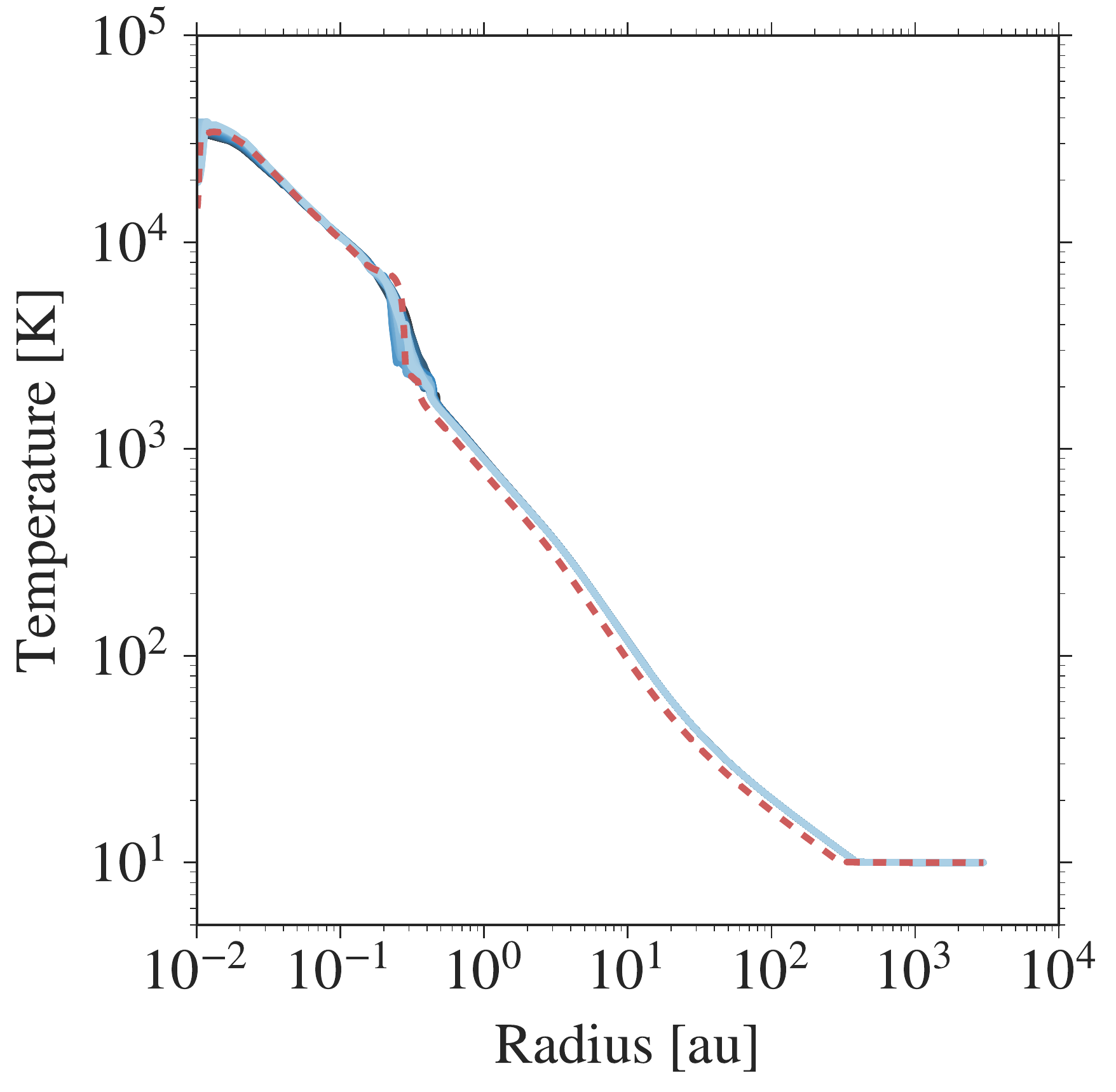}
	\end{subfigure}
	\hspace*{-1cm}	
	\begin{subfigure}{0.243\textwidth}
		\includegraphics[width= 1.2\textwidth]{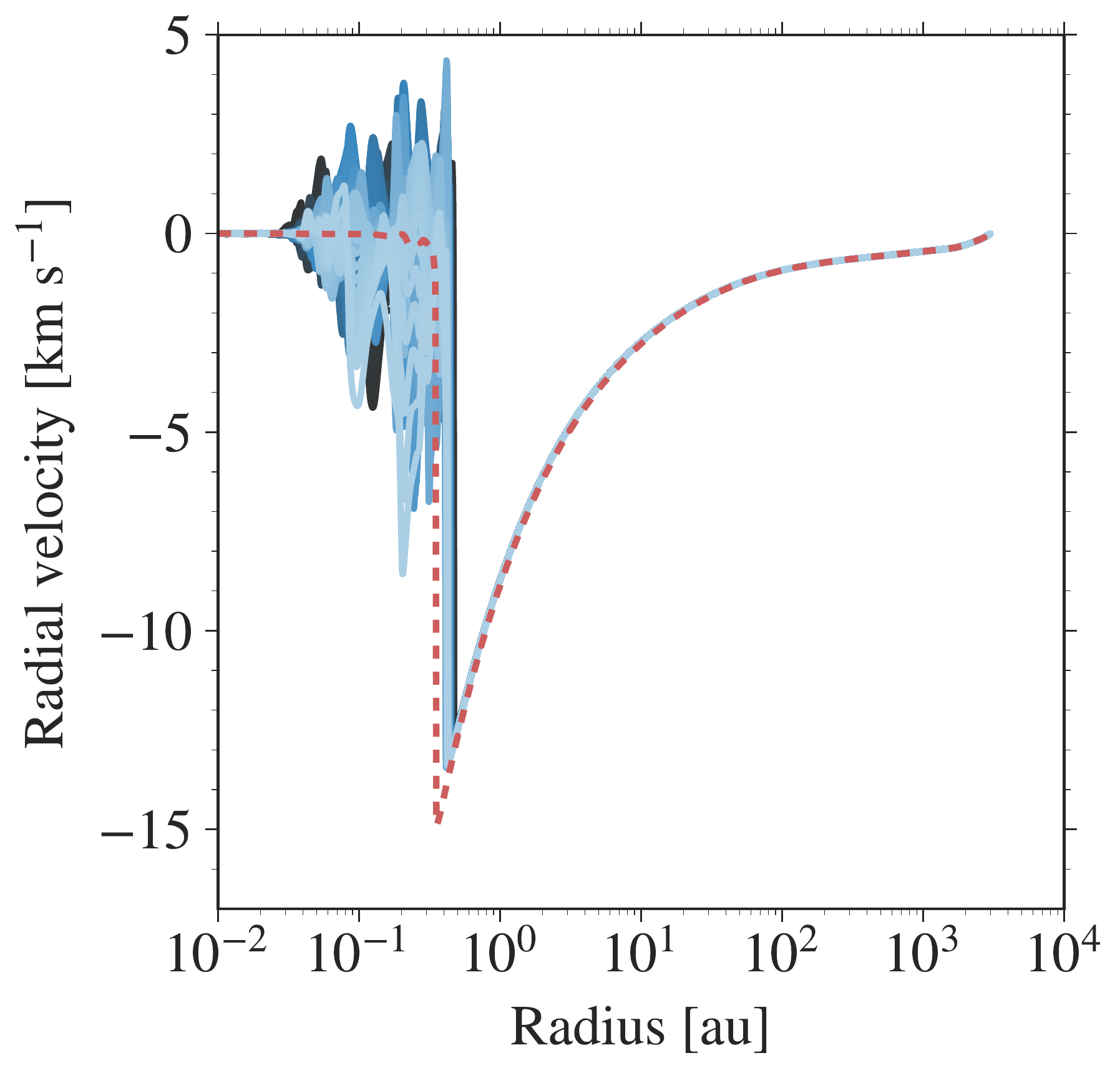}
	\end{subfigure}
	\hspace{0.5in}
	\begin{subfigure}{0.243\textwidth}
		\includegraphics[width= 1.2\textwidth]{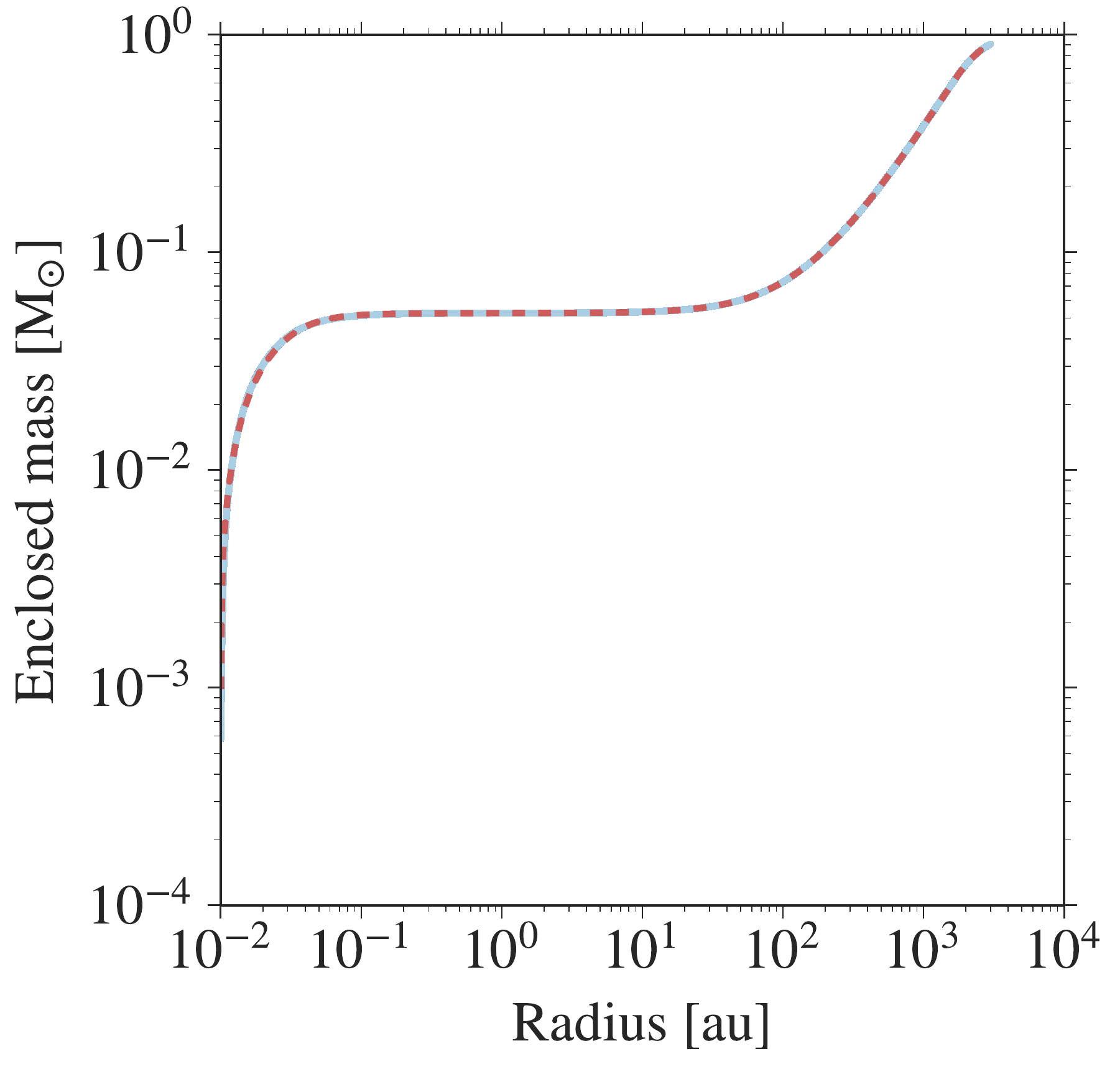}
	\end{subfigure}
	\hspace{0.5in}
	\begin{subfigure}{0.243\textwidth}
		\includegraphics[width= 1.2\textwidth]{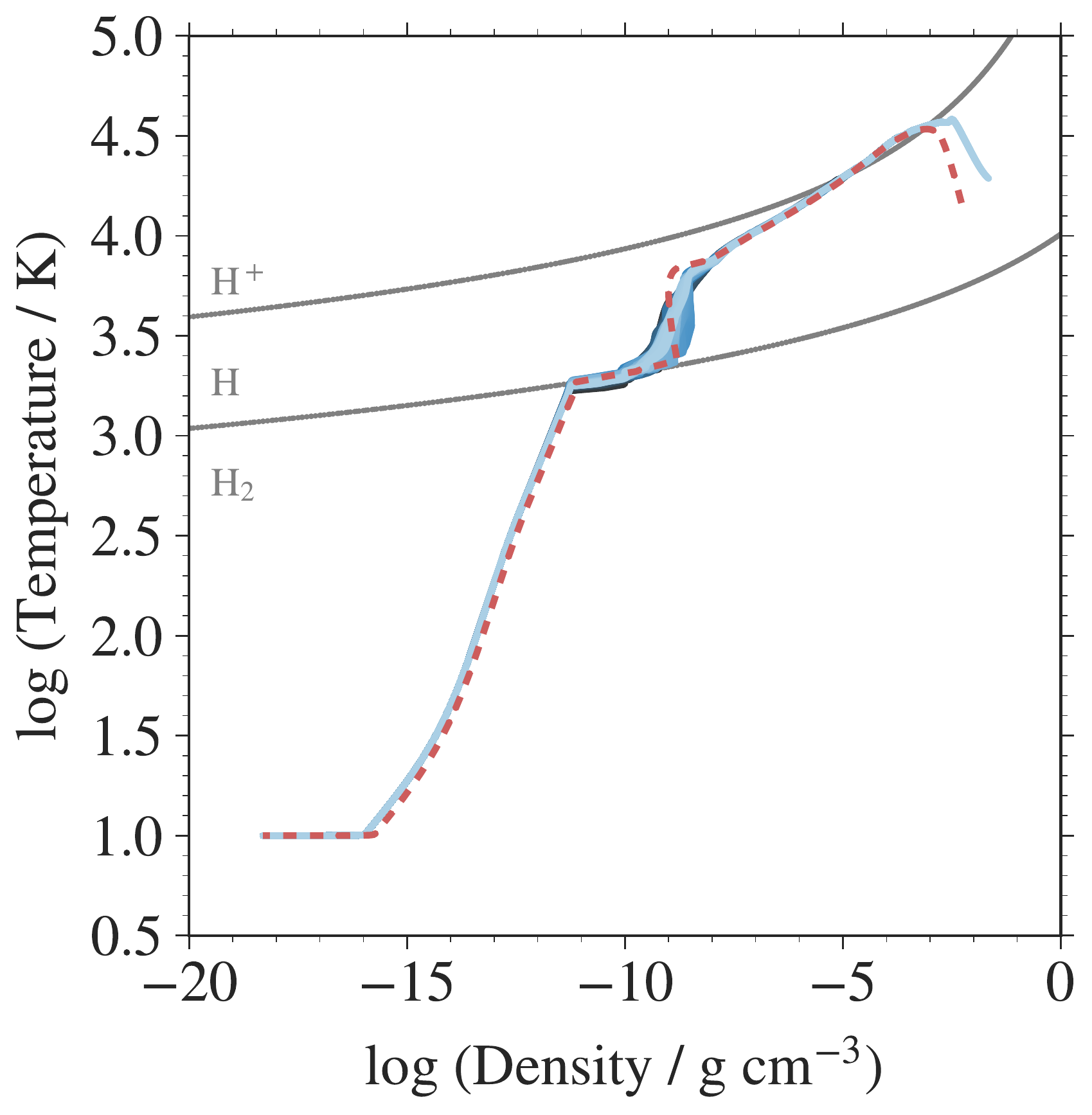}
	\end{subfigure}
	\hspace*{-1cm}	
	\begin{subfigure}{0.243\textwidth}
		\includegraphics[width= 1.2\textwidth]{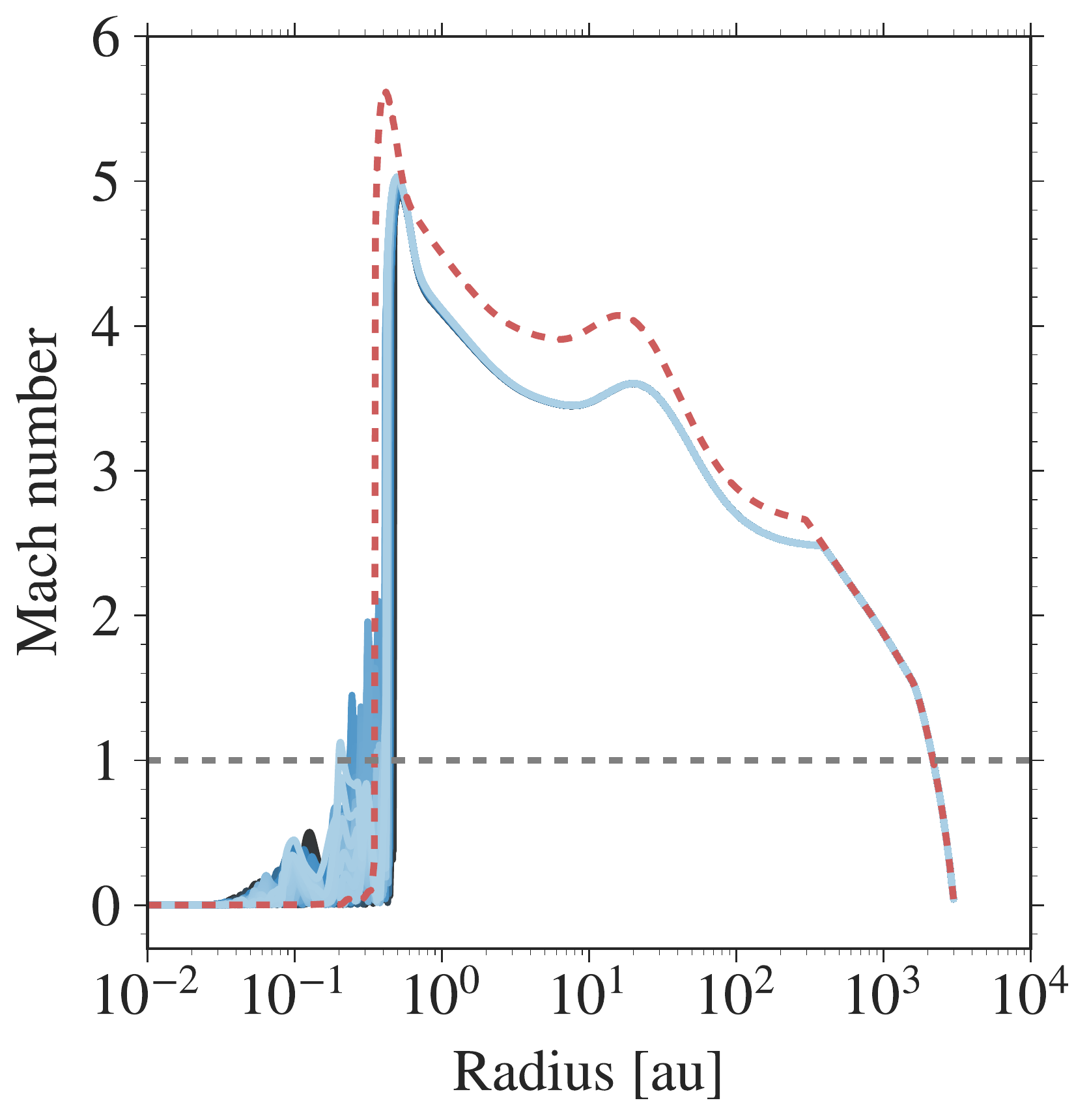}
	\end{subfigure}
	\hspace{0.5in}
	\begin{subfigure}{0.243\textwidth}
		\includegraphics[width= 1.2\textwidth]{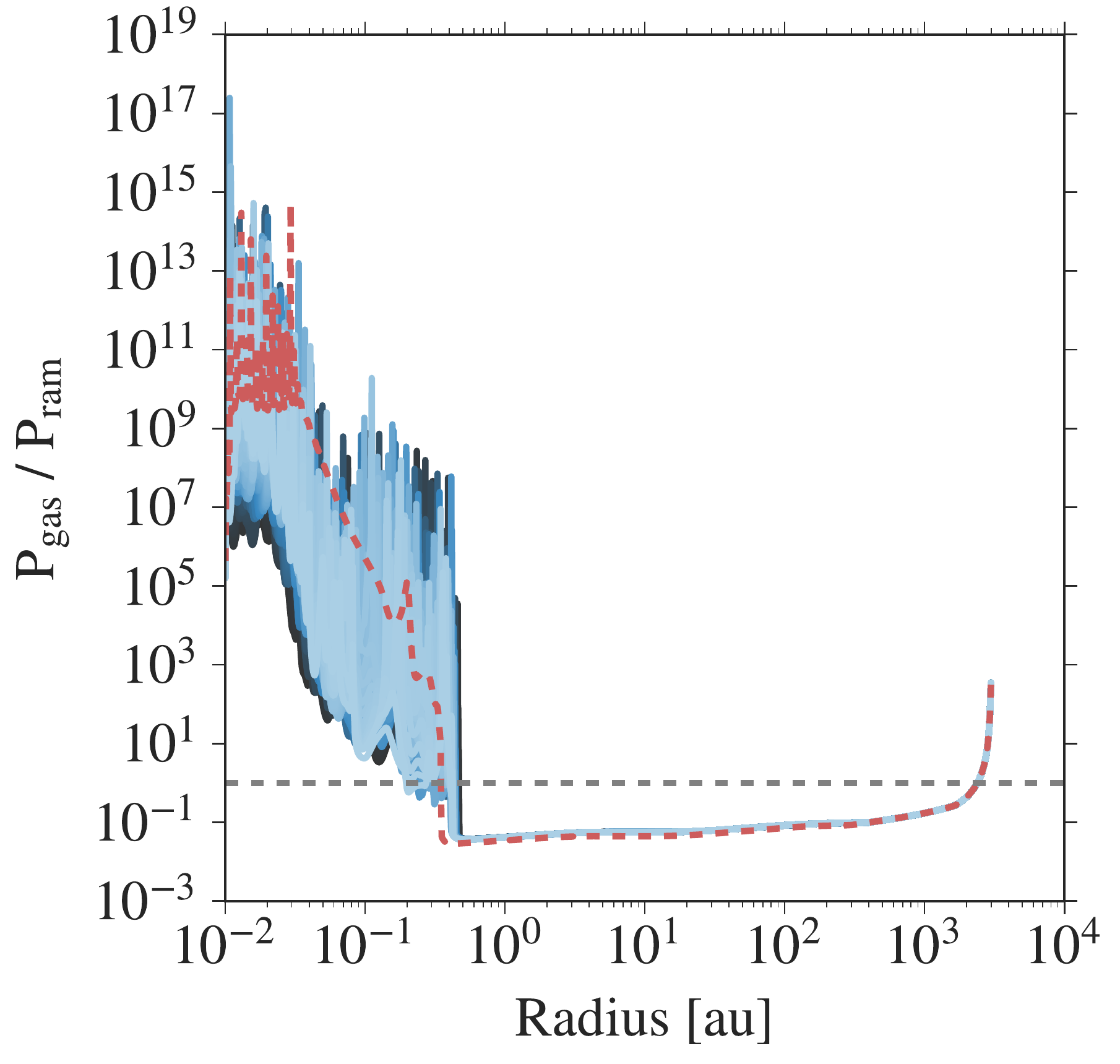}
	\end{subfigure}
	\hspace{0.5in}
	\begin{subfigure}{0.243\textwidth}
		\includegraphics[width= 1.2\textwidth]{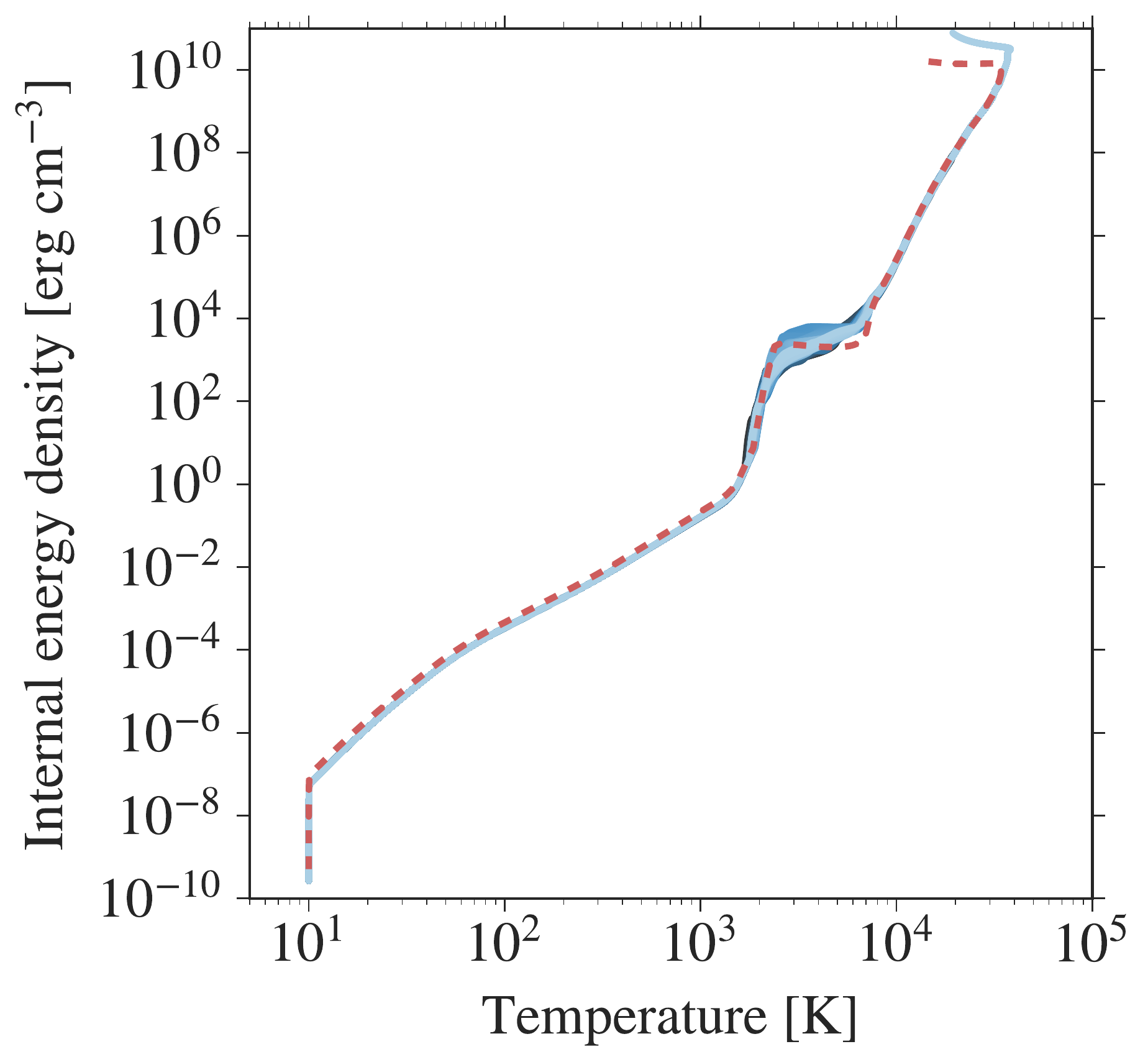}
	\end{subfigure}
	\hspace*{-1cm}	
	\begin{subfigure}{0.243\textwidth}
		\includegraphics[width=1.2\textwidth]{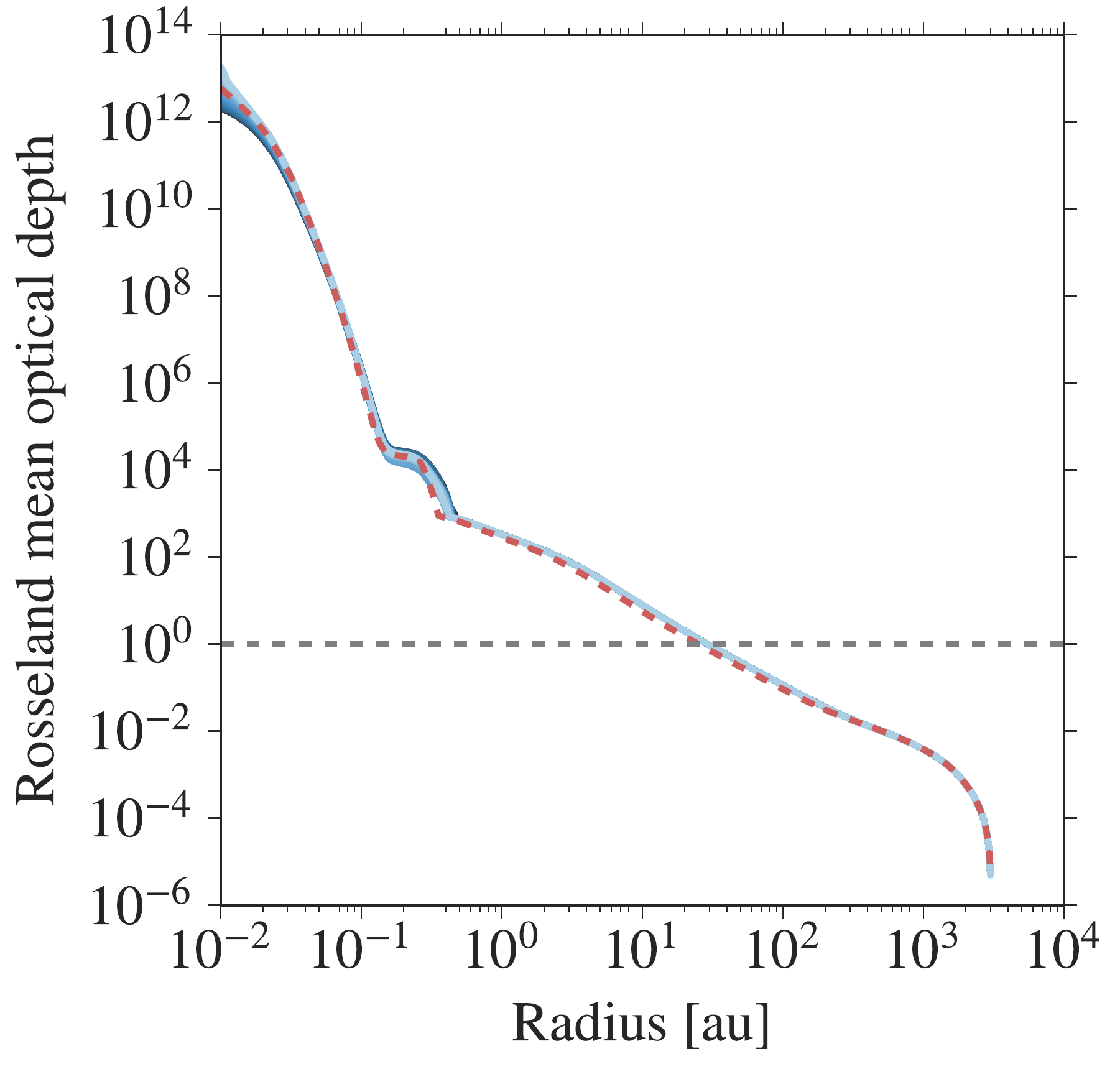}
	\end{subfigure}
	\hspace{0.5in}
	\begin{subfigure}{0.243\textwidth}
		\includegraphics[width= 1.2\textwidth]{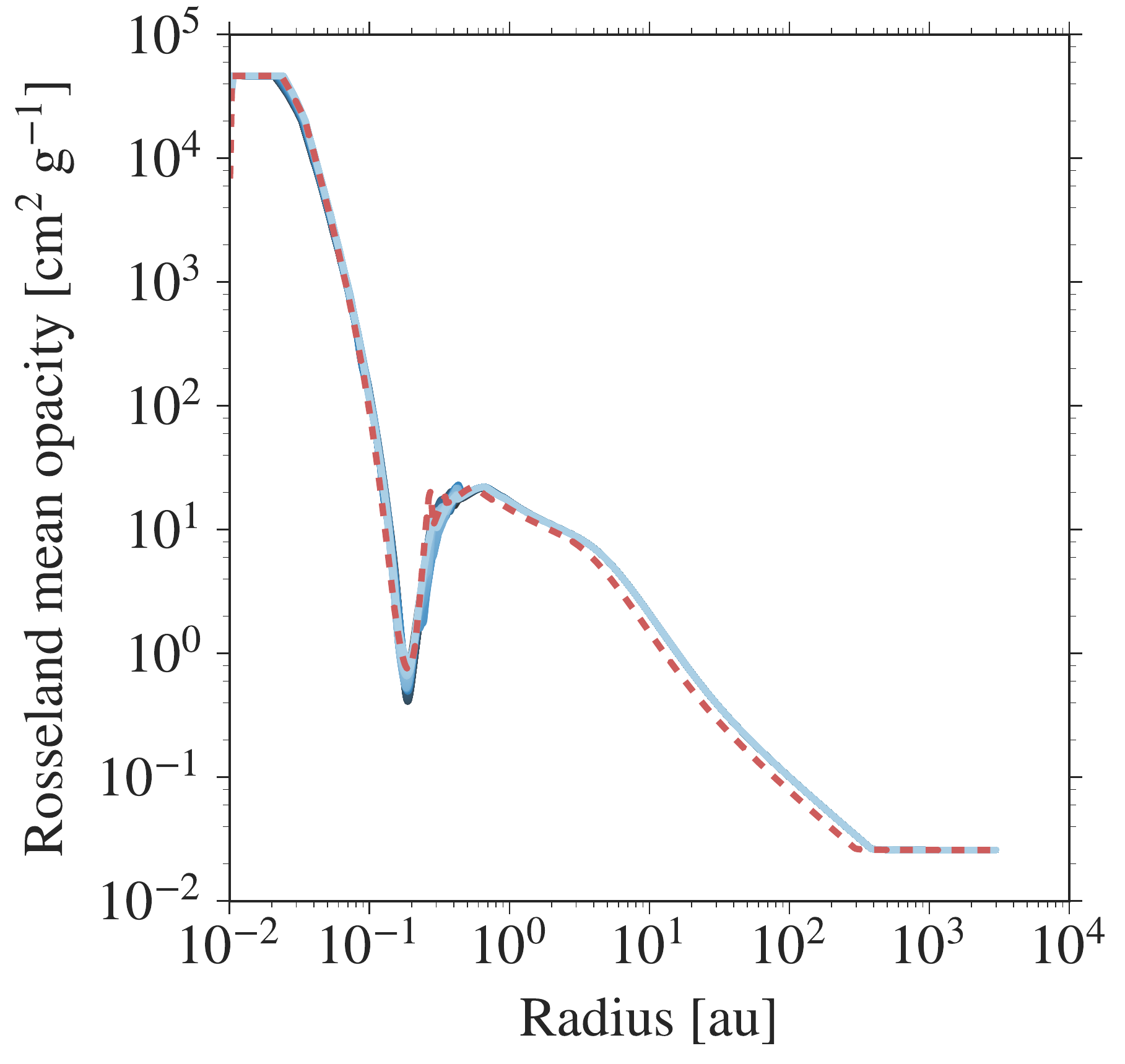}
	\end{subfigure}
	\hspace{0.5in}
	\begin{subfigure}{0.243\textwidth}
		\includegraphics[width= 1.2\textwidth]{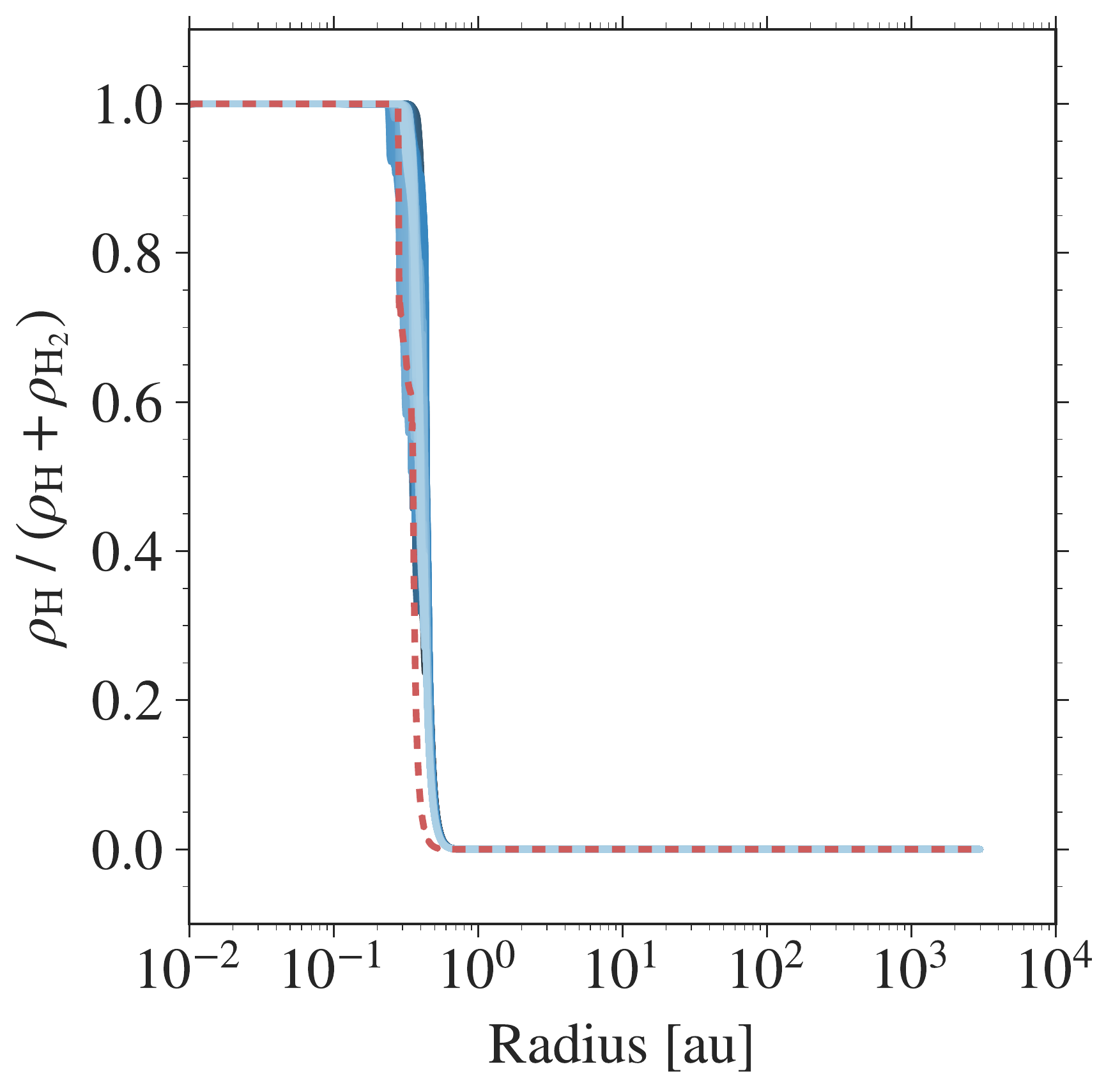}
	\end{subfigure}
	\caption{Radial profiles (across and down) at 312 years after formation of the second core, formed due to the collapse of a 1~$M_{\odot}$ cloud core with an outer radius of 3000~au and an initial temperature of 10~K. The different subplots show the radial profiles (across and down) of \mbox{\bf a)}~density, \mbox{\bf b)}~pressure, \mbox{\bf c)}~gas temperature, \mbox{\bf d)}~radial velocity, and \mbox{\bf e)}~enclosed mass as well as the \mbox{\bf f)}~thermal structure, \mbox{\bf g)}~Mach number, \mbox{\bf h)}~ratio of gas to ram pressure $P_\mathrm{ram} = \rho u^2$, \mbox{\bf i)}~internal energy density as a function of temperature, \mbox{\bf j)}~optical depth, \mbox{\bf k)}~Rosseland mean opacity, and \mbox{\bf l)}~dissociation fraction. The colour gradient from light to dark blue spans the polar angle from the midplane ($\theta = 90^\circ$) to the pole ($\theta = 0^\circ$). The grey lines in the thermal structure plot show the 50\,\%\ dissociation and ionisation curves. The radial profiles from the 1D collapse simulation for the same initial conditions and resolution are over-plotted as a dashed red line in all the subplots. }
	\label{fig:radialprofile}
\end{figure*}

\subsection{Evolution of a fiducial 1~$M_{\odot}$ pre-stellar core}
\label{sec:2Dsims}

\begin{figure*}[!htp]
	\centering
	\includegraphics[width=0.9\hsize]{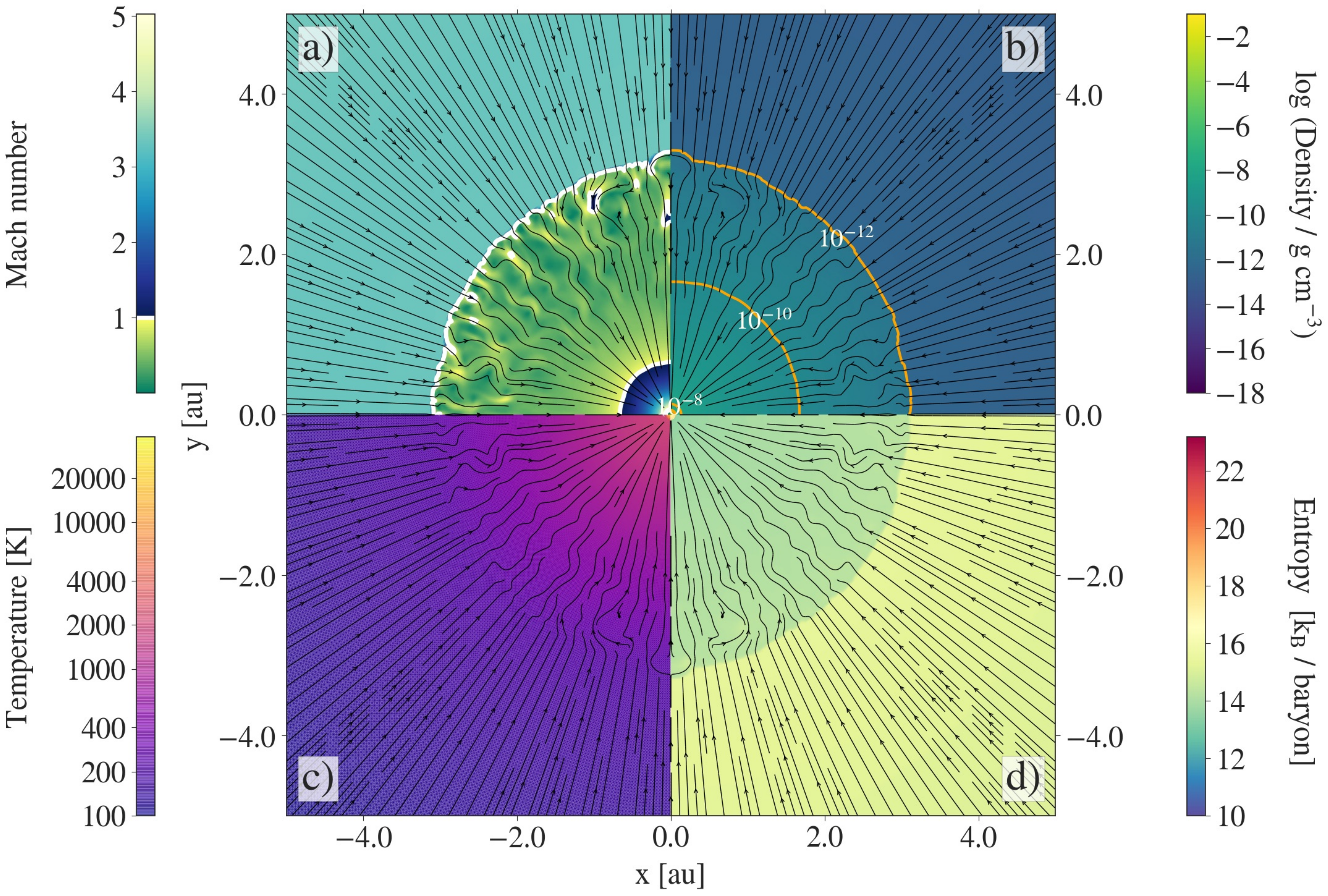}%
	\hspace{-0.9\hsize}%
	\begin{ocg}{fig:FC_off}{fig:FC_off}{0}%
	\end{ocg}%
	\begin{ocg}{fig:FC_on}{fig:FC_on}{1}%
		\includegraphics[width=0.9\hsize]{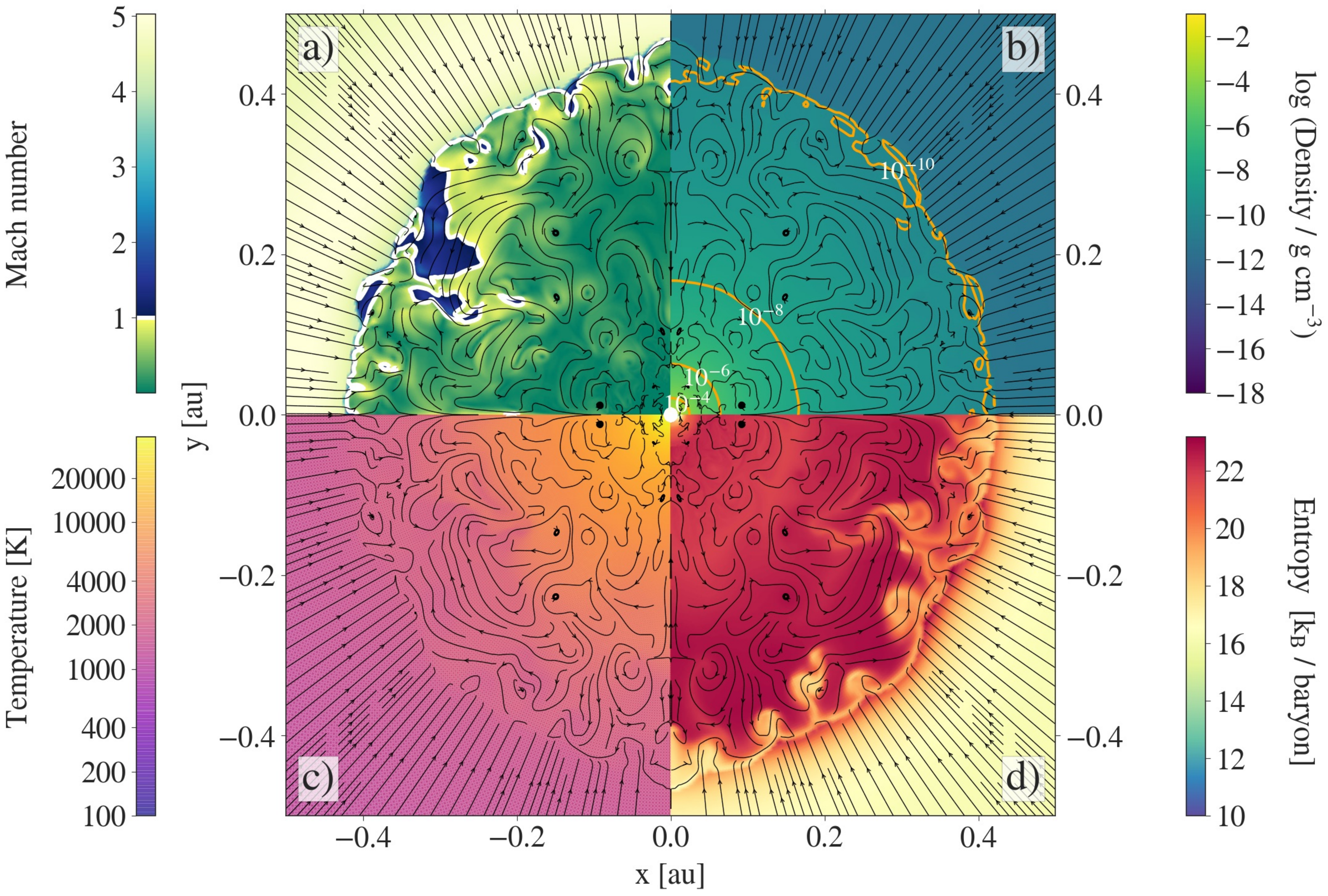}%
	\end{ocg}%
	\caption{2D view of the second hydrostatic core at 312 years after its formation as a result of the collapse of a 1 $M_{\odot}$ cloud core with an initial temperature of 10 K. The four panels show the \mbox{\bf a)}~Mach number, \mbox{\bf b)}~density, \mbox{\bf c)}~temperature, and \mbox{\bf d)}~entropy within the inner 0.5~au of the 3000~au collapsing cloud core. The velocity streamlines in black indicate the material falling on to the second core and the mixing inside the convective second core. The white contour in panel (a) showing Mach = 1.0 separates the sub- and supersonic regions. The different contour lines in panel (b) mark the increase in density towards the centre. When displayed in Adobe Acrobat, it is possible to switch to view the properties of the  \ToggleLayer{fig:FC_off,fig:FC_on}{\protect\cdbox{first core}} at the snapshot of the onset of the second core formation. A movie\protect\footnotemark~of the entire collapse is available online. } 
	\label{fig:2Dcores}
\end{figure*}

In this section, we first focus on the fiducial 1~$M_{\odot}$ case, which evolves through the two stages of the first and second collapse. Figure~\ref{fig:snapshots} shows temperature snapshots at different stages of the evolution zooming into a 3000~au cloud core down to $10^{-2}$~au (i.e.~sub-au scales), thus covering five orders of magnitude in spatial scale.

We follow the evolution of the second core for $\approx$ 312 years after its formation where the onset $t=0$ is defined when a prominent second accretion shock is seen in the velocity profile. The central temperature at this stage is around 40000~K. The gas and radiation temperatures are equal everywhere in our simulations.

Figure~\ref{fig:radialprofile} shows the radial profiles of different properties of the cloud core at the final time snapshot (312 years after second core formation) of our simulation. The gradient from light to dark blue covers the polar angle range from the midplane ($\theta = 90^\circ$) to the pole ($\theta = 0^\circ$), respectively. We also compare this to the results from our 1D collapse studies of a 1~$M_{\odot}$ cloud core, which is indicated by the dashed red line in all the subplots in Fig.~\ref{fig:radialprofile}. The drop seen in the innermost part of the radial temperature profile is dependent on the inner radius and hence is affected by the inner boundary conditions. However, as discussed further in Appendix~\ref{sec:Rin}, we confirm that this does not affect the second core properties nor violates energy conservation at the inner boundary. 

As expected, the same initial conditions lead to similar evolution in 1D and 2D. In both cases, the cloud core has evolved through the two phases of the first and second collapse, until a stage where the first core is no longer present and only the second core is visible, as indicated by the accretion shock in the radial velocity profile. 

The four panels in Fig.~\ref{fig:2Dcores}, showing the 2D view of the second hydrostatic core (zooming into the inner 0.5~au), indicate the Mach number, density, temperature, and entropy structure of the second core. The infalling gas flow and internal mixing are indicated by the black velocity streamlines. The white contour in Fig.~\ref{fig:2Dcores} (panel~a) indicates Mach number equal to 1.0. This sets a clear separation between the supersonic outer region and the subsonic second core. This transition is also seen as the strong jump in the radial Mach profile in Fig.~\ref{fig:radialprofile} at the second core radius. The contour lines in the density panel are labelled with density values at the different radial positions, marking an increase towards the centre, which confirms the non-homologous behaviour of the collapsing cloud core. Similar behaviour is seen in the temperature panel with the core centre having the highest value.

\begin{figure*}[!htp]
	\hspace*{0.2cm}	
	\begin{subfigure}{0.42\textwidth}
		\includegraphics[width=1.2\textwidth]{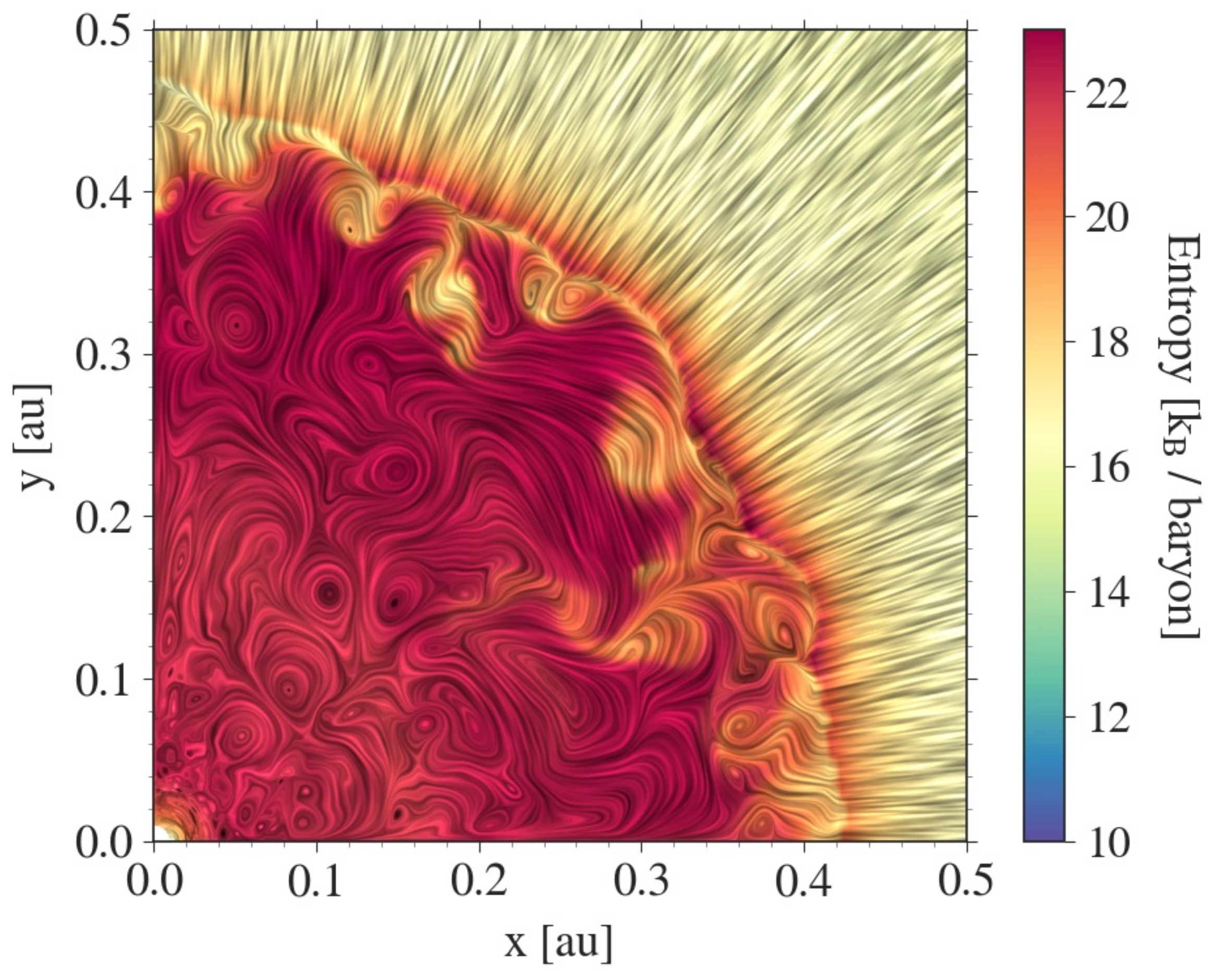}
	\end{subfigure}
	\hspace{2.2cm}
	\begin{subfigure}{0.34\textwidth}
		\includegraphics[width=1.2\textwidth]{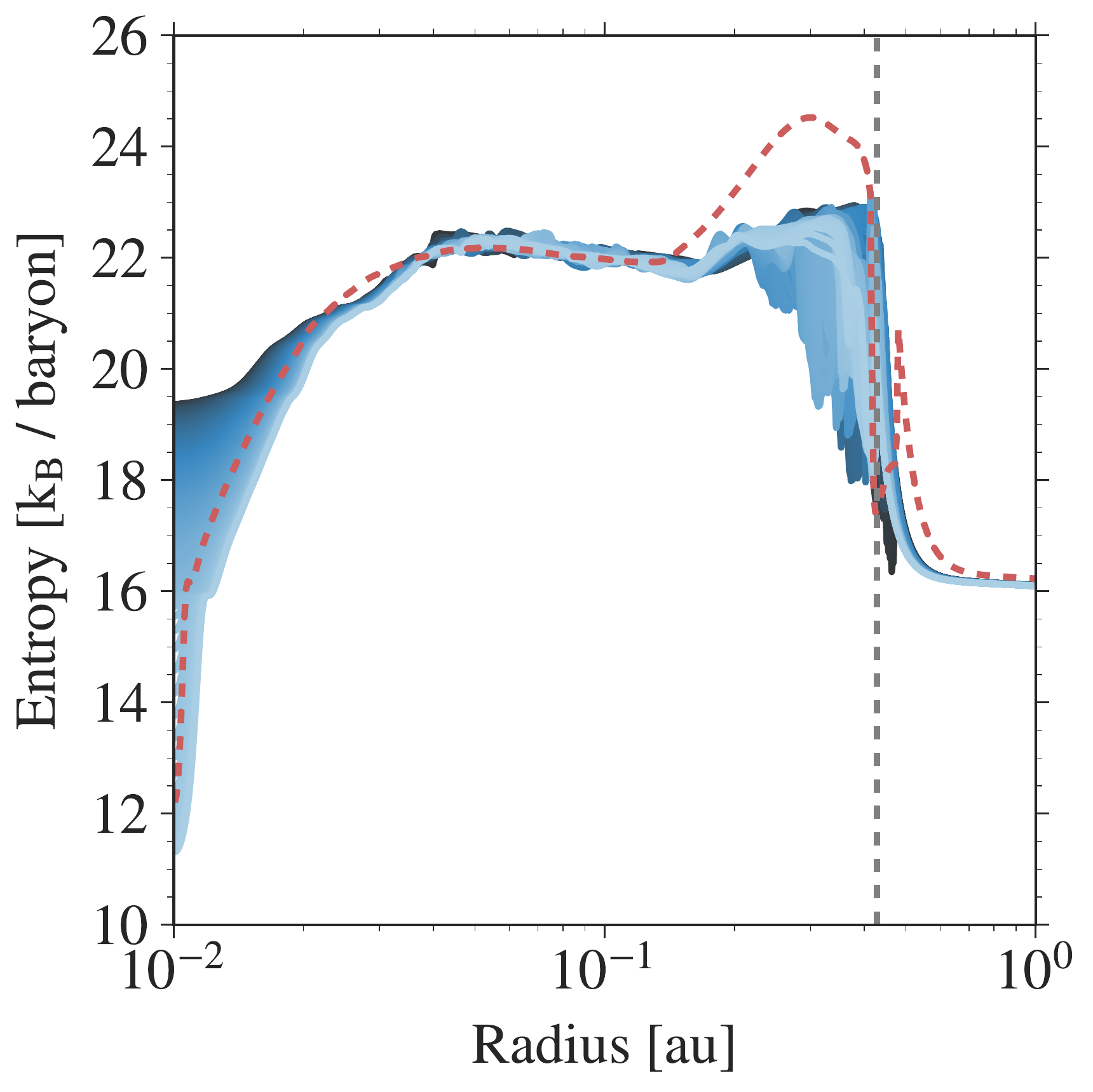}
	\end{subfigure}
	\caption{\textit{Left:} Line integral convolution visualisation of the entropy behaviour indicating the presence of eddies within the second core. Shown here is the inner 0.5~au of the 3000~au collapsing 1~$M_{\odot}$ cloud core at 312 years after the formation of the second core. 
	\textit{Right:} Radial entropy profile for the 1~$M_{\odot}$ case, within the inner 1.0~au of the 3000~au collapsing cloud core at 312 years after the formation of the second core. The vertical grey dashed line indicates the radius of the second core. The colour gradient from light to dark blue spans the polar angle from the midplane ($\theta = 90^\circ$) to the pole ($\theta = 0^\circ$). The dashed red line shows the radial profile from the 1D collapse simulation for the same initial conditions and resolution, which by definition omits the effect of convection.}
	\label{fig:entropy_m1}
\end{figure*}

We show the entropy panel again in Fig.~\ref{fig:entropy_m1} and plot it there with Line Integral Convolution to highlight the turbulent features within the second core. The entropy is calculated by the Sackur--Tetrode equation, which is consistent with the \citet{Dangelo2013} EOS used in the simulation and takes into account the molecular, atomic, and ionised hydrogen as well as the contribution from the electrons. It thus represents a straightforward extension of the expressions in \citet[][Appendix~A]{Berardo2017}, and details will be given in Marleau et al.\ (in prep.). We verified that the two agree well where ionisation (not included in \citealt{Berardo2017}) is not important.

\begin{figure}[t]
	\centering
	\includegraphics[width=0.47\textwidth]{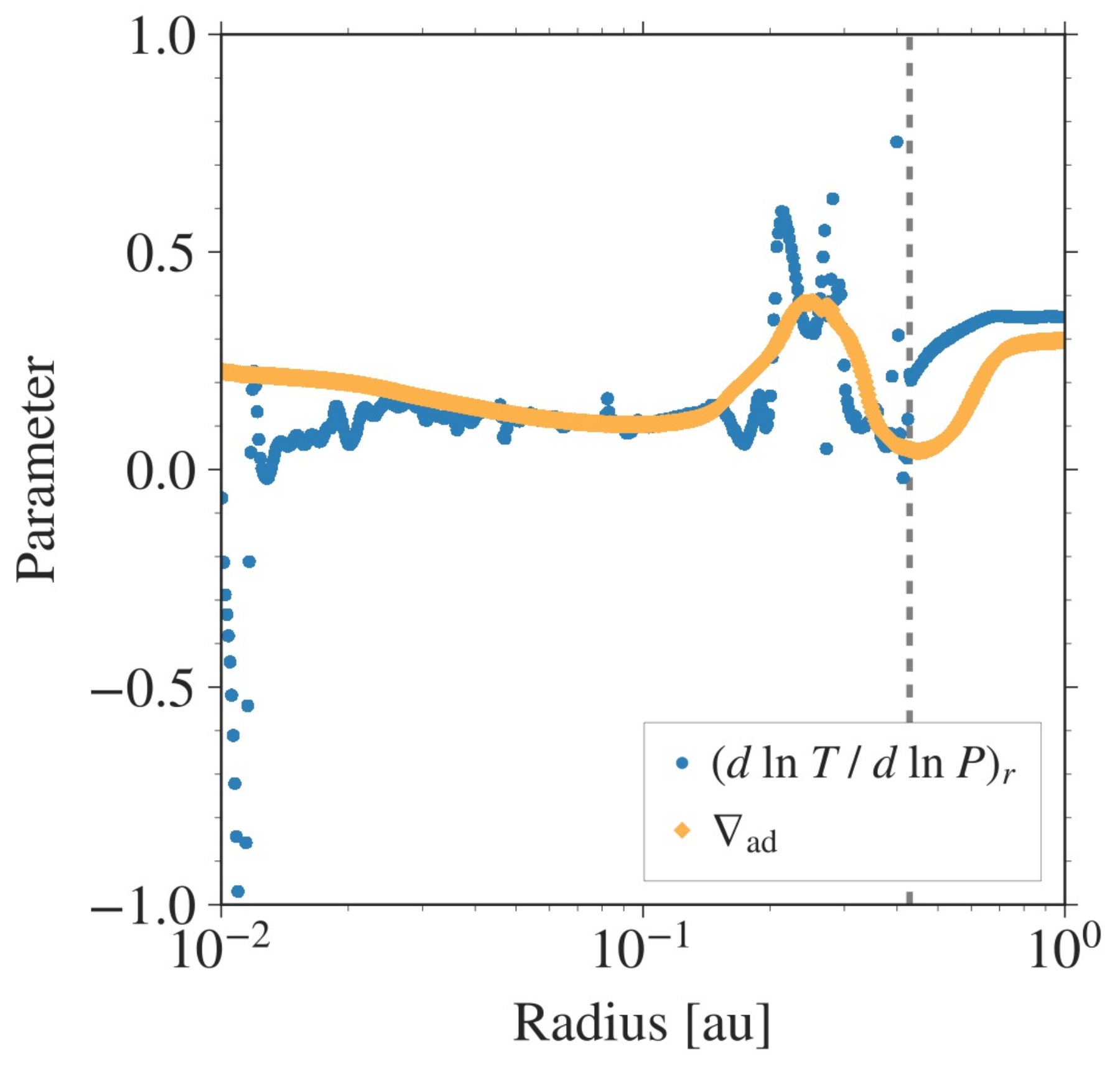}
	\caption{Polar-angle averaged actual temperature gradient $\delact(r)$ at 312~years after the formation of the second core in the 1~$M_{\odot}$ case compared to the polar-angle averaged adiabatic gradient $\delad(\rho(r),T(r))$ (see Eq.~\ref{eq:convection}). Classically, regions with $\delact > \delad$ are convectively unstable. The vertical dashed line indicates the radius of the second core.}
	\label{fig:m1convectivegrads}
\end{figure}

The entropy gradient is also seen in the radial profile in Fig.~\ref{fig:entropy_m1}, zoomed into the inner 1~au. A convective instability is known to occur when a lower-entropy fluid lies above a higher-entropy fluid as seen in the region below the second accretion shock or the second core radius (see also left panel in Fig.~\ref{fig:entropy_m1}). We interpret our results to mean that the instability is generated by the shock and grows radially inwards as the second core evolves over time (visualised in the \footnotetext{See \href{https://keeper.mpdl.mpg.de/f/f04abdeabdf3472fb56d/}{https://keeper.mpdl.mpg.de/f/f04abdeabdf3472fb56d/}.\label{movieref}}movie\footref{movieref}). The entropy is generated at the accretion shock and this yields a gradient from the high entropy at the shock towards the second core interior. Thus, convection is triggered in the outer layers of the second core at the accretion shock and drives the eddies inwards.

In Fig.~\ref{fig:entropy_m1}, we also compare the radial entropy profile from our 1D (dashed red line) and 2D studies. Since there is no convection in 1D, entropy is generated and increases at the shock position. In comparison, since energy generated at the accretion shock is transported due to convection in the 2D case, the entropy profile flattens.

In order to further investigate this behaviour within the second core, we compare in Fig.~\ref{fig:m1convectivegrads} the actual ratio of the temperature and pressure gradients $\delact$ to the adiabatic gradient $\delad$, which is the gradient at constant entropy. In a (quasi-)static fluid, convective motions are expected if
\begin{align}
\mbox{$\delact \equiv \Bigg(\dfrac{d \ln T}{d\ln P} \Bigg)_{r} > \Bigg(\dfrac{d \ln T}{d\ln P} \Bigg)_{\mathrm{ad}} \equiv \delad $}.
\label{eq:convection}	
\end{align}
We calculate $\delad$ according to \citet[][Eq.~3.98]{Hansen2004} using simple differentiation of the $P(\rho,T)$ function provided by the \citet{Vaidya2015} implementation of the \citet{Dangelo2013} EOS.

The dashed grey line in Fig.~\ref{fig:m1convectivegrads} indicates the radius of the second core. The region interior to this radius is convectively unstable. Due to the high resolution, the grid cell size in our simulations is roughly an order of magnitude smaller than the pressure scale height, thus allowing us to resolve the convection. A comparison between different resolutions is shown in Appendix~\ref{sec:resolution}, which indicates the need to use such high resolution in order to resolve the eddies.

Convection allows mixing within a star and contributes by being an efficient means of heat transport. We find that at this stage in evolution, the energy flux is still dominated by radiation, however the convective flux can become stronger at later stages.
 
\subsection{Dependence on initial cloud core mass}
\label{sec:highmass}

\begin{figure}[!htp]
	\centering
	\hspace*{-1.5cm}	
	\begin{subfigure}{0.4\textwidth}
		\includegraphics[width=1.2\textwidth]{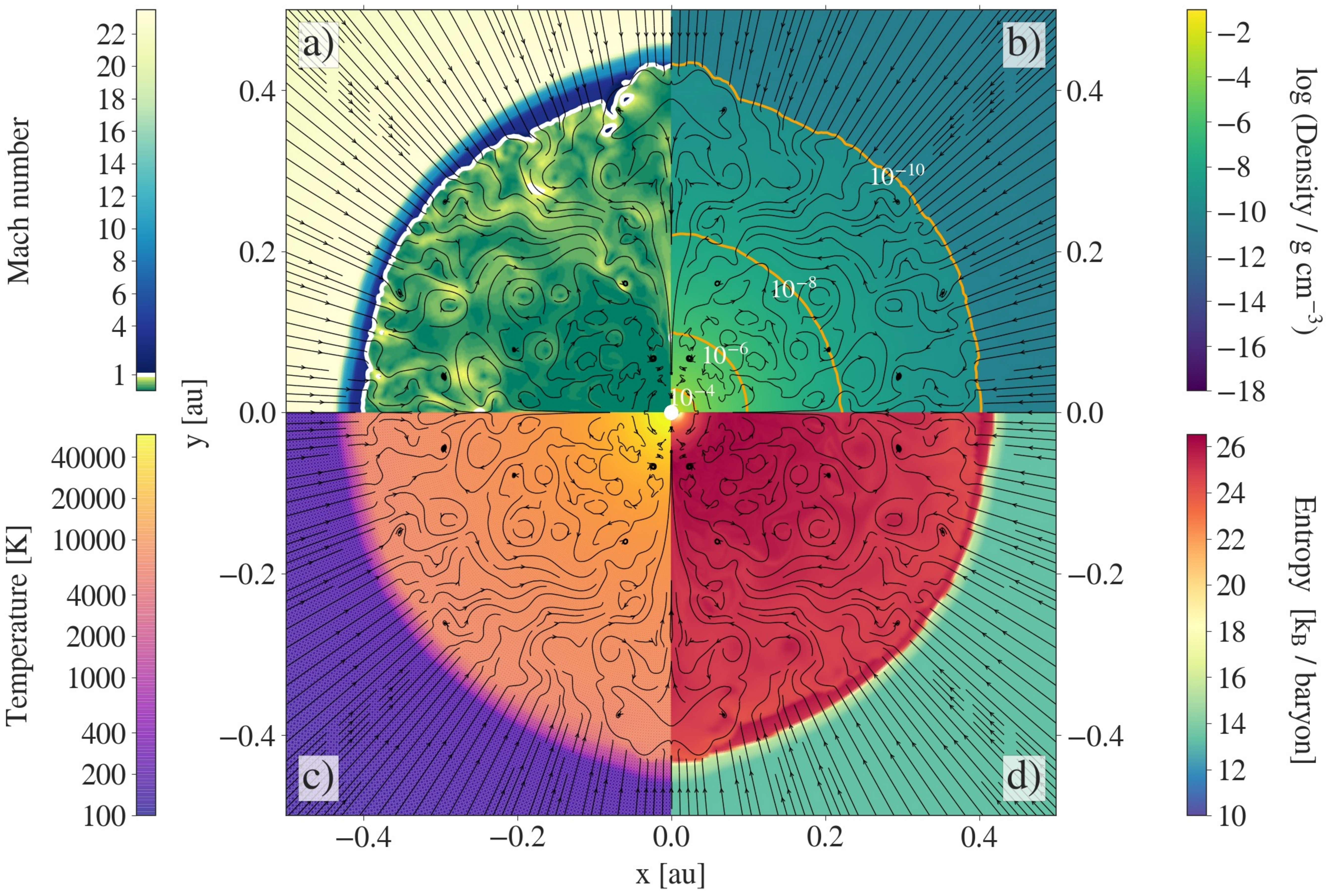}
		\vspace{0.5cm}
	\end{subfigure}
	\hspace*{-1.5cm}	
	\begin{subfigure}{0.4\textwidth}
		\includegraphics[width=1.2\textwidth]{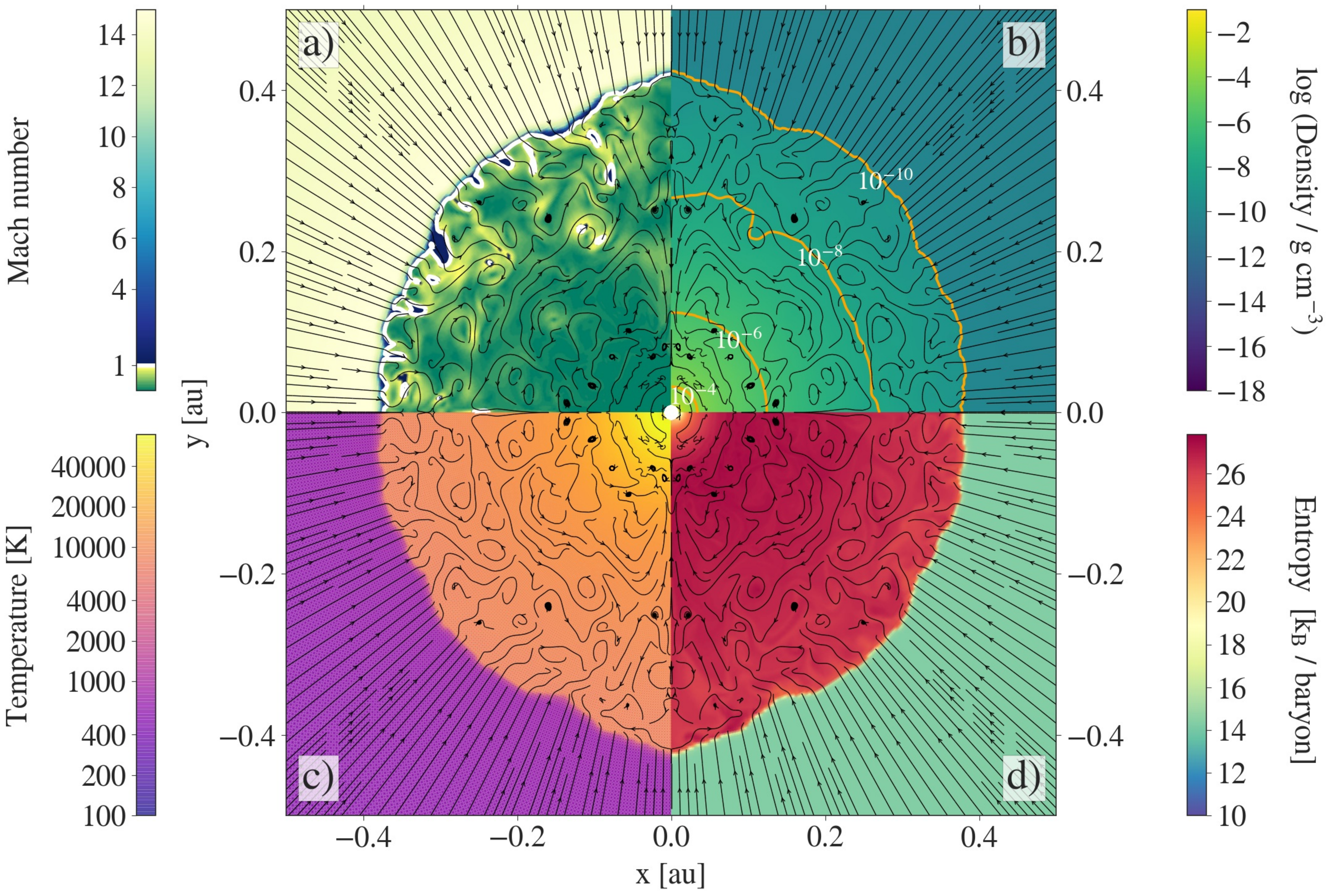}
		\vspace{0.5cm}
	\end{subfigure}
	\hspace*{-1.5cm}	
	\begin{subfigure}{0.4\textwidth}
		\includegraphics[width=1.2\textwidth]{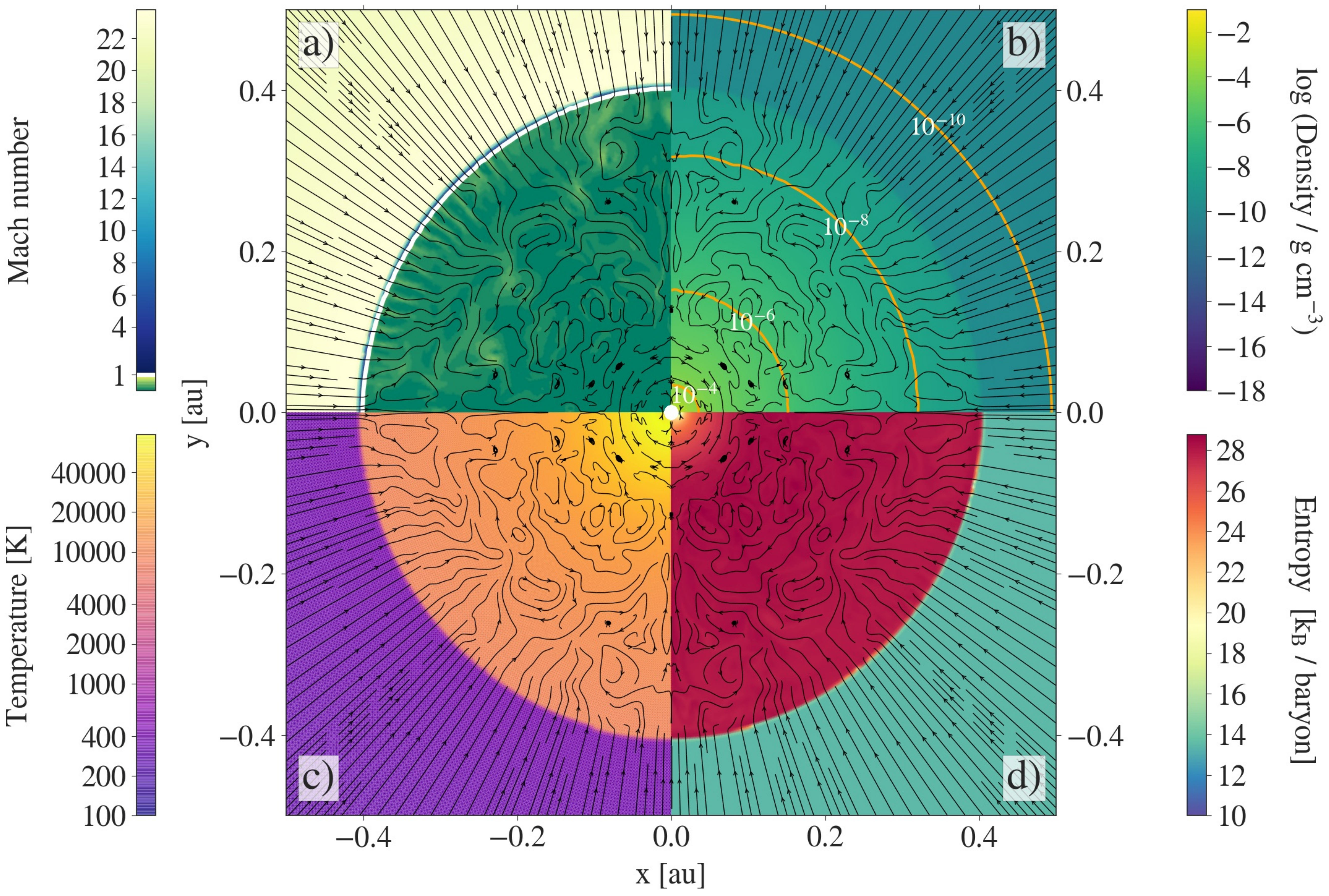}
	\end{subfigure}
	\caption{2D view of the second hydrostatic core formed as a result of the collapse of a 5~$M_{\odot}$ (top), 10~$M_{\odot}$ (middle), and 20~$M_{\odot}$ (bottom) cloud core with an initial temperature of 10 K. The four panels in each of the subplots show the \mbox{\bf a)}~Mach number, \mbox{\bf b)}~density, \mbox{\bf c)}~temperature, and \mbox{\bf d)}~entropy within the inner 0.5~au of an initial 3000~au cloud core. The velocity streamlines in black indicate the material falling onto the second core and the mixing inside the core. The white contour in panel (a) showing Mach = 1.0 separates the sub- and supersonic regions. The different contour lines in panel (b) show the increase in density towards the centre. The plots are shown at 128~years, 91.4~years, and 86.4~years after the second core formation, for the 5~$M_{\odot}$, 10~$M_{\odot}$, and 20~$M_{\odot}$ cases, respectively.}
	\label{fig:highmass}
\end{figure}

\begin{figure}[!htp]
	\centering
	\hspace*{-1.7cm}	
	\begin{subfigure}{0.39\textwidth}
		\includegraphics[width=1.2\textwidth]{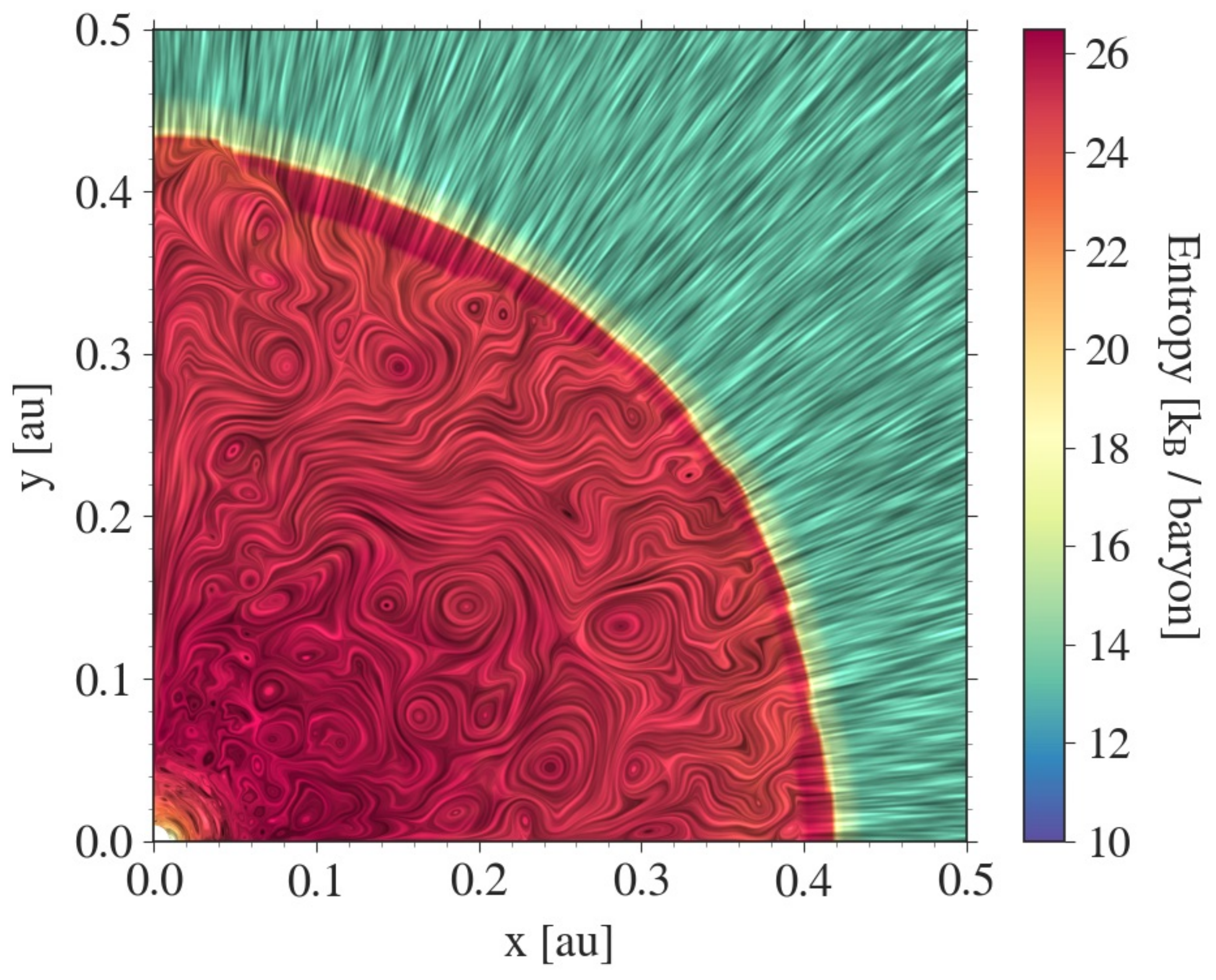}
		\vspace{0.1cm}
	\end{subfigure}
	\hspace*{-1.7cm}	
	\begin{subfigure}{0.39\textwidth}
		\includegraphics[width=1.2\textwidth]{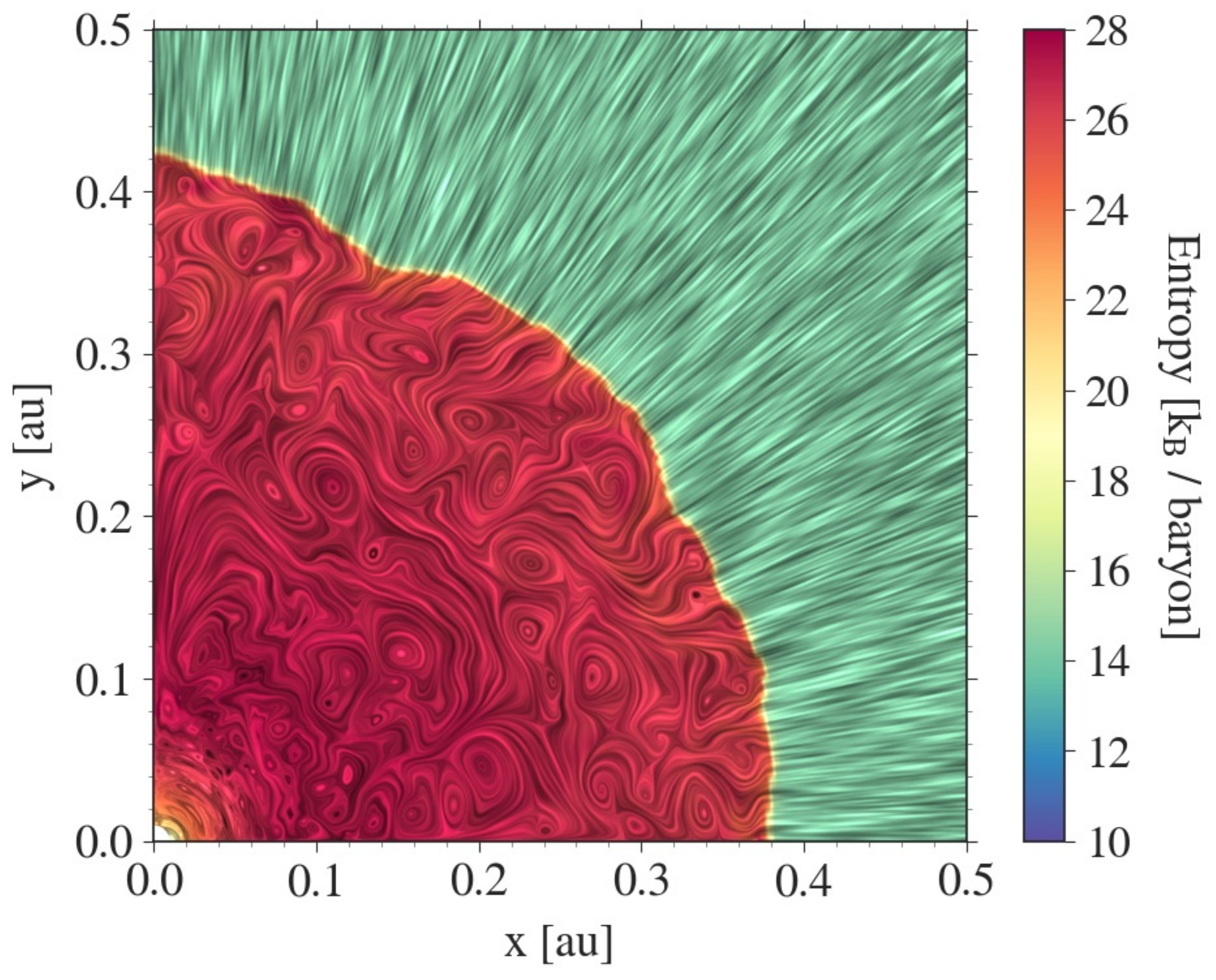}
		\vspace{0.1cm}
	\end{subfigure}
	\hspace*{-1.7cm}	
	\begin{subfigure}{0.39\textwidth}
		\includegraphics[width=1.2\textwidth]{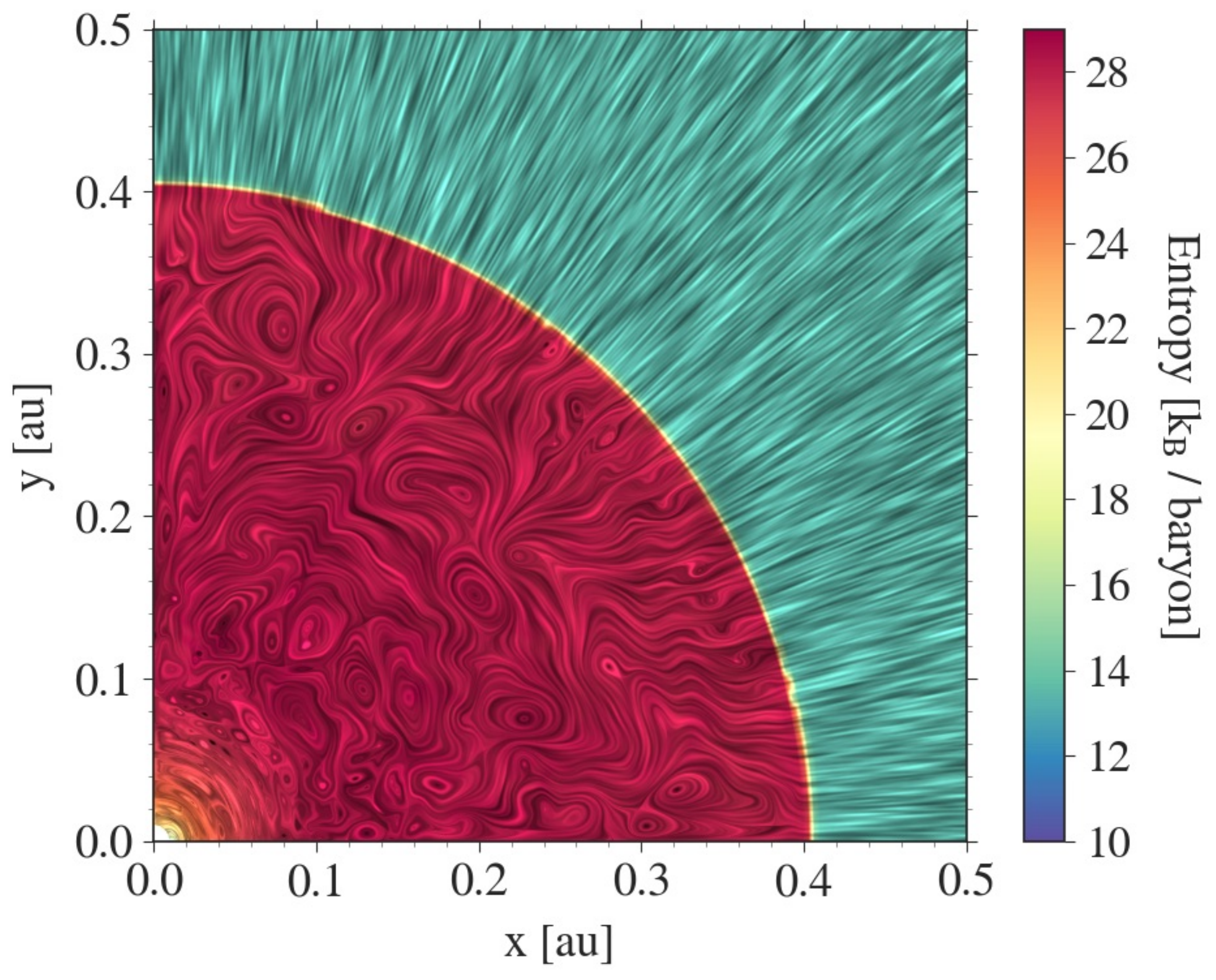}
	\end{subfigure}
	\caption{Line integral convolution visualisation of the entropy behaviour indicating the presence of eddies within the second core. Shown here is the inner 0.5~au of 3000~au collapsing cloud cores with different masses of 5~$M_{\odot}$ (top), 10~$M_{\odot}$ (middle), and 20~$M_{\odot}$ (bottom). The plots are shown at 128~years, 91.4~years, and 86.4~years after the second core formation, for the 5~$M_{\odot}$, 10~$M_{\odot}$, and 20~$M_{\odot}$ cases, respectively. }
	\label{fig:LIC_highmass}
\end{figure}

We further investigate the evolution of collapsing cloud cores with initial masses of 5~$M_{\odot}$, 10~$M_{\odot}$, and 20~$M_{\odot}$, thus covering a few cases in the intermediate- and high-mass regimes. The same initial temperature of 10~K and outer cloud core radius of 3000~au are used as in the 1~$M_{\odot}$ case. We study the effects of initial cloud core mass on the convective instability discussed in Sect.~\ref{sec:2Dsims}. Similar to the 1~$M_{\odot}$ case, for the 5~$M_{\odot}$, 10~$M_{\odot}$, and 20~$M_{\odot}$ runs, we find a turbulent pattern within the second core, indicated by the black velocity streamlines in Fig.~\ref{fig:highmass} and seen as the eddies in Fig.~\ref{fig:LIC_highmass}. The plots are shown at 128 years, 91.4 years, and 86.4 years after the second core formation, for the 5~$M_{\odot}$, 10~$M_{\odot}$, and 20~$M_{\odot}$ cases, respectively.  

\begin{figure}[!htp]
	\centering
	\hspace*{-1.4cm}		
	\begin{subfigure}{0.32\textwidth}
		\includegraphics[width=1.2\textwidth]{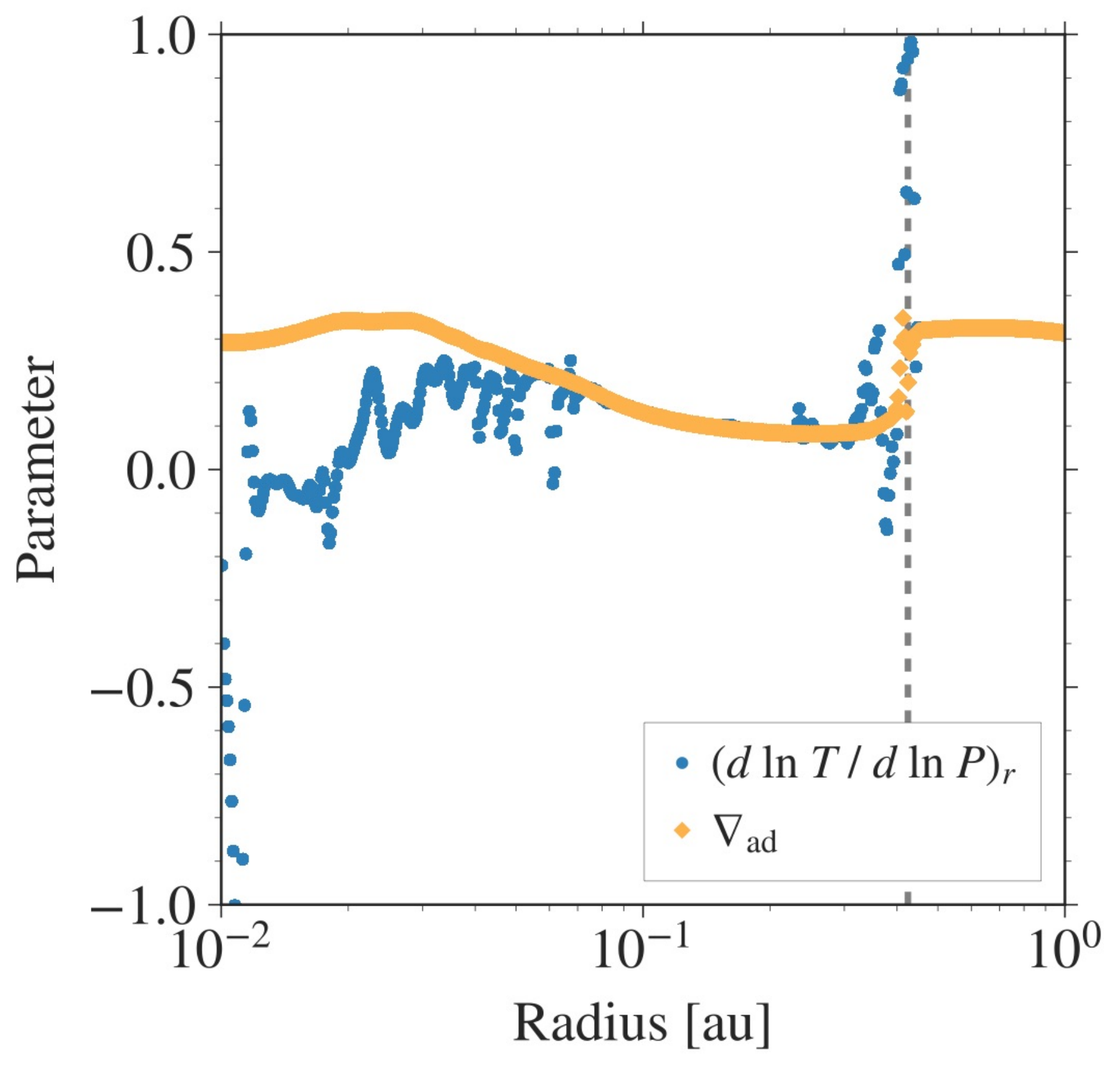}
		\vspace{0.1cm}
	\end{subfigure}
	\hspace*{-1.4cm}	
	\begin{subfigure}{0.32\textwidth}
		\includegraphics[width=1.2\textwidth]{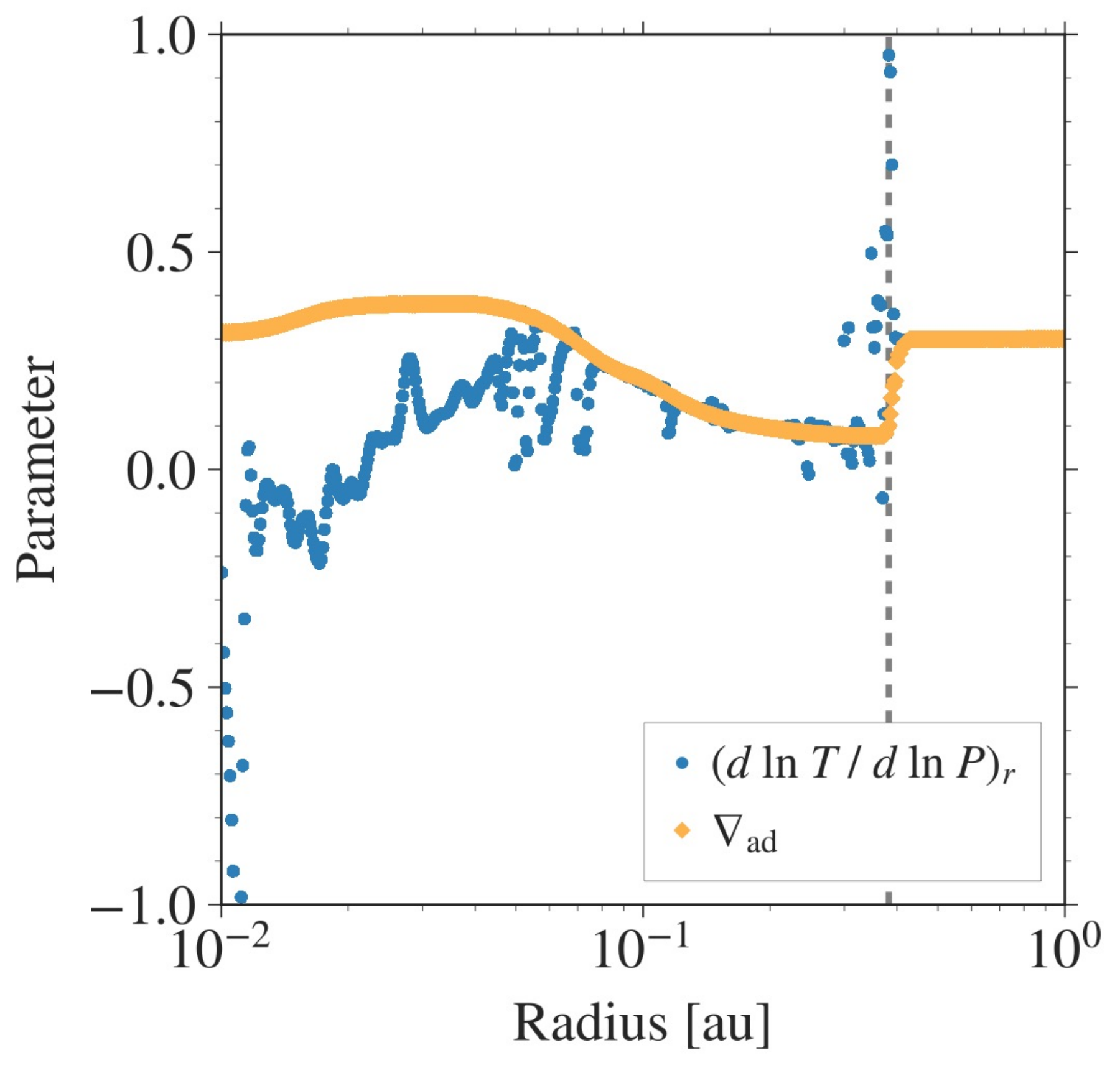}
		\vspace{0.1cm}
	\end{subfigure}
	\hspace*{-1.4cm}	
	\begin{subfigure}{0.32\textwidth}
	\includegraphics[width=1.2\textwidth]{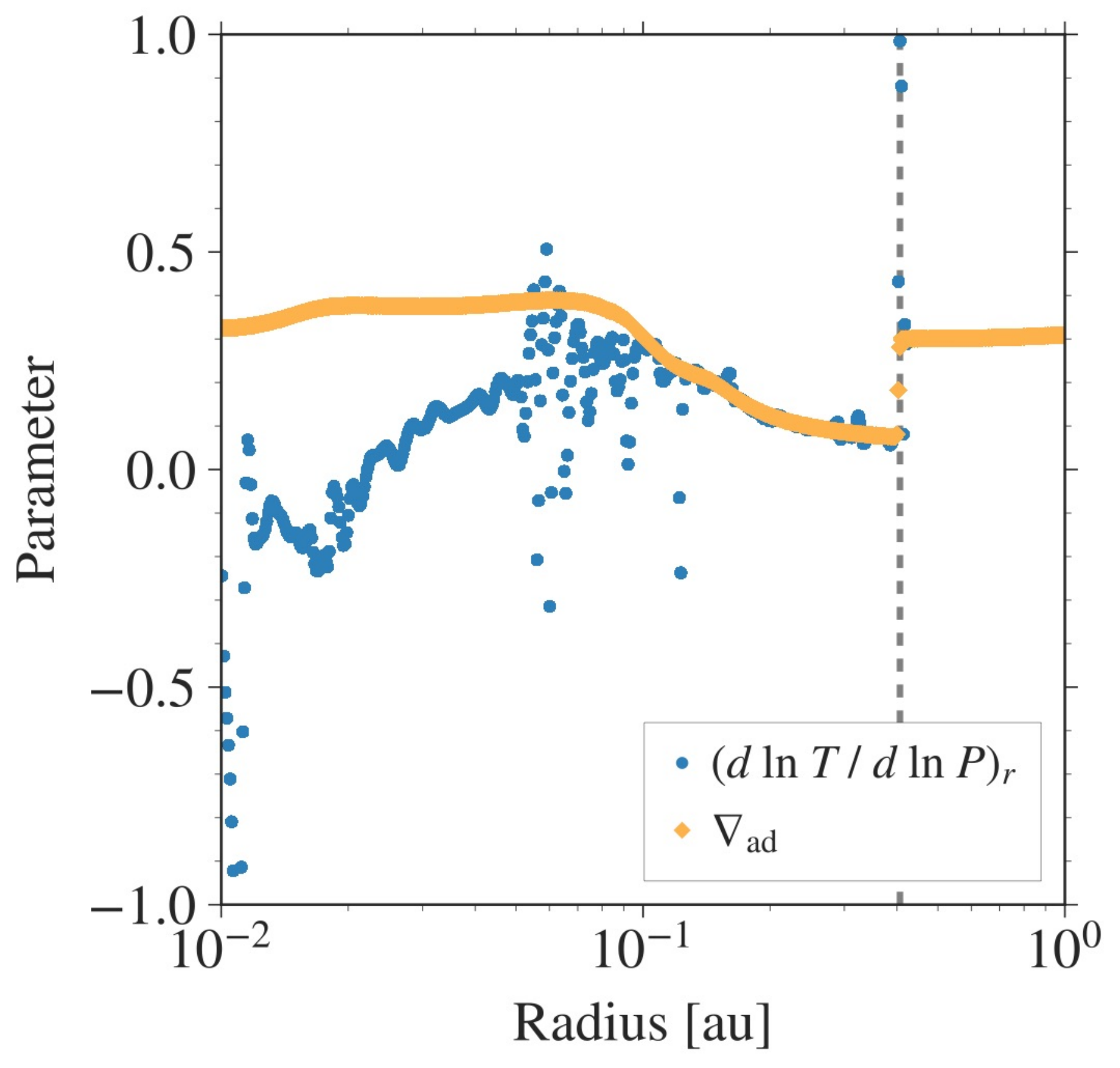}
	\end{subfigure}
	\caption{Polar-angle averaged actual temperature gradient $\delact(r)$
	compared to the polar-angle averaged adiabatic gradient $\delad(\rho(r),T(r))$ (see Eq.~\ref{eq:convection}) for the 5~$M_{\odot}$ (top), 10~$M_{\odot}$ (middle), and 20~$M_{\odot}$ (bottom) initial cloud core masses. The indication for convective instability is not as prominent as in the 1~$M_{\odot}$ case (see Fig.~\ref{fig:m1convectivegrads}). The vertical grey dashed line indicates the radius of the second core. The plots are shown at 128~years, 91.4~years, and 86.4~years after the second core formation, for the 5~$M_{\odot}$, 10~$M_{\odot}$, and 20~$M_{\odot}$ cases, respectively.}
	\label{fig:convective_grads}
\end{figure}

\begin{figure}[!htp]
	\centering
	\hspace*{-1.5cm}		
	\begin{subfigure}{0.33\textwidth}
		\includegraphics[width=1.2\textwidth]{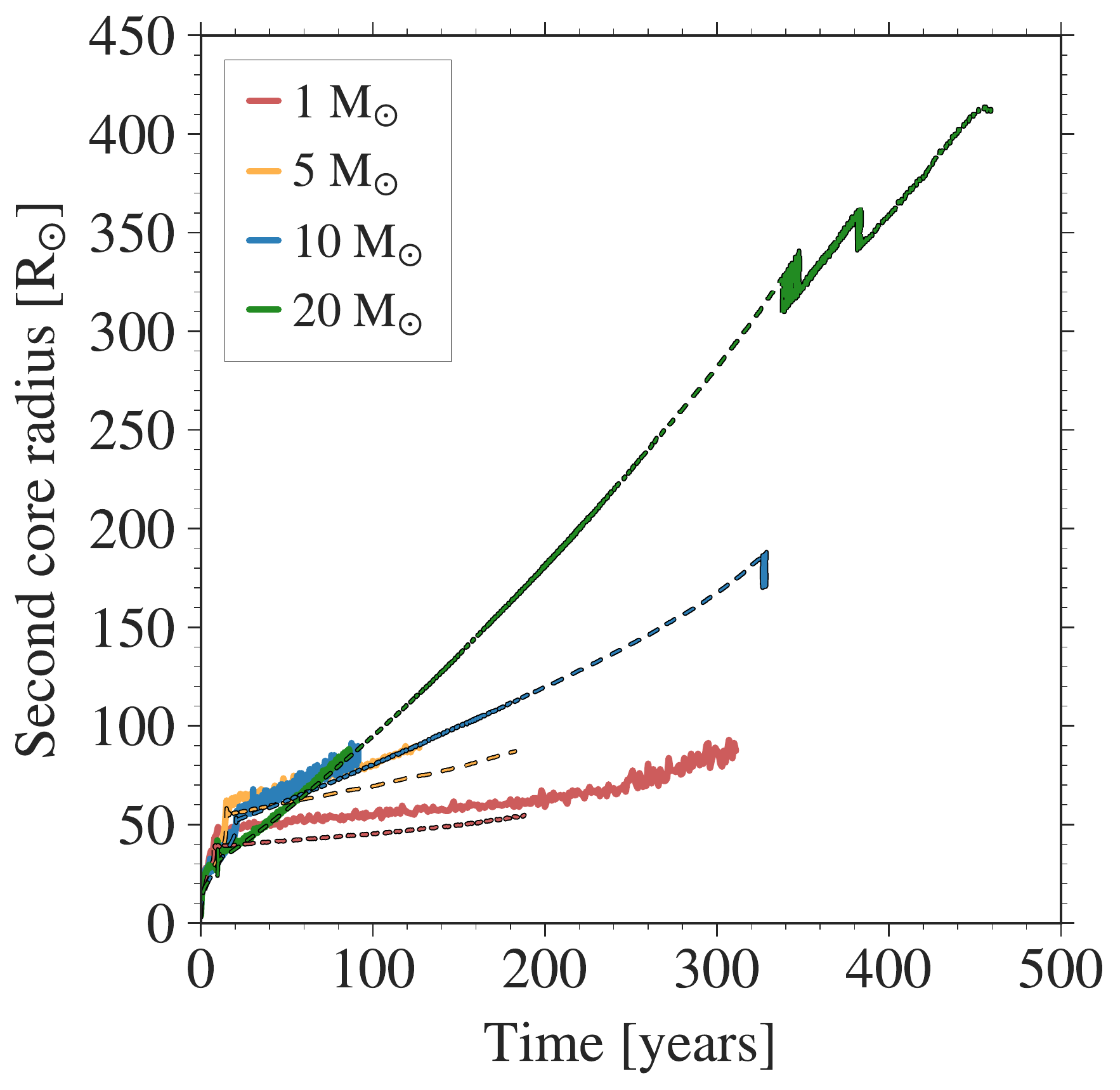}
		\vspace{0.1cm}
	\end{subfigure}
	\hspace*{-1.5cm}	
	\begin{subfigure}{0.33\textwidth}
		\includegraphics[width=1.2\textwidth]{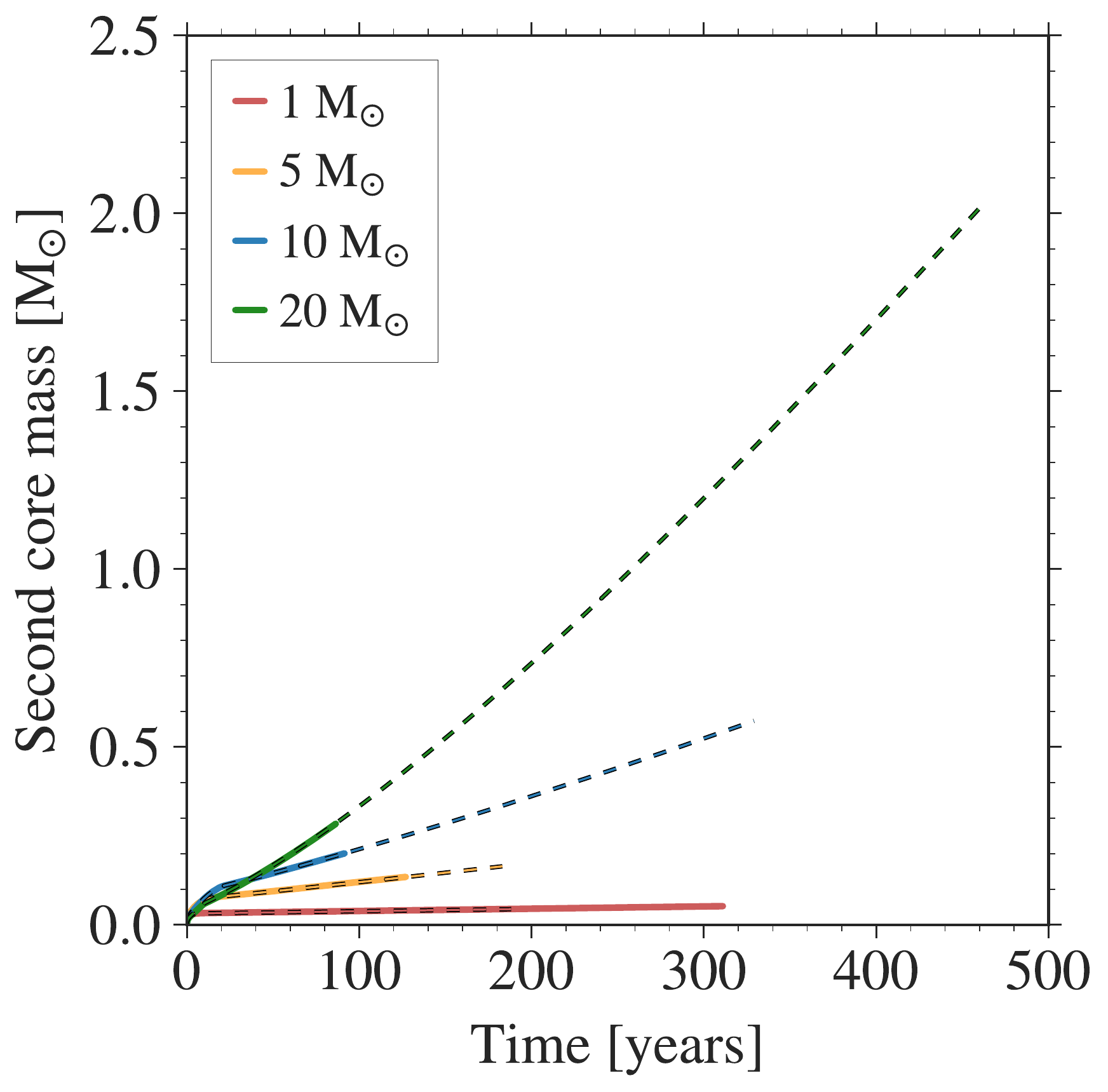}
		\vspace{0.1cm}
	\end{subfigure}
	\hspace*{-1.5cm}	
	\begin{subfigure}{0.33\textwidth}
	\includegraphics[width=1.2\textwidth]{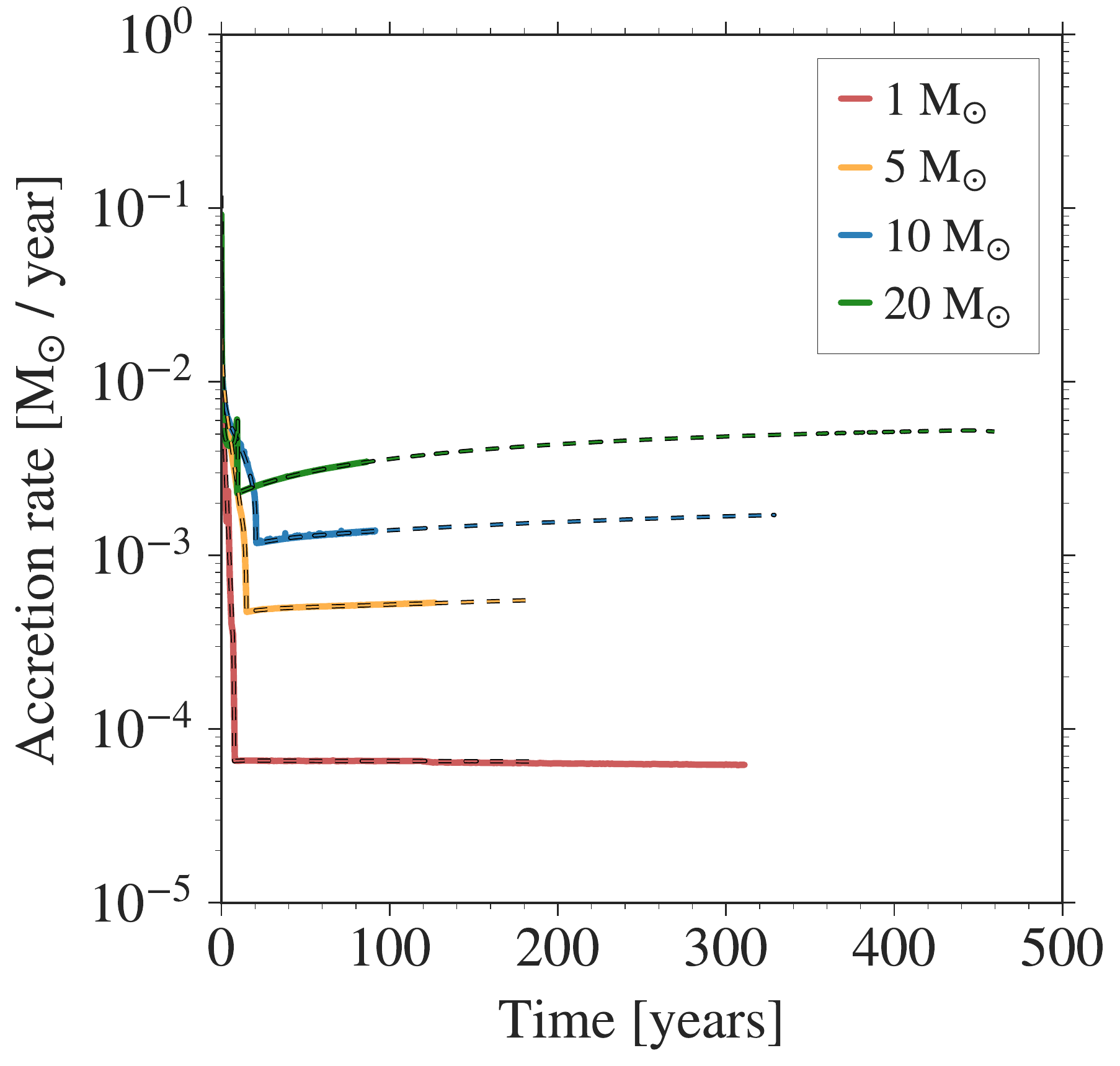}
	\end{subfigure}
	\caption{Time evolution of the second core radius (top), mass (middle), and accretion rate (bottom) for the different core collapse scenarios with initial cloud core mass of 1~$M_{\odot}$ (red), 5~$M_{\odot}$ (yellow), 10~$M_{\odot}$ (blue), and 20~$M_{\odot}$ (green). The time $t=0$ marks the onset of the second core formation. The overplotted dashed lines indicate the evolution from the high-resolution 1D simulations discussed in Sect.~\ref{sec:1Dsecondcore}.}
	\label{fig:2DSCproperties}
\end{figure}

Figure~\ref{fig:convective_grads} shows the comparison between the adiabatic index and the ratio of temperature and pressure gradients. On testing the criterion for convective instability stated in Eq.~\eqref{eq:convection}, we do not find a strong indication as seen in the 1 $M_{\odot}$ case. However, this may change as the second core evolves further for these cases. We are currently unable to further follow the evolution of the second core due to high computational expenses and this remains to be tested as part of future studies.

Figure~\ref{fig:2DSCproperties} shows the evolution of the second core radius, mass, and accretion rate for all the cases with different initial cloud core mass. The time $t=0$ marks the onset of the second core formation, which is indicated by a prominent second accretion shock as per the definition used herein. Higher initial cloud core mass leads to a faster collapse. Following the evolution, our results indicate that the 1~$M_{\odot}$ and 5~$M_{\odot}$ cloud cores will eventually form a protostar with mass less than 0.5~$M_{\odot}$. Stars within this mass range are known to be fully convective throughout their life.

\subsection{Standing accretion shock instability}
\label{sec:sasi}

In this section, we further investigate the source of the turbulence visible in the post-shock regions. The standing (or spherical) accretion shock instability (SASI), known to play a crucial role in the explosion mechanism of core collapse supernovae, is seen to induce large-scale non-spherical oscillations of the shock \citep{Foglizzo2000, Foglizzo2002, Blondin2003, Foglizzo2006, Guilet2012}. There are two proposed mechanisms that lead to the linear growth of this instability, one being the interplay between advected entropy--vorticity perturbations and acoustic waves \citep{Foglizzo2007, Foglizzo2009}, whereas the second is a purely acoustic mechanism, which assumes that the trapped acoustic waves can be amplified by the shock \citep{Blondin2006}.

During the evolution of the second core in our 2D collapse simulations, we observed some non-spherical, large-scale oscillations of the accretion shock front in the 1~$M_{\odot}$, 5~$M_{\odot}$, and 10~$M_{\odot}$ cases and comparatively small-scale oscillations of the accretion shock front for the 20~$M_{\odot}$ case. In their study, \citet{Scheck2008} have reported that large-amplitude SASI oscillations produce strong variations in the entropy, which can drive convective instability in the supernova core. 

\begin{figure*}[!htp]
	\centering
	\hspace*{-1.5cm}	
	\begin{subfigure}{0.4\textwidth}
		\includegraphics[width=1.2\textwidth]{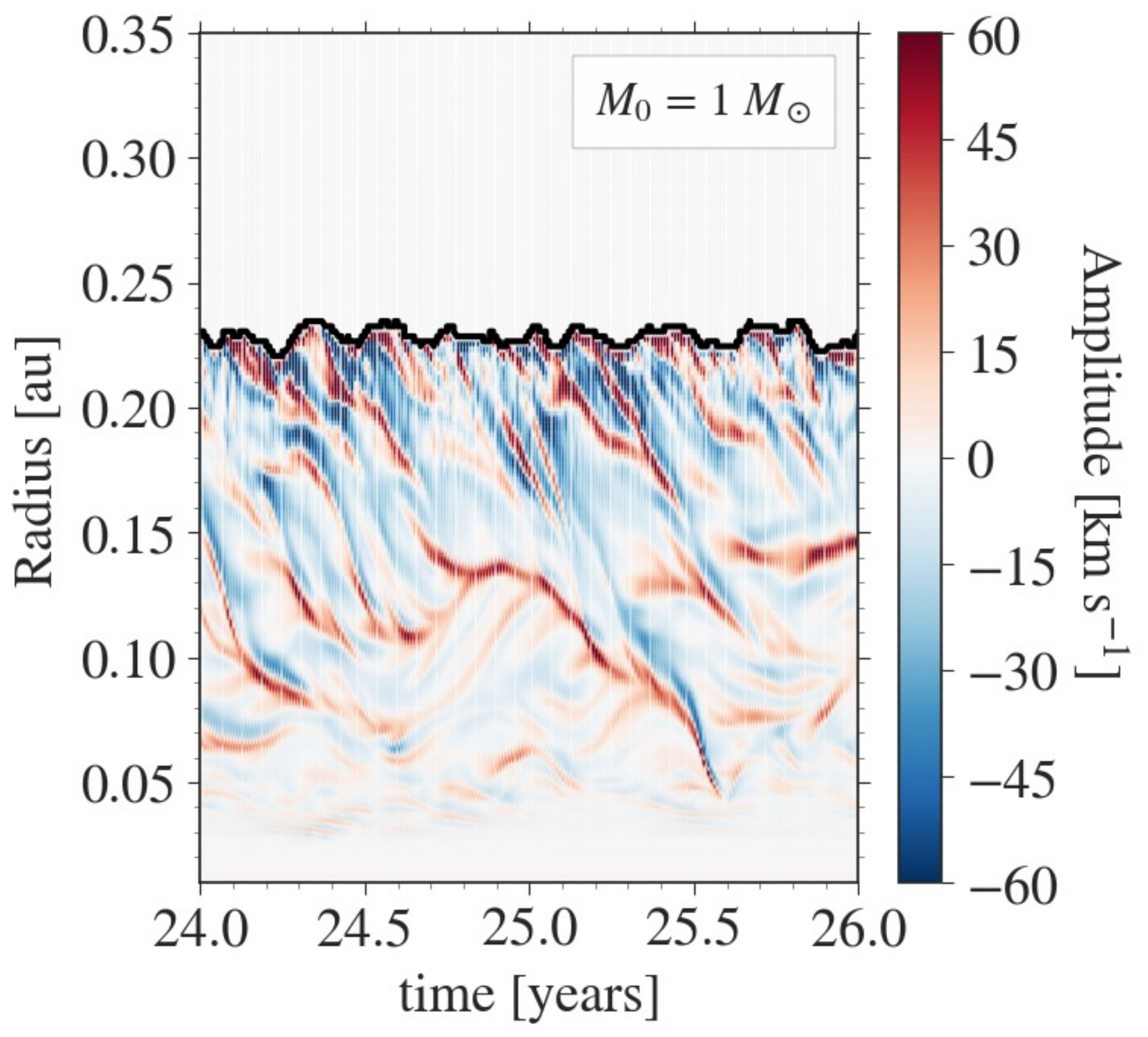}
		\vspace{0.2cm}
	\end{subfigure}
	\hspace{1.7cm}
	\begin{subfigure}{0.4\textwidth}
		\includegraphics[width=1.2\textwidth]{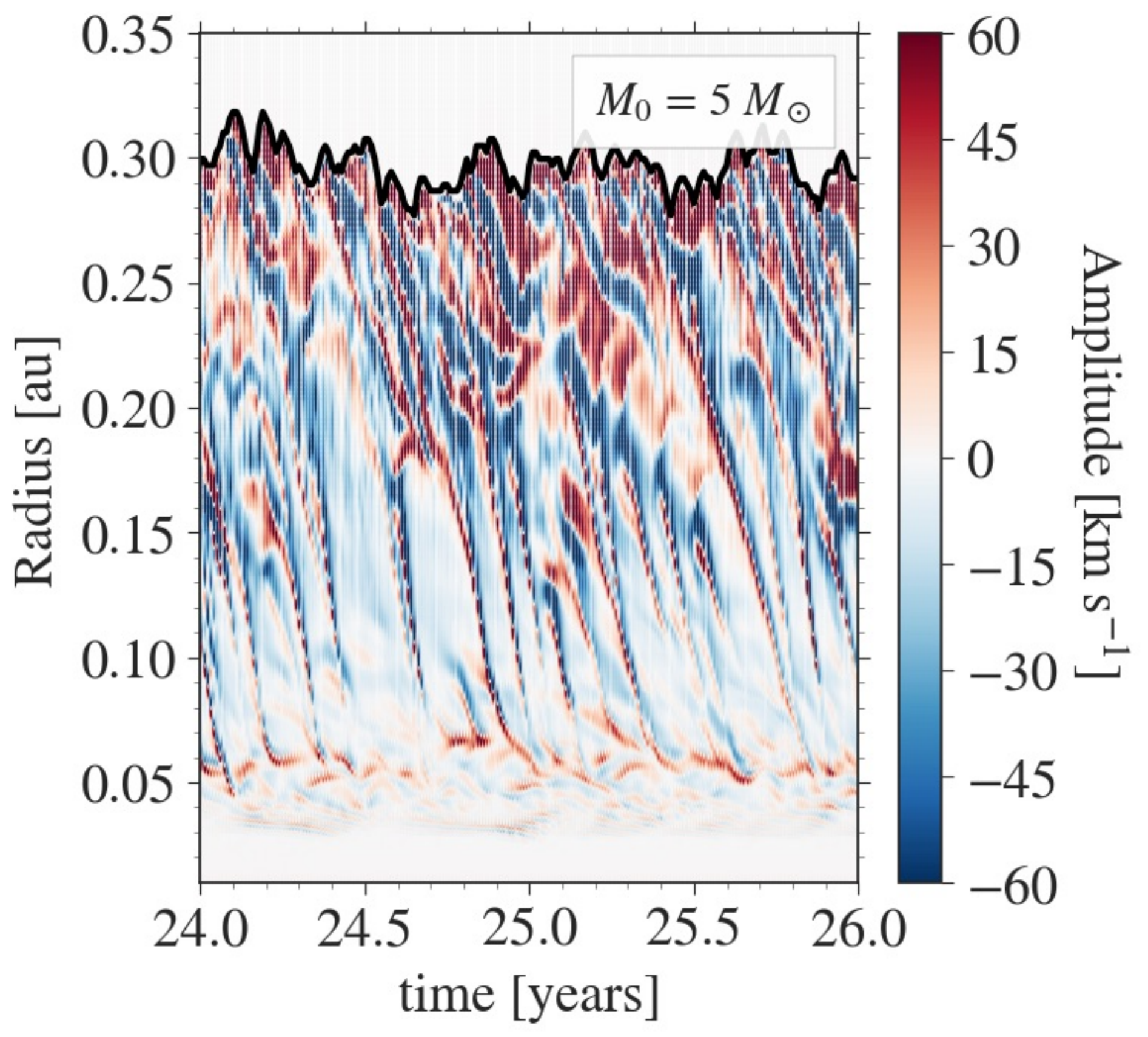}
		\vspace{0.2cm}
	\end{subfigure}
	\hspace*{-1.5cm}	
	\begin{subfigure}{0.4\textwidth}
		\includegraphics[width=1.2\textwidth]{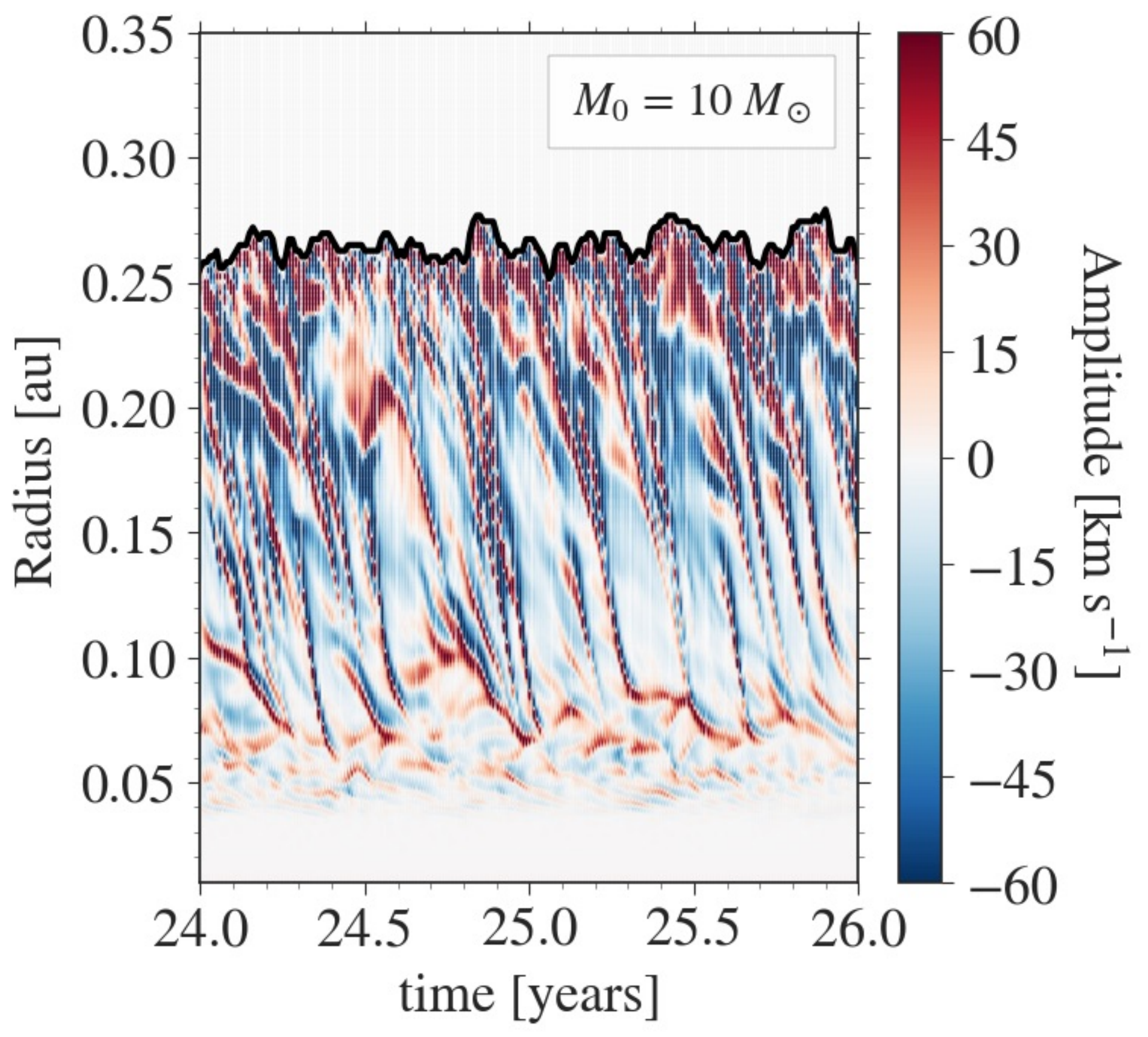}
	\end{subfigure}
	\hspace{1.7cm}
	\begin{subfigure}{0.4\textwidth}
		\includegraphics[width=1.2\textwidth]{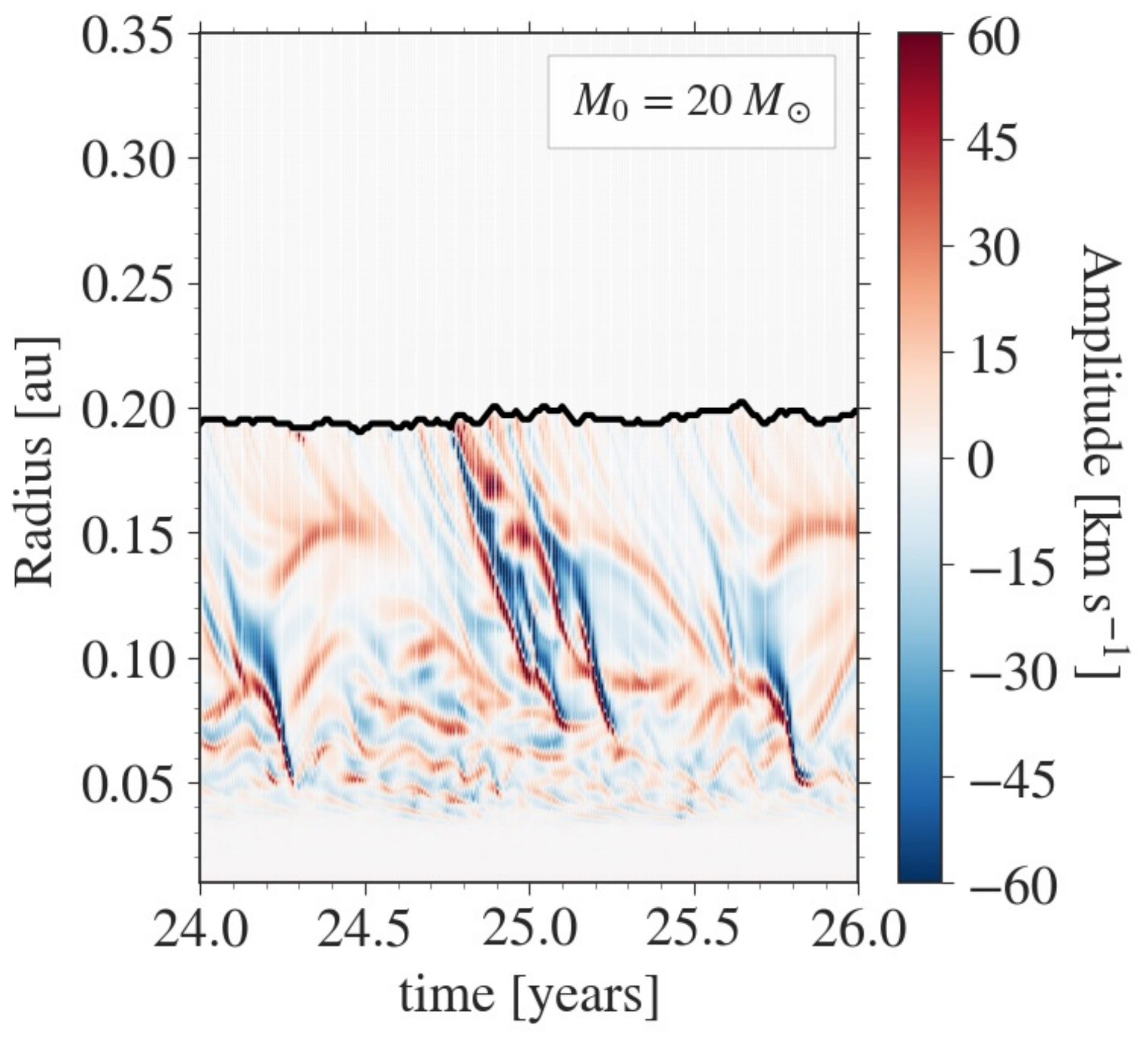}
	\end{subfigure}
	\caption{Time evolution of the amplitude of the dominant spherical harmonics mode of the quantity $A(t, r, \theta)$ from Eq.~\eqref{eq:m1SASI}. The second core radius is shown by the black line. The time $t=0$ indicates the onset of the second core formation for the different collapse scenarios. Shown here is a small interval in time for the 1~$M_{\odot}$ (top left), 5~$M_{\odot}$ (top right), 10~$M_{\odot}$ (bottom left), and 20~$M_{\odot}$ (bottom right) cases.}
	\label{fig:SASI}
\end{figure*}

We further investigate the presence of the SASI in the case of protostellar cores and its role to generate turbulence behind the shock for all the different collapse scenarios. Large-scale, non-spherical oscillations of the second accretion shock front are indicated by the black line in Fig.~\ref{fig:SASI}. For a quantitative analysis of the SASI, following \citet{Scheck2008}, Fig.~\ref{fig:SASI} shows the advected perturbations in terms of the amplitudes of the largest modes of the spherical harmonics of the quantity $A(t, r, \theta)$ given by
\begin{align}
\mbox{$A(t, r, \theta) = \dfrac{1}{\mathrm{sin} \theta} \dfrac{\partial}{\partial \theta} (v_\theta (t, r, \theta) ~\mathrm{sin} \theta) $}.
\label{eq:m1SASI}
\end{align}

The term $r^{-1}A$ is the divergence of the lateral velocity component. Several works have shown that the SASI can be measured more easily by determining \textit{A} even for lower amplitudes of the instability \citep{Scheck2008,Blondin2006}. For all the cases with different initial cloud core masses, we plot the amplitude for a small time interval since it helps to view the perturbations better. Although the amplitudes show more complex patterns than in \citet[see their Fig.~12]{Scheck2008}, there are some noticeable advected trajectories as well as some acoustic feedback. The important acoustic feedback timescale is given by the sound crossing time from the centre of the second core to the accretion shock and back.

We thus conclude that the SASI may not be operating as strongly as seen in the supernovae core-collapse studies and the convective instability seems to be the main source of the turbulent cells seen in the post-shock regions. However, the SASI could still be responsible for the large scale oscillations of the accretion shock front seen during the evolution of the second core. Nonetheless, it is interesting to note that the SASI can operate in different regimes. 

\subsection{Limitations}
\label{sec:limitations}

The simulations discussed in this paper only include the effects of self-gravity, radiation transport, dissociation, and ionisation on the core properties, but not due to rotation. Our ongoing studies which account for the effects of initial cloud rotation and magnetic fields on the hydrostatic cores and early discs will be reported in a follow-up article. Magnetic fields would indeed affect the formation and evolution of the second core. However, convective instability within the second core, at least for the low-mass end, could still be generated during the core evolution.  

The 2D simulations do not stay spherically symmetric. Hence, the evolution of the convective second core will also not stay axially symmetric. Unfortunately, a 3D model achieving the same resolution as in the axial and midplane symmetric 2D simulations presented here is unfeasible at the moment.

\section{Comparisons with previous work}
\label{sec:comparison}

In this section we first compare the second core properties from our 1D simulations to some of the previous studies for the case of a collapsing 1~$M_{\odot}$ cloud core.

In their study, \citet{Masunaga2000} use a uniform initial density profile with $\rho = 1.415 \times 10^{-19} ~\mathrm{g ~cm^{-3}}$ and an outer radius of $10^{4}$~au. A few years\footnote{See Table~1 in \citet{Masunaga2000} for their simulation run-time.} after the formation of the second core when the central density reaches $\approx$~1~$\mathrm{g ~cm^{-3}}$, they find the second core radius to be $\approx$ 4~$R_{\odot}$ and the second core mass as 0.73~$M_{\odot}$. On the other hand, at central density greater than $\mathrm{10^{-1} ~g ~cm^{-3}}$, \citet{Tomida2013} report a bigger second core radius of $\approx$~10~$R_{\odot}$ enclosing a mass of 2~$\times 10^{-2} M_{\odot}$ within 0.7 years after its formation\footnote{\citet{Tomida2013} define the onset of second core formation at the time when the central density exceeds $10^{-3} ~\mathrm{g ~cm^{-3}}$ in their simulations.}. They suggest that the second core continues to expand during the main accretion phase. Their simulations are stopped once the central temperature reaches $10^{5}$~K. In their work, they adopt a Bonnor--Ebert sphere like density profile with an initial central density of 1.2~$\times 10^{-18} ~\mathrm{g ~cm^{-3}}$ and an outer radius of 8800~au. 

\begin{figure}[!htp]
	\centering
	\hspace*{-1.2cm}	
	\begin{subfigure}{0.33\textwidth}
		\includegraphics[width=1.2\textwidth]{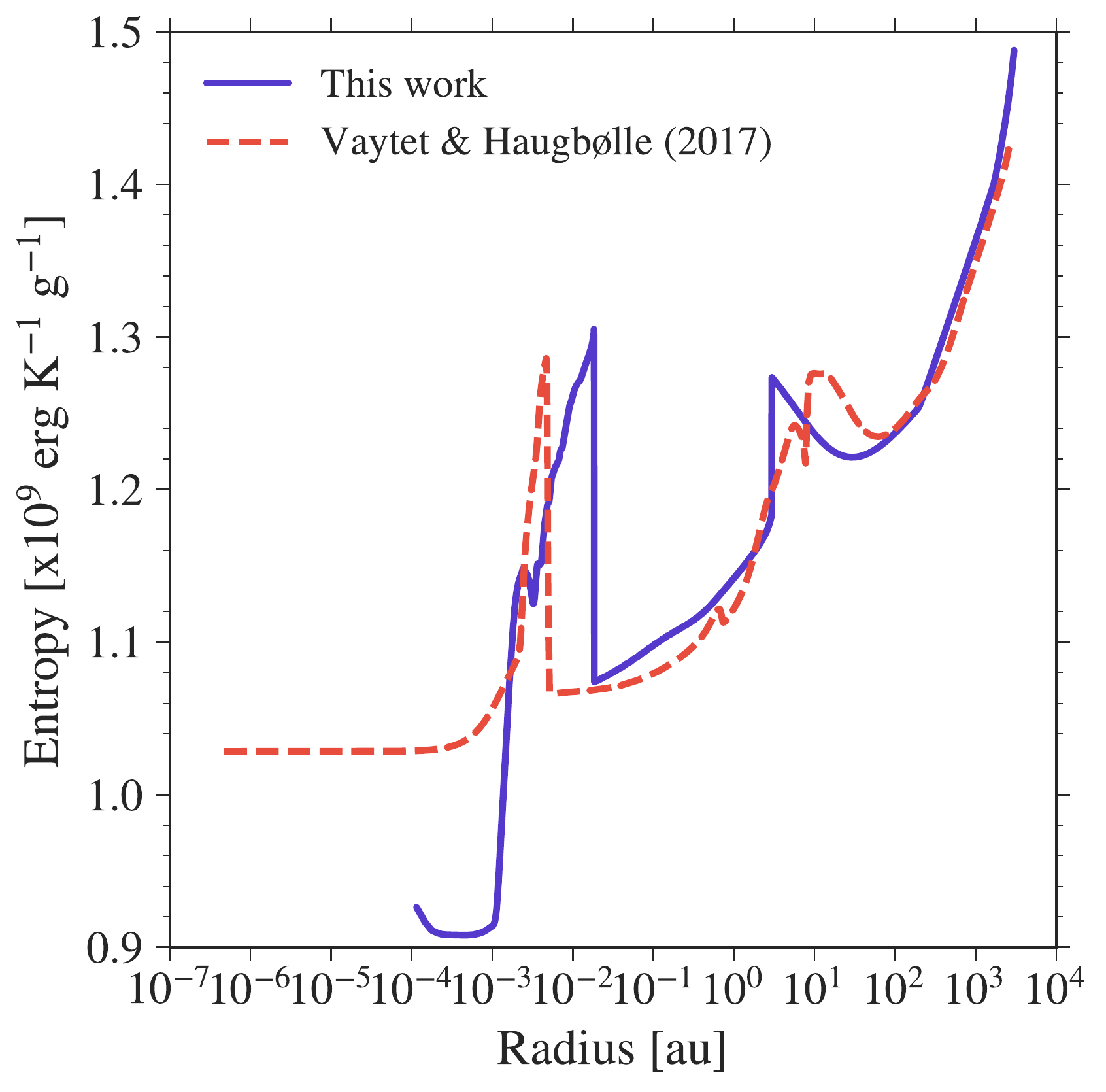}
		\vspace{0.1cm}
	\end{subfigure}
	\hspace*{-1.2cm}	
	\begin{subfigure}{0.33\textwidth}
		\includegraphics[width=1.2\textwidth]{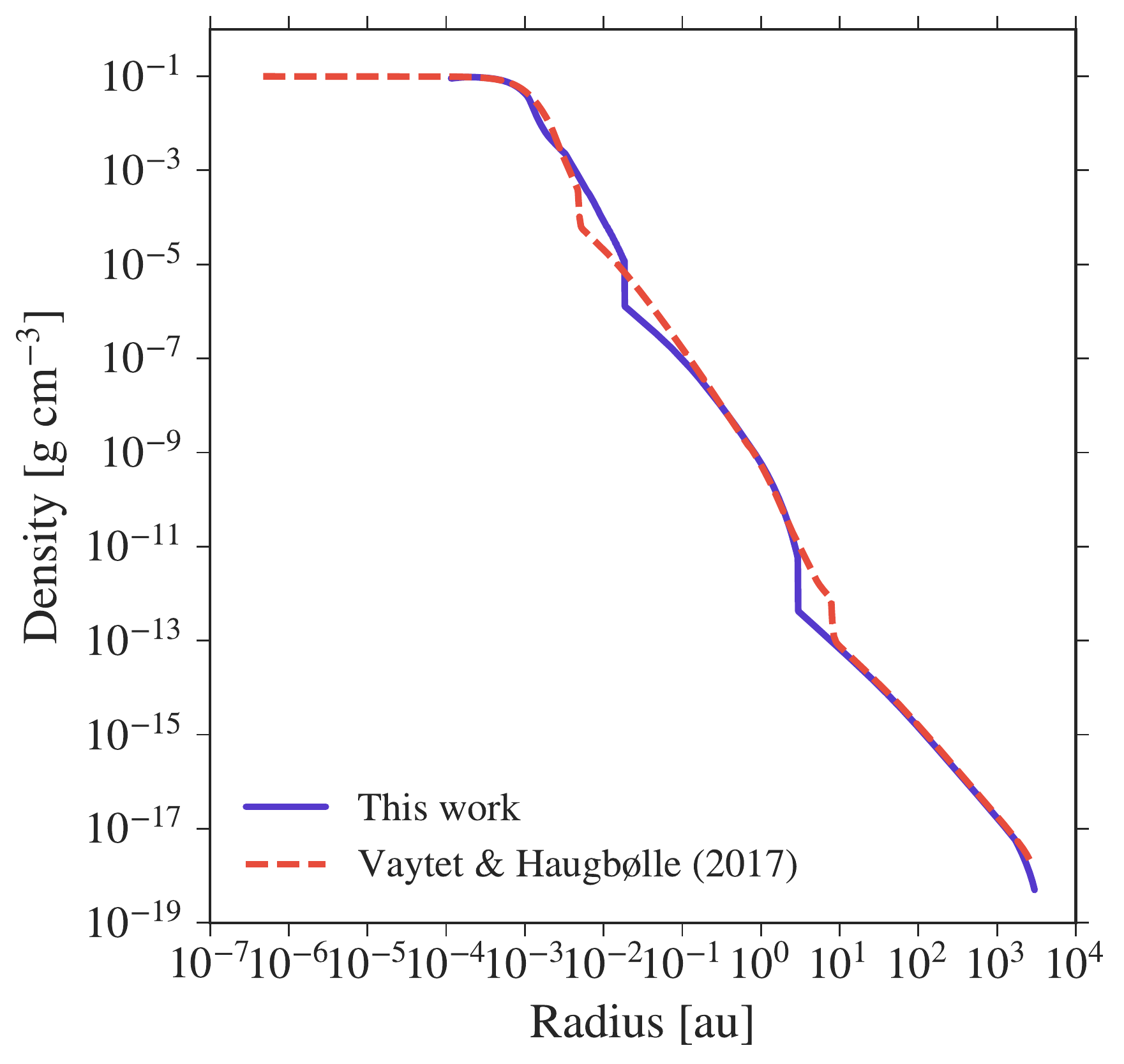}
		\vspace{0.1cm}
	\end{subfigure}
	\hspace*{-1.2cm}	
	\begin{subfigure}{0.33\textwidth}
	\includegraphics[width=1.2\textwidth]{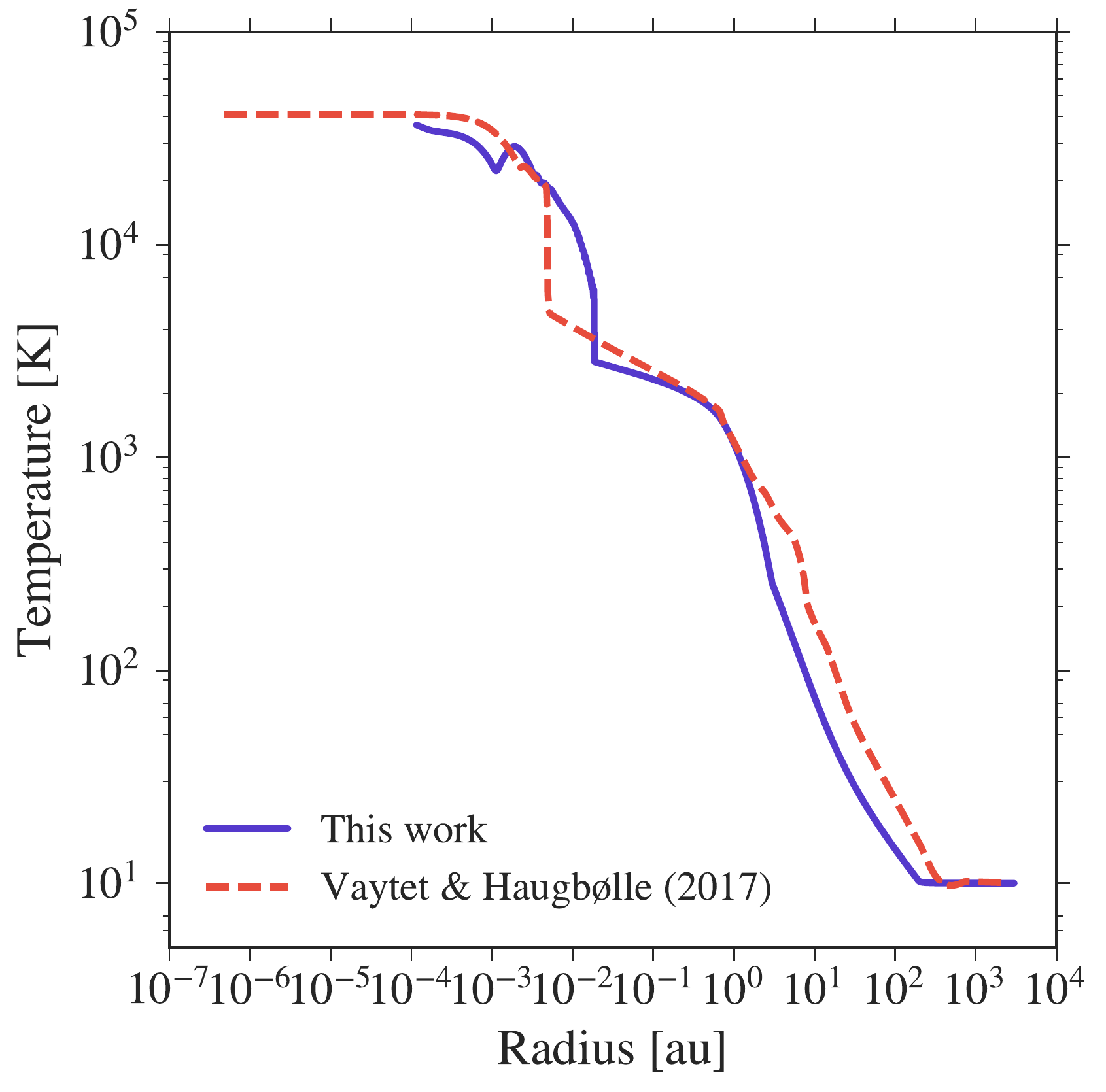}
	\end{subfigure}
	\caption{Comparisons of our results for an initial 1 $M_{\odot}$ cloud core indicated in bluish purple to those by \citet{Vaytet2017} shown using dashed red line. Radial entropy (top), density (middle), and temperature (bottom) profiles are compared at the time when central density $\rho_\mathrm{c}$ in both simulations reach roughly $10^{-1} ~\mathrm{g ~cm^{-3}}$.}
	\label{fig:entropy_comparison}
\end{figure}

In Fig.~\ref{fig:entropy_comparison}, we compare the behaviour of the density and temperature from our 1D simulations (bluish purple) to those by \citet[][dashed red]{Vaytet2017}, at a snapshot when the central density is roughly $10^{-1} ~\mathrm{g ~cm^{-3}}$, since their method is closest to ours. Both studies use the same initial conditions for the collapse of a 1~$M_{\odot}$ cloud core with an initial temperature of 10~K, outer radius of 3000~au, and an initial Bonnor--Ebert sphere like density profile. We note some differences in the profiles which arise due to the different gas EOS (\citealt{Saumon1995} used by \citealt{Vaytet2017} versus \citealt{Dangelo2013} used in this work), opacities, and grid schemes (Lagrangian versus Eulerian). Moreover, both simulations are not compared at the same time in evolution \citep[see also discussion in][Sect.~4.5]{Bhandare2018}. On comparing both studies, these differences in the numerical methods also lead to discrepancies in the second core properties, for example, in the second core radius and enclosed mass. \citet{Vaytet2017} report a smaller second core radius of $\approx$~1~$R_{\odot}$ with an enclosed mass of 2.62~$\times 10^{-3} M_{\odot}$. They expect the second core to grow in size due to further heating and mass accretion from the infalling envelope. In this work, at a similar central density of 9.6~$\times ~10^{-2} ~\mathrm{g ~cm^{-3}}$, we find the second core radius to be $\approx$ 4~$R_{\odot}$ with an enclosed mass of 5.12~$\times 10^{-3} M_{\odot}$.

In our studies, we follow the evolution of the second core for 188.2 years after the second core formation (see Fig.~\ref{fig:SCevolution}). At the final simulation time snapshot, our results indicate that as the central density reaches 0.85~$\mathrm{g ~cm^{-3}}$ the second core radius grows to be much bigger $\approx$~54~$R_{\odot}$ with an enclosed mass of 4.40~$\times 10^{-2} ~M_{\odot}$. As demonstrated in Fig.~\ref{fig:SCevolution}, the expansion and contraction of the evolving second Larson core is controlled by the timescale ratio of Kelvin--Helmholtz contraction versus accretion.

Figure~\ref{fig:entropy_comparison} also shows the comparison of the radial entropy profile. The two peaks seen in the entropy correspond to the positions of the first and second accretion shocks in both 1D studies. In Appendix~\ref{sec:entropy_convergence}, we discuss the dependence of entropy on the numerical resolution. Furthermore, in Appendix~\ref{sec:entropy} we discuss the change in the entropy profile as the cloud core evolves beyond the formation of the second core in our 2D simulations.

\citet{Schonke2011} followed the collapse of a 1~$M_{\odot}$ cloud core using grid-based 2D RHD simulations for up to 240 years after the formation of the second core. They included an initial uniform rotation of their cloud core and could hence also investigate the early phases of disc formation. In order to evolve the system for a longer duration, they replaced the physical domain within 0.7~au with a sink prescription once the second core reached a quasi-static state. Their main goal was to investigate the effect of hydrodynamically driven turbulence using a $\beta$-viscosity prescription. In their models with $\beta = 10^{-3}$ and $\beta = 10^{-2}$ they have described dynamically unstable layers as a consequence of dust evaporation in the central regions within the inner 3~au. They also indicate the occurrence of convection seen via temperature gradients and the presence of a strong vortex in this innermost region.

In the 2D studies presented herein, we observe some short-lived unstable regions within the first accretion shock during the evolution of the \textit{first} core but without any prominent vortices or a convective instability. However, as already discussed in Sect.~\ref{sec:2Dsims}, we observe convection in the outer layers of the \textit{second} core, which eventually evolves to become the protostar. The ability to follow the evolution for 312 years after the formation of the second core allows us to trace the evolution of the eddies and we find the convective instability to grow radially inwards from the second shock as the second core evolves over time. 

\section{Summary and conclusions}
\label{sec:Summary}

The collapse of a molecular cloud core proceeds through an initial isothermal phase leading to the formation of the first hydrostatic core, which undergoes an adiabatic contraction phase. This is followed by the second collapse phase triggered by the dissociation of $\mathrm{H_2}$ once the central temperature rises above 2000~K. The second hydrostatic core is formed as a result of this process once most of the $\mathrm{H_2}$ is dissociated. 

In this work, we investigate the gravitational collapse of molecular cloud cores using 1D and 2D RHD simulations. We include self-gravity and radiation transport. Additionally, the gas EOS takes into account rotational and vibrational degrees of freedom for the $\mathrm{H_2}$ molecules, which start being excited as the cloud core transitions from being effectively monatomic to diatomic, as well as their dissociation and ionisation.

For the 1D studies, we model cloud cores with an initial constant temperature of 10~K, a fixed outer radius of 3000~au, and initial cloud core masses that span a wide range from 0.5~$M_{\odot}$ to 100~$M_{\odot}$. We further expand our collapse studies to 2D with an identical initial setup as in the 1D runs. We model 3000~au non-rotating cloud cores with initial masses of 1~$M_{\odot}$, 5~$M_{\odot}$, 10~$M_{\odot}$, and 20~$M_{\odot}$, thus covering a few cases in the low-, intermediate-, and high-mass regime. We now summarise our key findings from both the 1D and 2D studies, which focus on the formation and evolution of the second Larson core. 

Our 1D studies indicate that the cloud cores with a higher initial mass, collapse faster and form bigger, more massive second cores. We describe the dependence of the second core properties such as the radius, mass, accretion rate, and accretion luminosity on the initial cloud core mass. We discuss the expansion and contraction of the evolving second Larson core, which is controlled by the timescale ratio of Kelvin--Helmholtz contraction versus accretion. The accretion rate in the high-mass regime is found to be much higher than in the low-mass end, as expected. Here we have investigated cases in which the higher-mass molecular cloud cores are gravitationally more unstable than in the low-mass regime. A parameter study using different initial cloud core temperatures and outer radii for various initial cloud masses can be found in our previous work \citep{Bhandare2018}. The results presented herein are consistent with previous core collapse studies in the low-mass regime.

Circumstellar discs form as a consequence of conservation of angular momentum around stars. Currently, evolving the disc for a longer time, that means until the Class 0 phase, is often hindered due to time step restrictions and hence most studies replace the central (second) core with a sink particle. The influence of a sink particle on disc formation has been extensively discussed in \citet{WursterLi2018}. Data from the 1D studies presented here can be used as a lookup-table to compute the evolution of the central object (i.e.~protostar) for a longer duration, within a sink-cell paradigm.

Using our 2D setup for the four non-rotating collapse cases, we follow the evolution of the second core for $\geq$ 100 years after its formation. For the 1~$M_{\odot}$ case, we follow the evolution of the second core for 312 years after its formation. Our 2D studies show that the accretion shock leads to a convective instability in the outer layers of the second core, which grows radially inward over time. Due to the high resolution in our simulations ($\approx 10$ cells per pressure scale height below the shock), that has not been achieved before, we can resolve the convective cells for the first time. For the 1~$M_{\odot}$ case, we find convection being driven from the accretion shock towards the interior of the second Larson core. In contrast to fully convective stars, here, the energy is not generated at the stellar centre, but is provided by the accretion energy from outside the core. Investigating the evolution from these early convective phases due to accretion up to fully convective low-mass stars due to hydrogen burning remains a challenging task for future research in stellar physics.

The origin of magnetic fields in low-mass stars is still a matter of debate. Several studies speculate that the fields are either dominated by primordial or fossil-fields or are replaced by dynamo-generated fields within the first 100 years of evolution. Since young low-mass stars are observed to have strong ($>$~kilogauss) magnetic field strengths, the likelihood of a fossil field could be excluded for cases where the magnetic field amplitude in the second core at birth is found to be less than a kilogauss. In this work, since we already observe convection in the outer layers of the second hydrostatic core, further evolution may enable the generation of a convective-dynamo \citep{Chabrier2006}. This would support the interesting possibility that dynamo-driven magnetic fields may be generated during this very early phase of low-mass star formation.  

We note that the simulations presented here do not account for the effects due to initial cloud rotation and magnetic fields. Even though both of these will affect the evolution of the second core and its properties, we predict that convection seen in our studies should still be generated during this collapse phase in the low-mass regime.

\begin{acknowledgements} 

We thank the anonymous referee for constructive and insightful comments that helped improve the quality of this manuscript. A.B.~would like to thank Beno\^{i}t Commer{\c{c}}on and Hubert Klahr for useful discussions during this work. We would like to thank A.~Mignone and collaborators for their work in developing the open-source PLUTO code. A.B.~is grateful to Thomas M\"uller (Haus der Astronomie, MPIA) for creating the 1~$M_{\odot}$ core collapse movie available online at \url{https://keeper.mpdl.mpg.de/f/f04abdeabdf3472fb56d/} and for his help with the line integral convolution visualisation. R.K.~acknowledges financial support via the Emmy Noether Research Group on Accretion Flows and Feedback in Realistic Models of Massive Star Formation funded by the German Research Foundation (DFG) under grant no.~KU 2849/3-1 and KU 2849/3-2. T.H.~acknowledges support from the European Research Council under the Horizon 2020 Framework Program via the ERC Advanced Grant Origins 83 24 28. G-D.M.~acknowledges the support of the DFG priority program SPP 1992 ``Exploring the Diversity of Extrasolar Planets'' (KU 2849/7-1). Parts of this work have been carried out within the framework of the NCCR PlanetS supported by the Swiss National Science Foundation. The simulations were performed on the ISAAC cluster at the Max Planck Center for Data and Computing in Garching. 
		
\end{acknowledgements} 

\bibliographystyle{yahapj}

\bibliography{Bibliography}

\begin{appendix}

\section{Comparisons for different inner radii}
\label{sec:Rin}

We used an inner radius $R_{\mathrm{in}}$ of $10^{-2}$~au for the simulations discussed in this paper. We note the decrease in temperature at the inner boundary seen in Fig.~\ref{fig:radialprofile} for both the 1D and 2D studies. Due to the high computational expenses, we currently cannot perform tests with an inner radius less than $10^{-2}$~au in 2D. However, given that there are no significant differences in the second core properties between the 1D and 2D simulations with the same initial setup, we compare the 1D results for runs with two different inner radii of $10^{-4}$~au and $10^{-2}$~au. We do not see a drop in innermost regions for the collapse case with $R_{\mathrm{in}} = 10^{-4}$~au. We thus conclude that the decrease seen in case of $R_{\mathrm{in}} = 10^{-2}$~au could be a numerical artefact due to the inner boundary being much closer to the second accretion shock. Besides the temperature decrease in the innermost central region, we do not find any significant differences in the second core properties when comparing both $R_{\mathrm{in}}$ cases. We also confirm that energy conservation is not violated at the inner boundary.

\begin{figure}[!htp]
	\centering
	\begin{subfigure}{0.4\textwidth}
		\includegraphics[width=\textwidth]{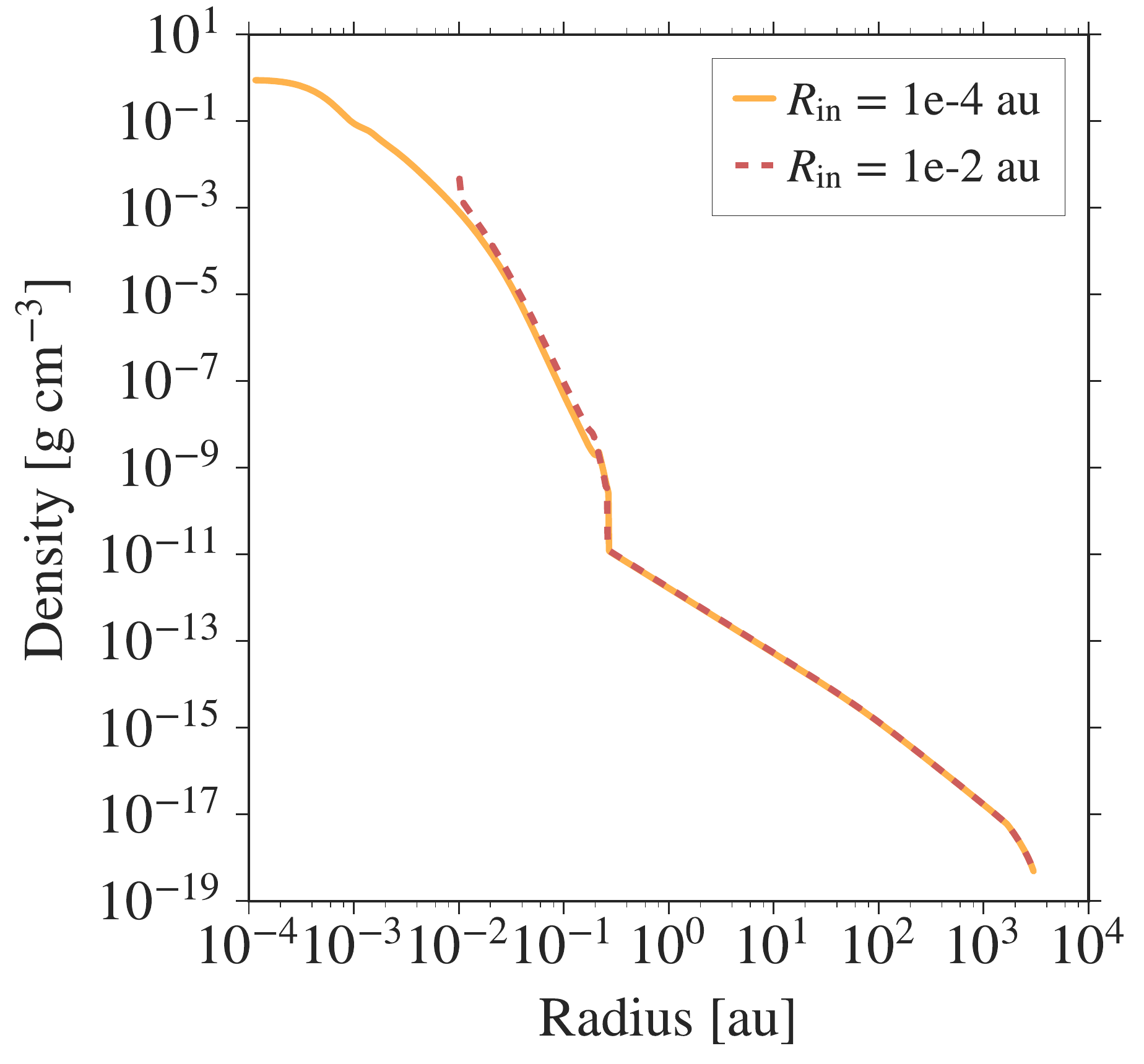}
	\end{subfigure}
	\begin{subfigure}{0.4\textwidth}
		\includegraphics[width=\textwidth]{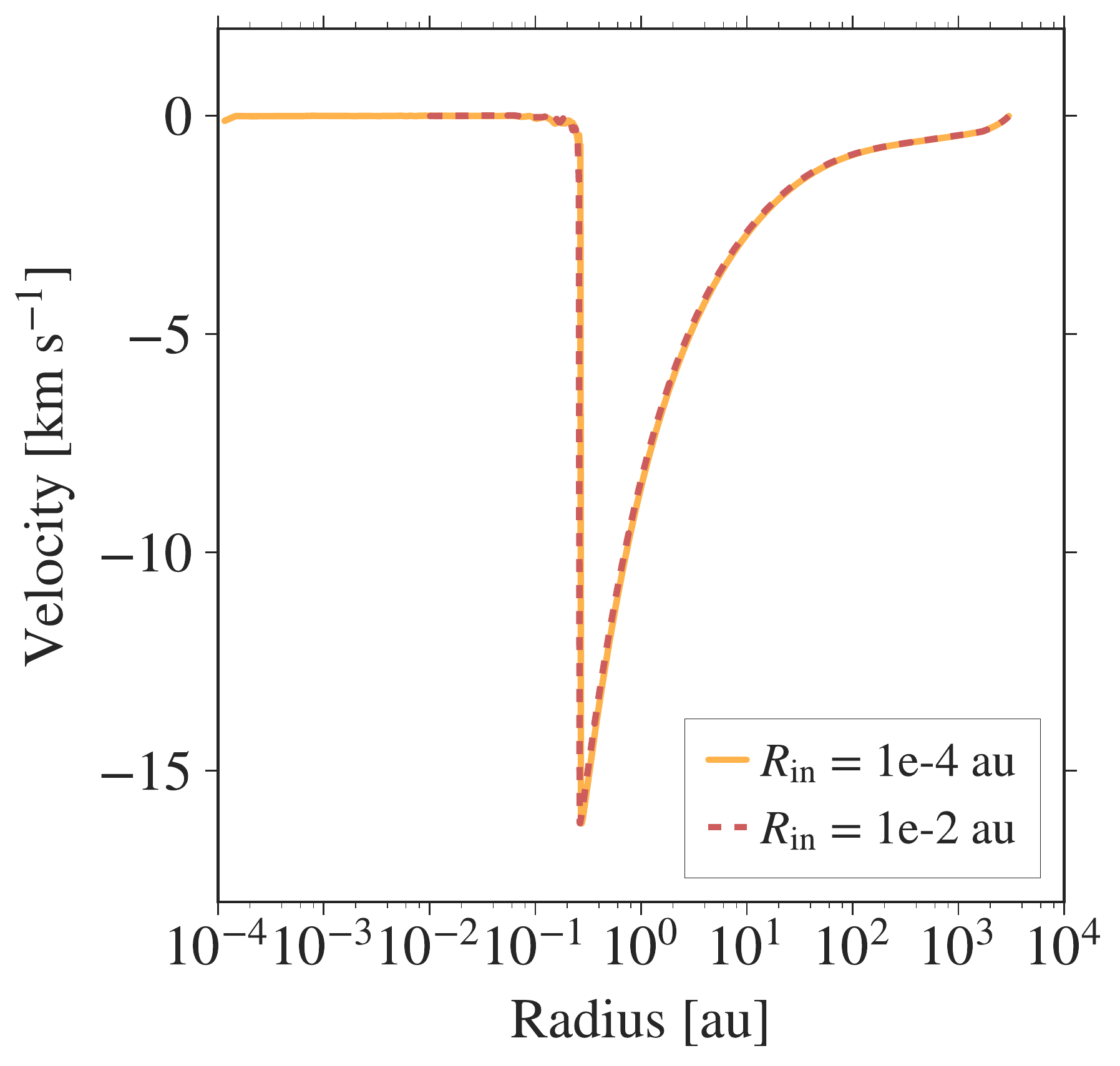}
	\end{subfigure}
	\begin{subfigure}{0.4\textwidth}
		\includegraphics[width=\textwidth]{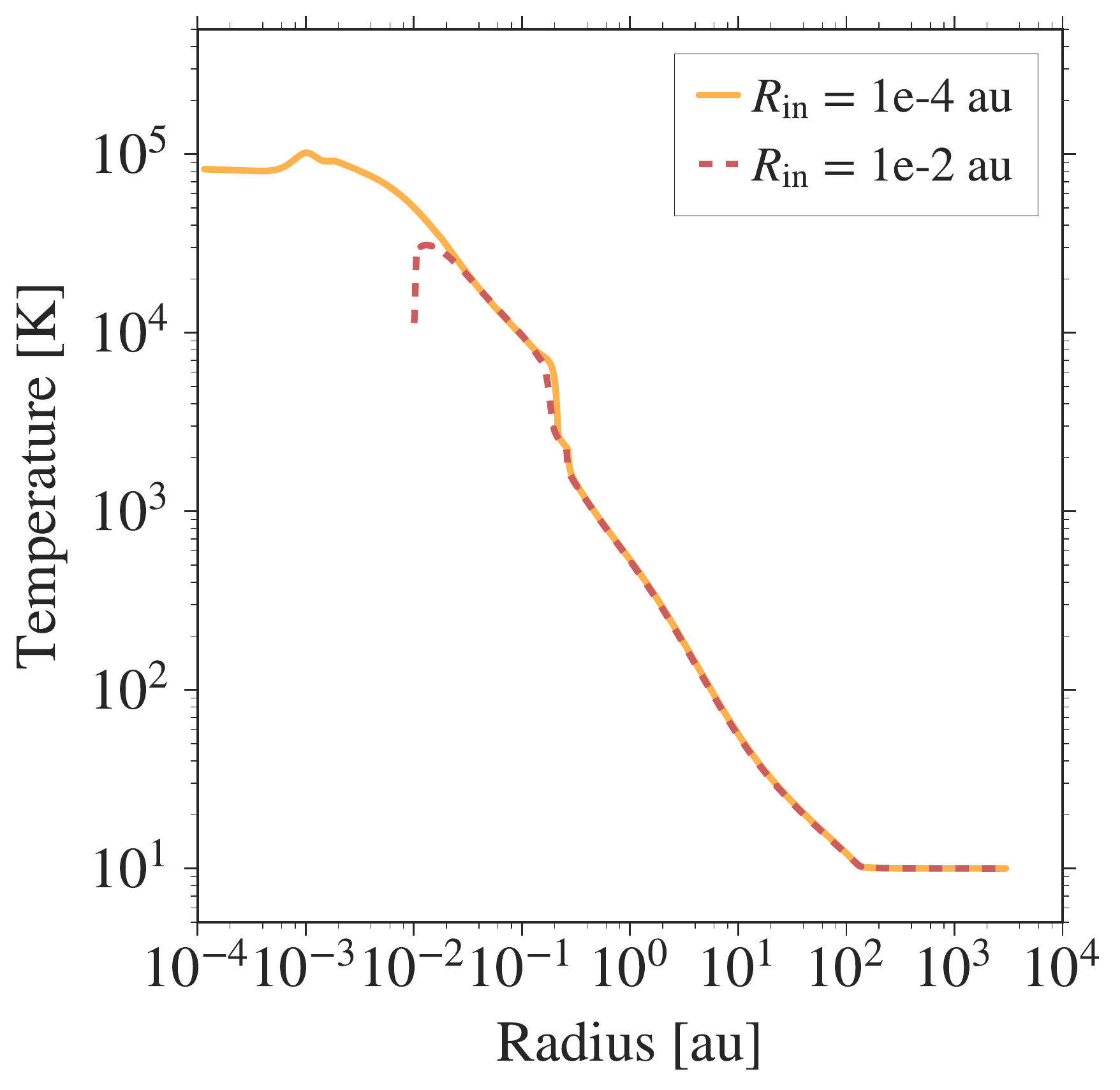}
	\end{subfigure}
	\caption{Radial profiles of the density (top), velocity (middle), and gas temperature (bottom) for an initial 1~$M_{\odot}$ collapsing cloud core from the 1D simulations are shown at a time step after the second core formation. The yellow line indicates an inner radius of $10^{-4}$~au and the dashed red line indicates an inner radius of $10^{-2}$~au. }
	\label{fig:rin}
\end{figure}

\section{Convergence tests}
\label{sec:convergence}

\subsection{Resolution tests for two-dimensional simulations}
\label{sec:resolution}

\begin{figure}[!htp]
	\centering
	\begin{subfigure}{0.4\textwidth}
		\includegraphics[width=\textwidth]{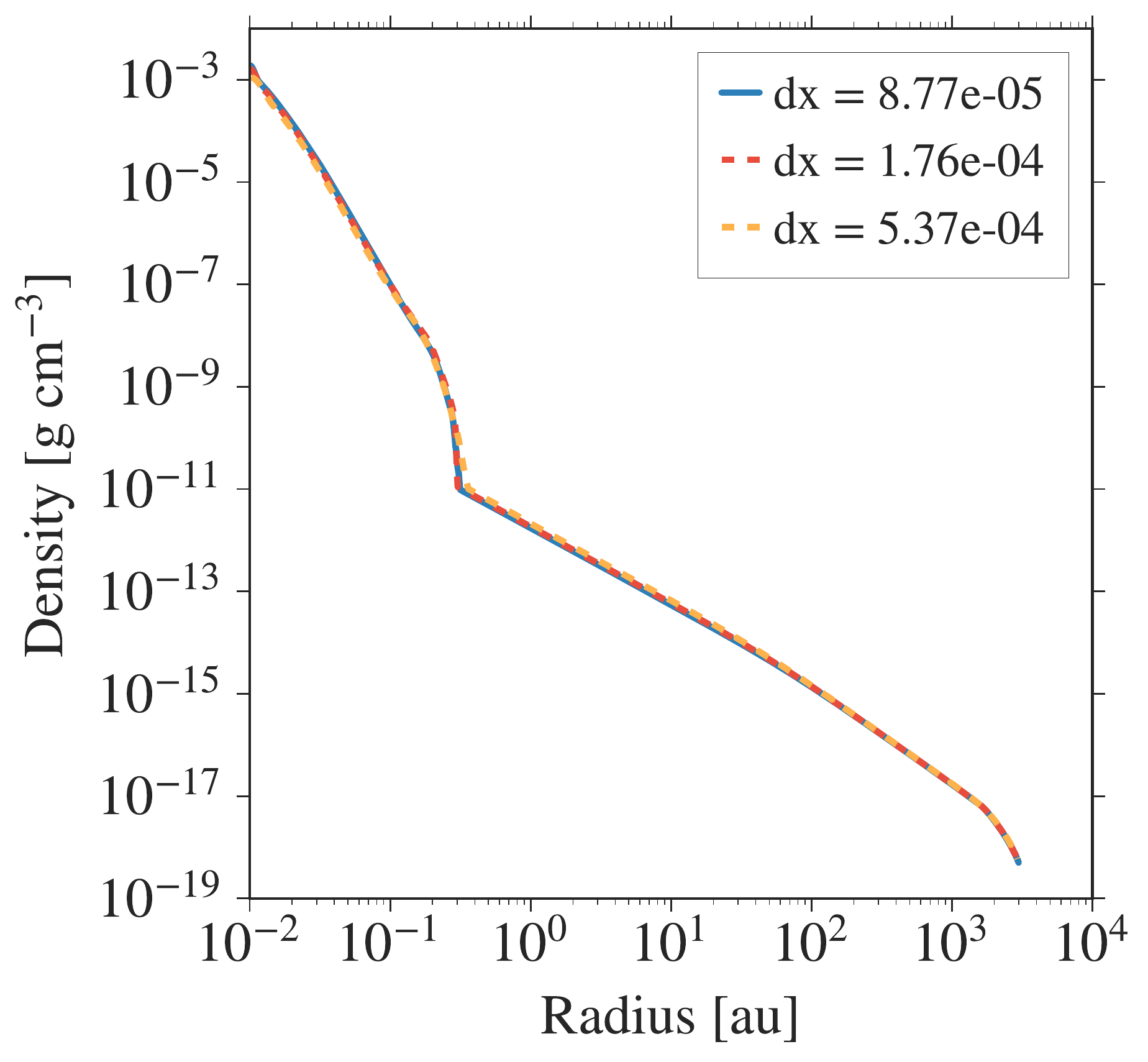}
	\end{subfigure}
	\begin{subfigure}{0.4\textwidth}
		\includegraphics[width=\textwidth]{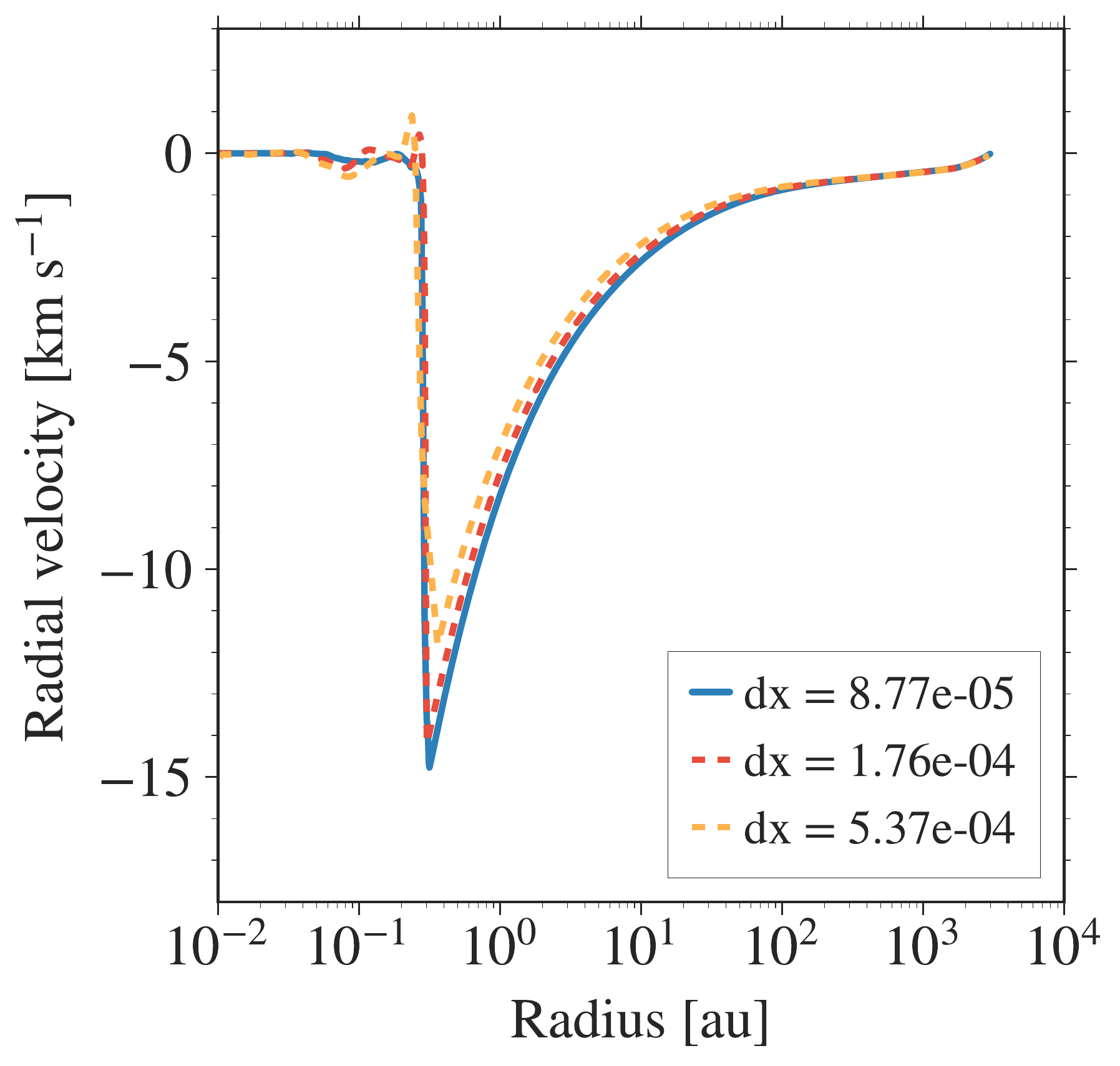}
	\end{subfigure}
	\begin{subfigure}{0.4\textwidth}
		\includegraphics[width=\textwidth]{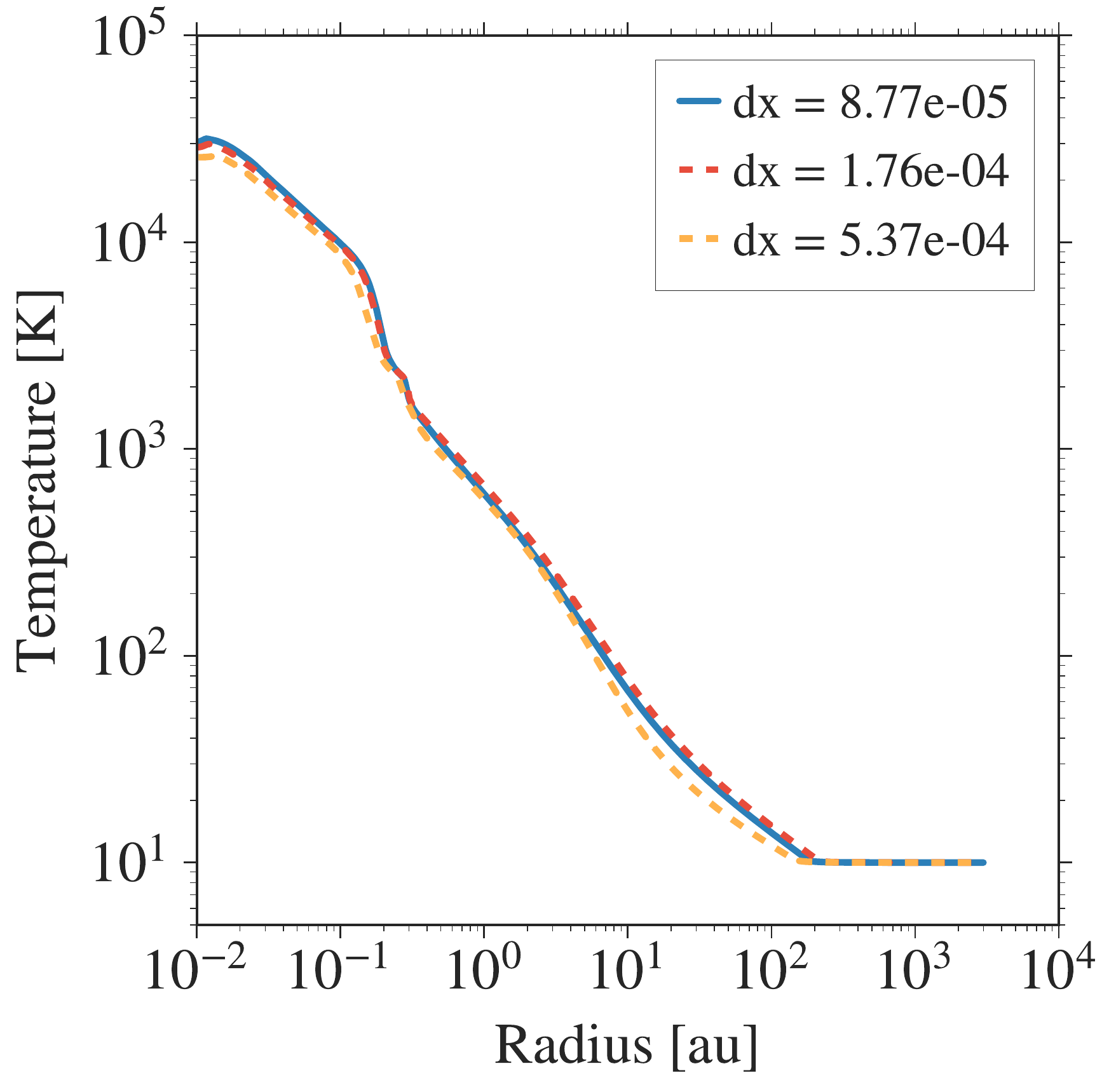}
	\end{subfigure}
	\caption{Polar-angle averaged radial profiles of the density (top), velocity (middle), and gas temperature (bottom) for an initial 1~$M_{\odot}$ collapsing cloud core at an initial temperature $T_\mathrm{0}$ of 10~K are shown at a time step after the second core formation. The different lines indicate the results using various grid resolutions in the 2D simulations.}
	\label{fig:restest}
\end{figure}

\begin{figure}[htp]
	\centering
	\hspace*{-1.5cm}	
	\begin{subfigure}{0.37\textwidth}
		\includegraphics[width=1.2\textwidth]{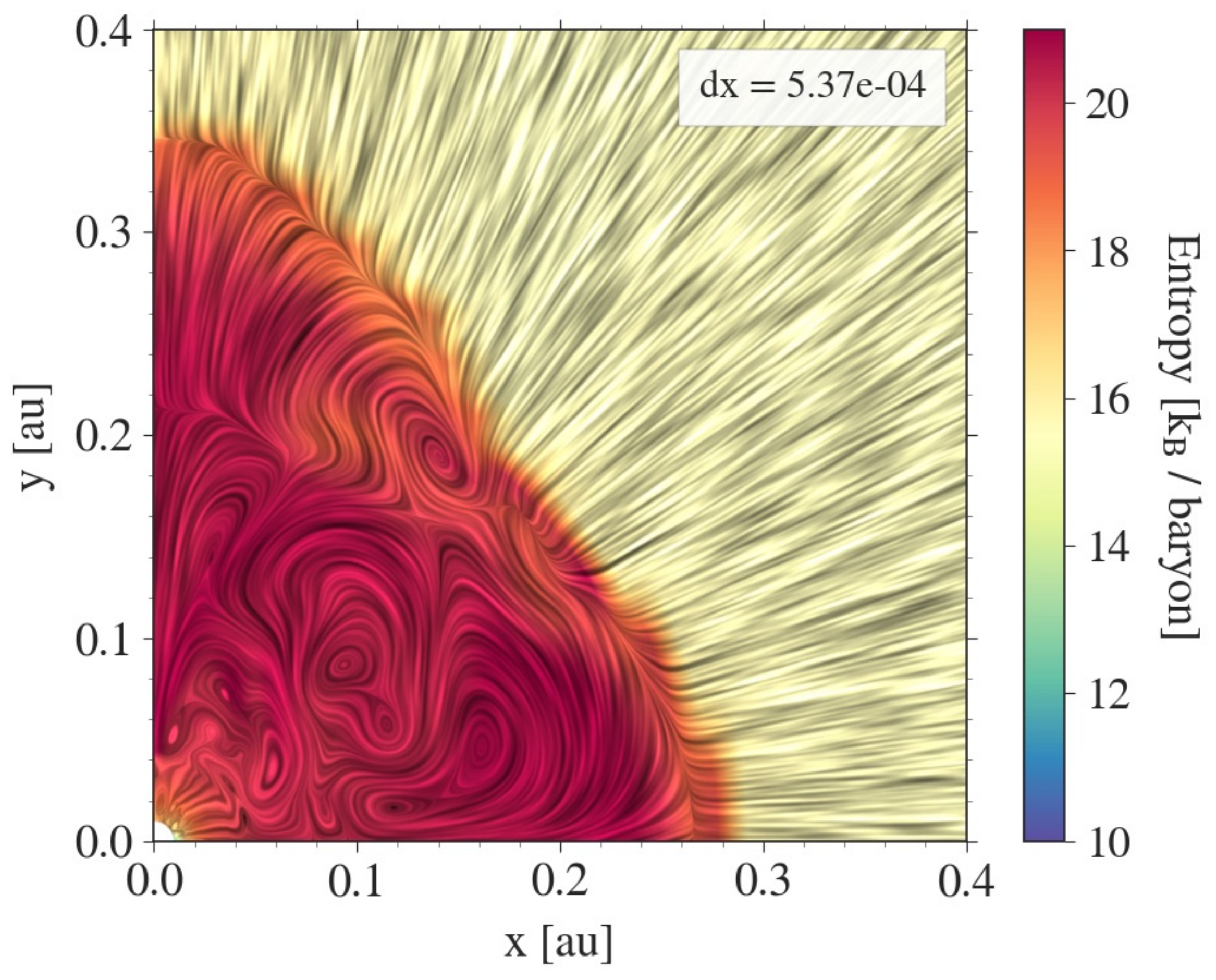}
		\vspace{0.2cm}
	\end{subfigure}
	\hspace*{-1.5cm}	
	\begin{subfigure}{0.37\textwidth}
		\includegraphics[width=1.2\textwidth]{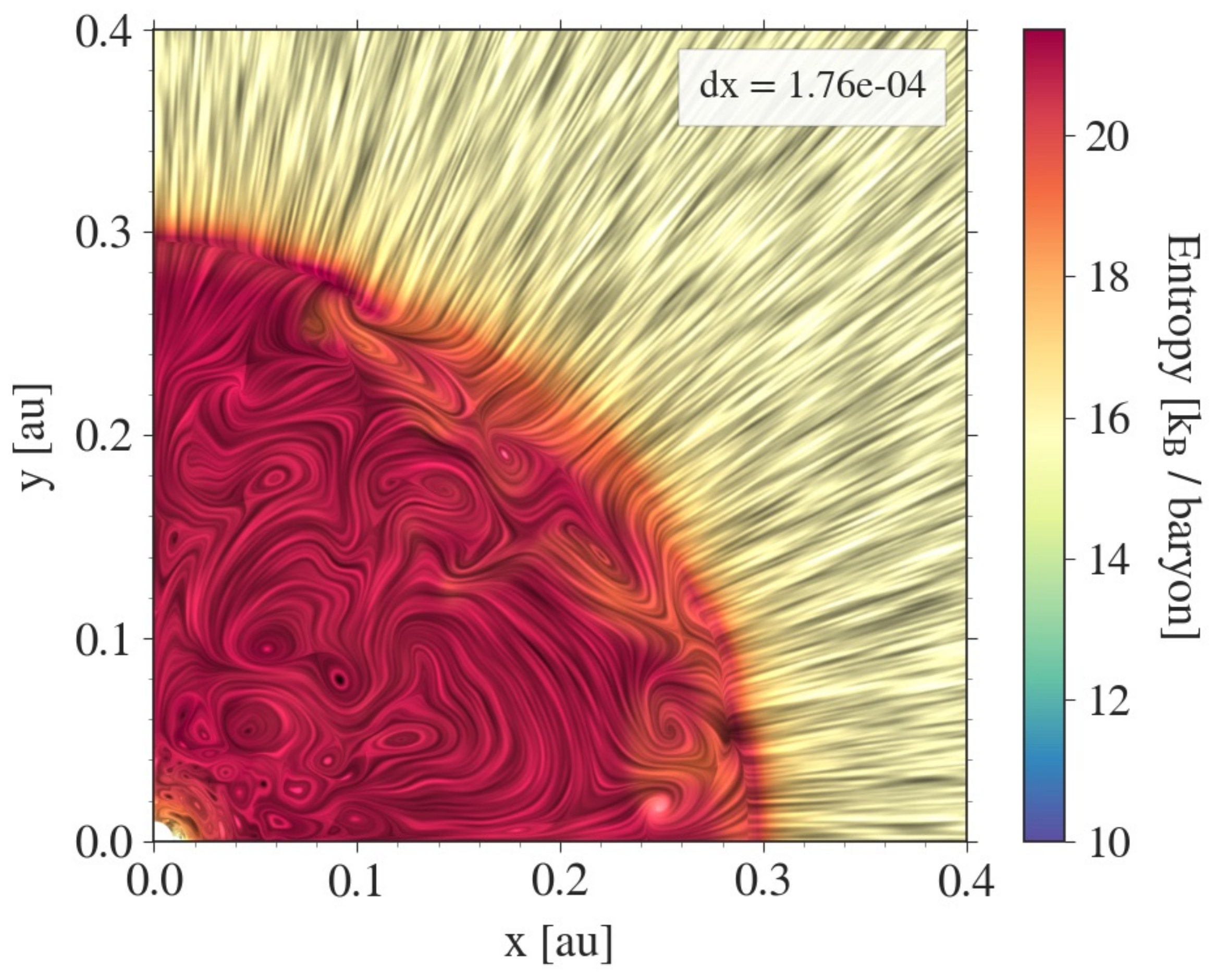}
		\vspace{0.2cm}
	\end{subfigure}
	\hspace*{-1.5cm}	
	\begin{subfigure}{0.37\textwidth}
		\includegraphics[width=1.2\textwidth]{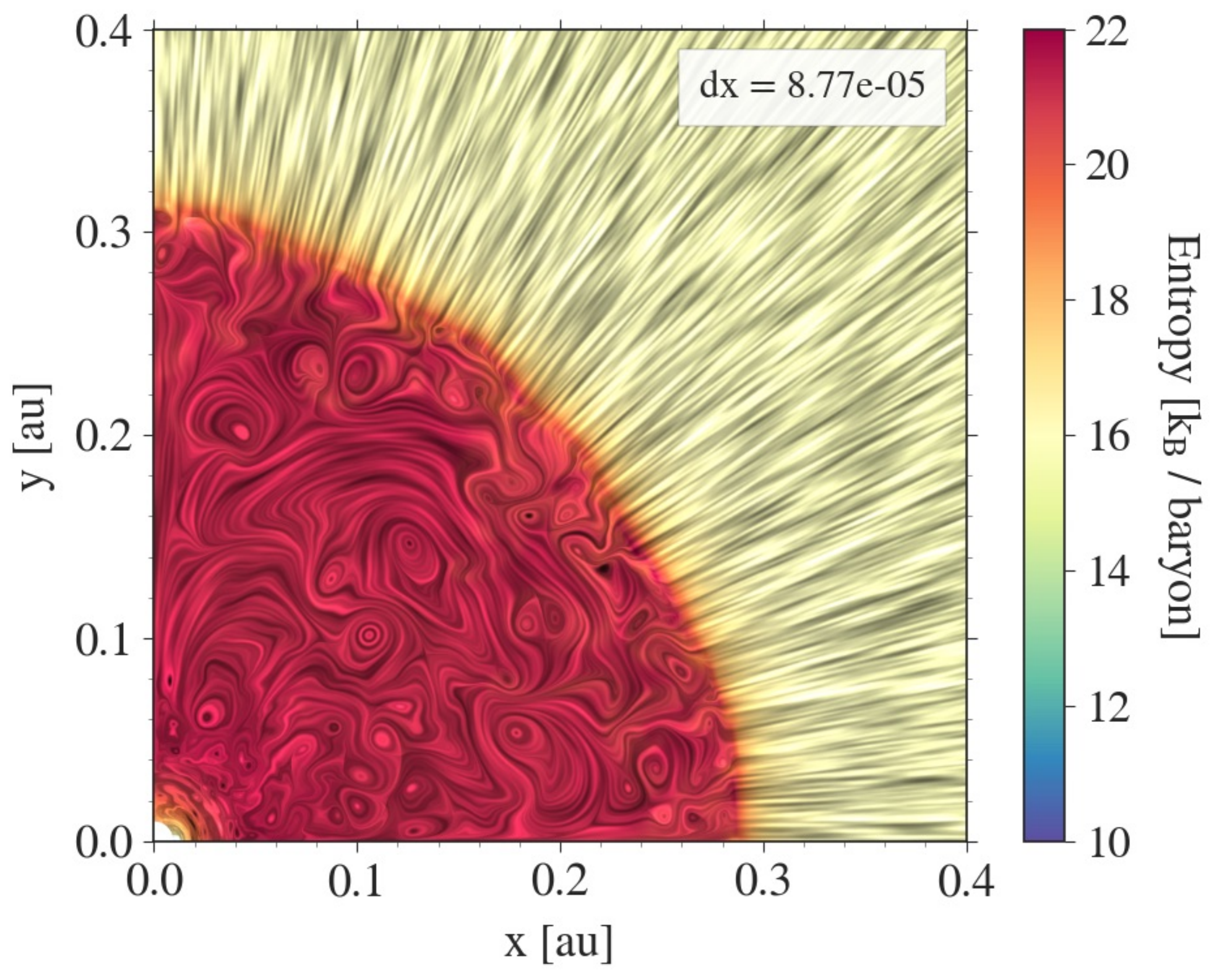}
	\end{subfigure}
	\vspace{0.2cm}
	\caption{Line integral convolution visualisation of the second core formed from the collapse of a 1~$M_{\odot}$ cloud core at an initial temperature $T_\mathrm{0}$ of 10~K and an outer radius of 3000~au. The panels (top to bottom) show results from runs using different resolutions in an increasing order as indicated in the legends. The entropy behaviour is shown at a snapshot when the central densities are roughly similar. For the results presented in this paper, we use the highest resolution, which allows us to resolve the convective eddies within the second core.}
	\label{fig:res_entropy}
\end{figure}

For an initial 1~$M_{\odot}$ cloud core, we perform collapse simulations using three different resolutions shown in Figs.~\ref{fig:restest} and \ref{fig:res_entropy}. In both the figures, we indicate the results at a snapshot in time when the central densities are roughly similar. Polar-angle averaged radial density, velocity, and temperature profiles for the three different resolution runs in Fig.~\ref{fig:restest} do not show any significant differences in the second core properties.    

The main aim of this test was to highlight the importance of using a higher resolution in order to better resolve the eddies indicating a convective instability as discussed in Sect.~\ref{sec:2Dsims}. Resolution differences within the second core are more prominent in the 2D entropy plots seen in Fig.~\ref{fig:res_entropy}. Computational time restrictions prevent us from using an even higher resolution to test the convergence. 

\subsection{Dependence of entropy on resolution}
\label{sec:entropy_convergence}

\begin{figure}[!htp]
	\centering
	\begin{subfigure}{0.4\textwidth}
		\includegraphics[width=\textwidth]{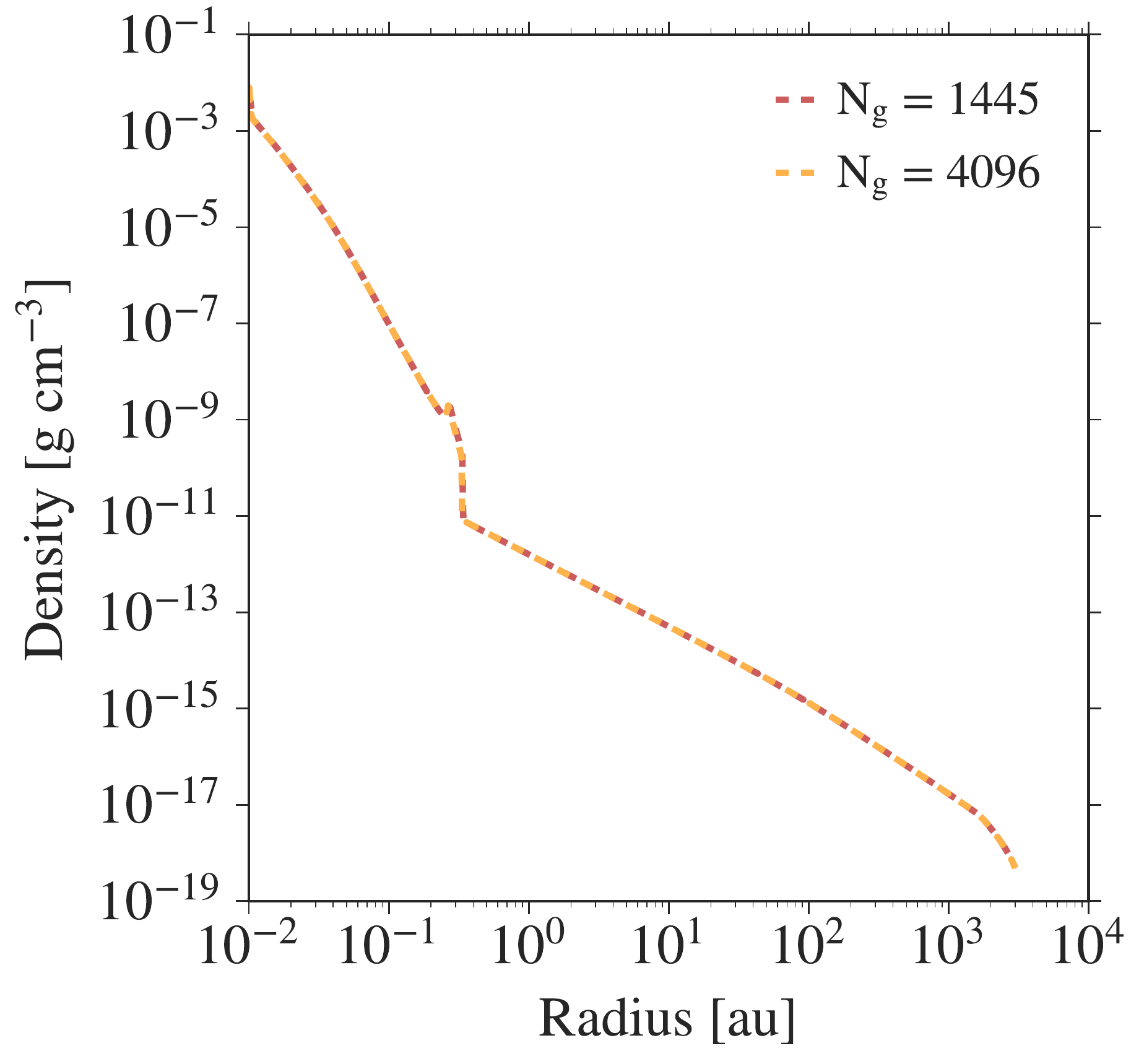}
	\end{subfigure}
	\begin{subfigure}{0.4\textwidth}
		\includegraphics[width=\textwidth]{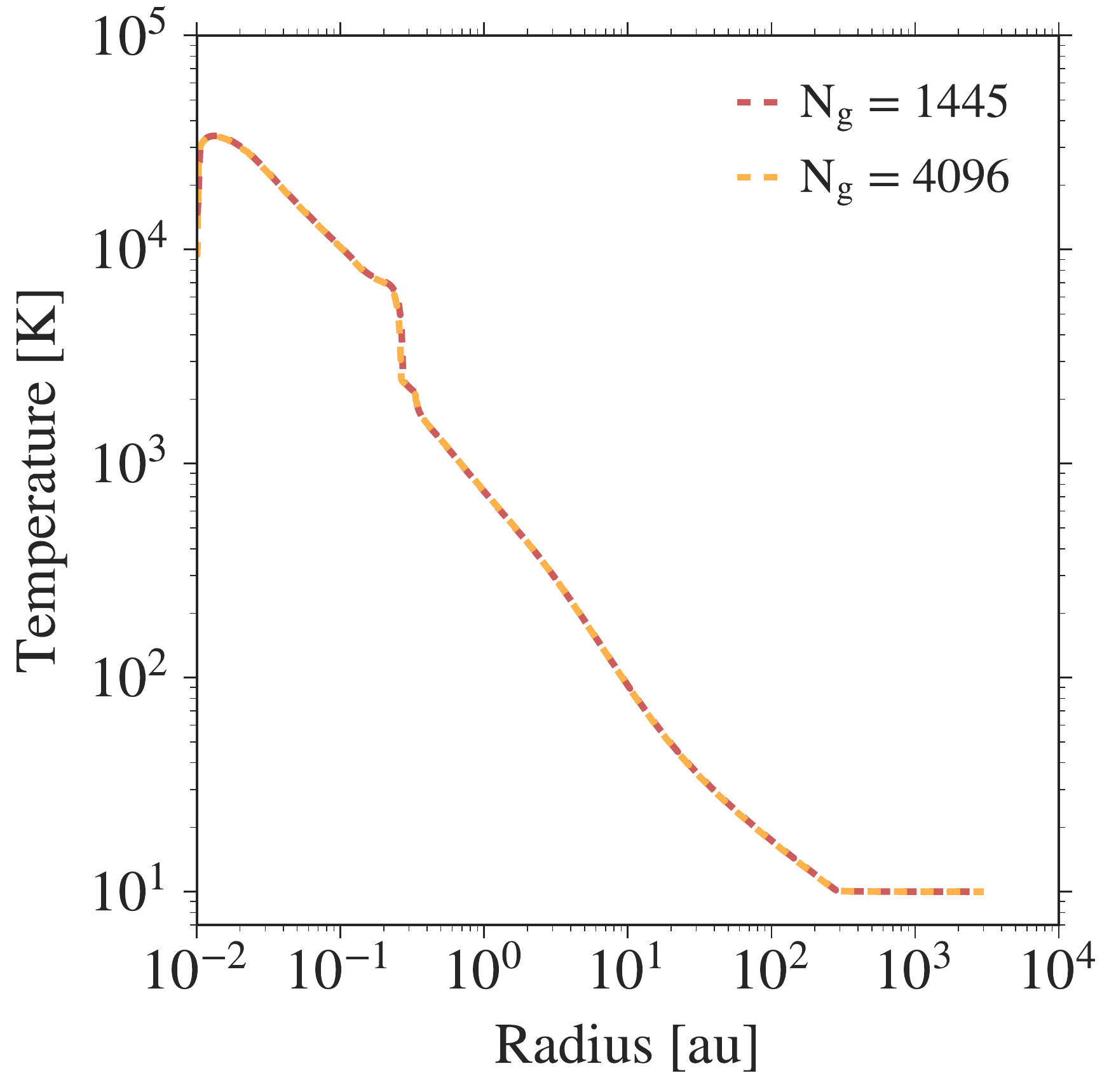}
	\end{subfigure}
	\begin{subfigure}{0.4\textwidth}
		\includegraphics[width=\textwidth]{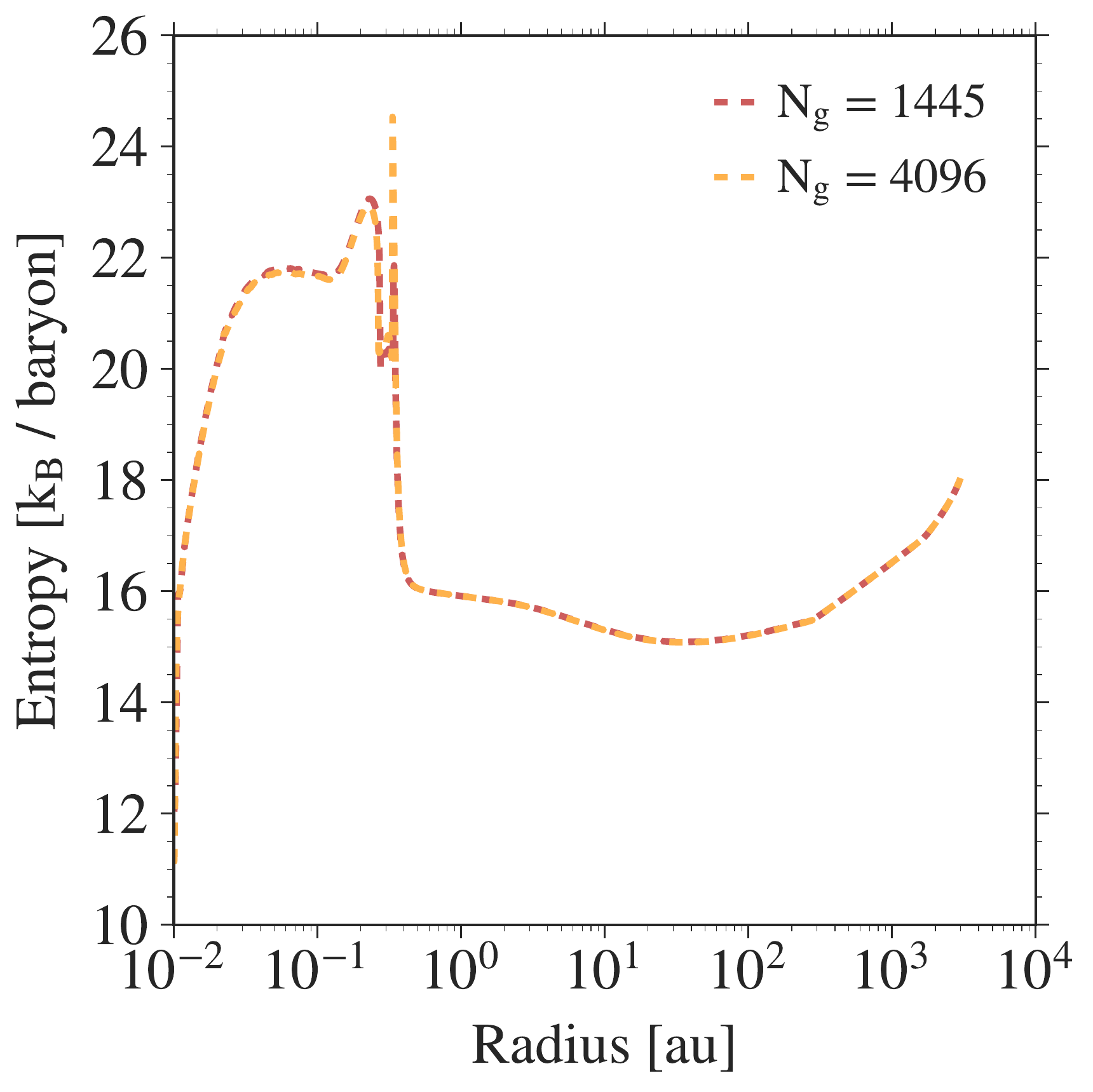}
	\end{subfigure}
	\caption{Radial profiles of the density (top), gas temperature (middle), and entropy (bottom) for an initial 1~$M_{\odot}$ collapsing cloud core at an initial temperature $T_\mathrm{0}$ of 10~K and an outer radius of 3000~au are shown. The yellow and red lines indicate the results using two different grid resolutions in the 1D simulations.}
	\label{fig:1D convergence}
\end{figure}

In Figure~\ref{fig:1D convergence} we compare, for two different resolutions, the radial density, temperature, and entropy profiles from our 1D simulations of the collapse of a 1~$M_{\odot}$ cloud core at an initial temperature $T_\mathrm{0}$ of 10~K and outer radius of 3000~au. The peak in the entropy profile corresponds to the position of the second accretion shock. The two runs show convergence with no significant differences in the behaviour. This suggests that a radial resolution of 1445 cells used in this study is sufficient.  

\section{Entropy}
\label{sec:entropy}

\begin{figure}[t]
	\centering
	\includegraphics[width=0.48\textwidth]{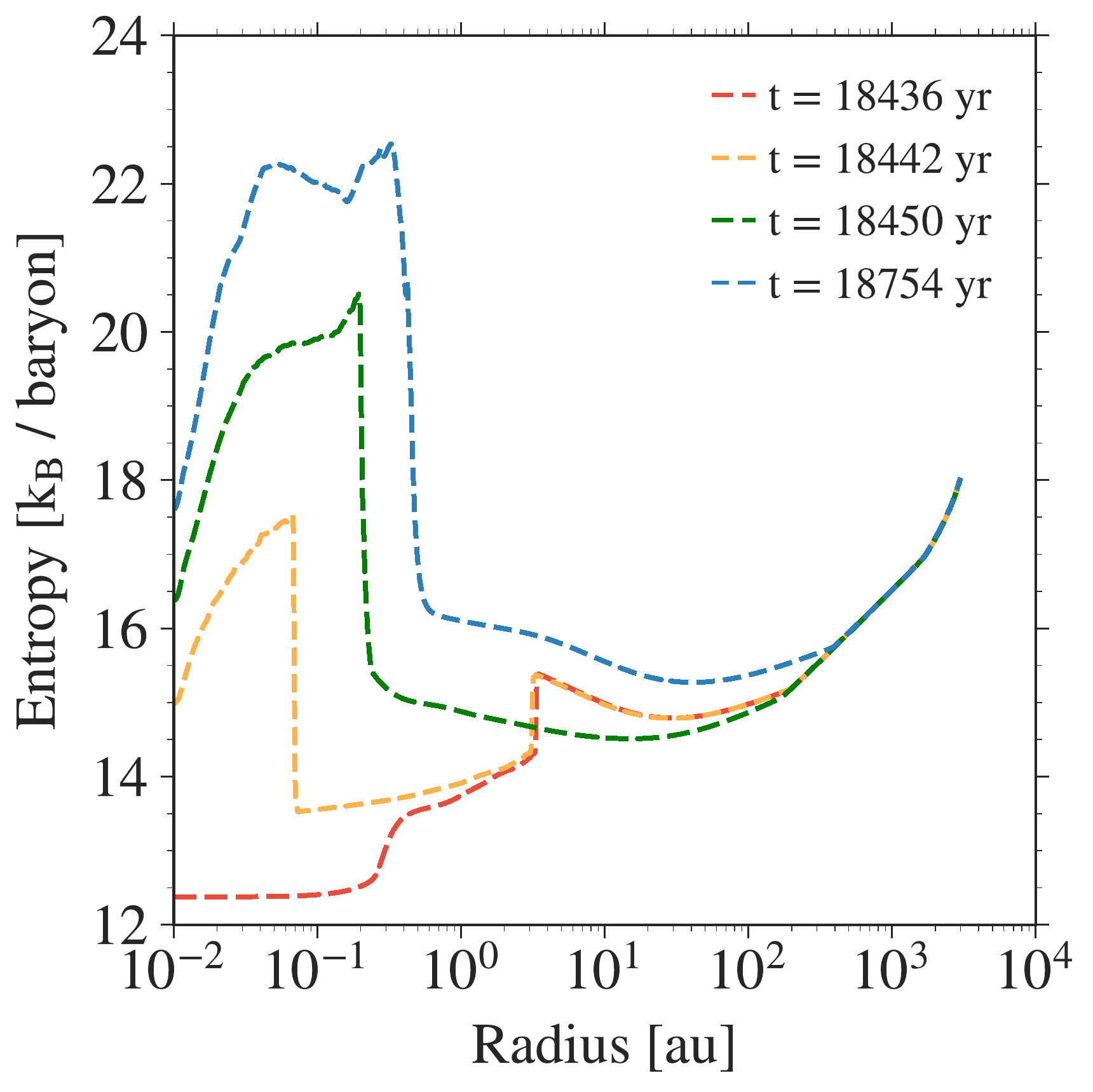}
	\caption{Polar-angle averaged radial entropy profiles are shown at four different time snapshots as a 1~$M_{\odot}$ cloud core transitions through the formation and evolution of the first and second hydrostatic cores. }
	\label{fig:Time_evolution}
\end{figure}

Figure~\ref{fig:Time_evolution} shows the spatial and temporal evolution of the polar-angle averaged radial entropy profile from the 2D simulation of a 1~$M_{\odot}$ collapsing cloud core with an initial temperature $T_\mathrm{0}$ of 10~K and an outer radius of 3000~au. The two peaks in the entropy profile at earlier time snapshots (dashed red and yellow lines) are seen at the positions of the first and second accretion shocks. As the cloud evolves further, material from the first core is accreted onto the second core and the first shock disappears. Figure~\ref{fig:entropy_comparison} shows the comparison of the entropy profile, at an earlier time snapshot, from the 1D studies presented herein with those by \citet{Vaytet2017}.

\end{appendix}

\end{document}